\documentclass[11pt]{article}%
\usepackage[a4paper,margin=3cm,footskip=1cm]{geometry}

\usepackage{layouts}

\usepackage{graphicx} %

\usepackage{xcolor}
\definecolor{light-gray}{gray}{1}
\definecolor{dark-gray}{gray}{0.5}

\usepackage{tikz}
\usetikzlibrary{patterns}
\usetikzlibrary{spy}
\usetikzlibrary{decorations.pathreplacing}

\def\widthVirus{0.9}
\def\heightVirus{0.65}
\def\shiftViruses{1.2} %

\def\heightHost{2.5}
\def\hostMargin{0.2}

\def\gapRandomMaps{0.15}
\def\widthThickLines{1.5pt}

\newcounter{mycounter}
\def\virus#1#2#3{
  \setcounter{mycounter}{0} %

  \foreach \value in {#3}{
  \draw[fill=\value] (#1 + \themycounter * \widthVirus/3, #2) rectangle ++(\widthVirus/3, \heightVirus);
  \addtocounter{mycounter}{1};
  }

  \fill[pattern=north west lines, pattern color=black] (#1, #2) rectangle ++(\widthVirus/3, \heightVirus);
  \fill[pattern=grid, pattern color=black] (#1 + \widthVirus/3, #2 + 0) rectangle ++(\widthVirus/3, \heightVirus);
}

\def\host#1#2#3#4#5#6#7#8#9{

    \draw (#1 - \hostMargin, #2 + \hostMargin) -- ++(0, - 2 * \hostMargin) -- ++(2 * \hostMargin + 4*\shiftViruses + \widthVirus, 0) -- ++(0, 2 * \hostMargin);
    
    \ifnum#8=1
    \draw[->] (#1 - \hostMargin*2, #2 + \heightHost + \heightVirus + \hostMargin) -- (#1 -\hostMargin * 2, #2 - \hostMargin*2);

    \draw (#1 - \hostMargin*2.5, #2 + \heightHost + \heightVirus * 0.5) -- ++(\hostMargin, 0) \ifnum#9=1node[left, font=\small, xshift=-0.2 cm] {past}
    \fi;
    
    \draw (#1 -\hostMargin * 2.5, #2 + \heightVirus * 0.5) -- ++(0.2, 0) \ifnum#9=1node[left, font=\small, xshift=-0.2 cm] {present}
    \fi;
    \fi
    
    \virus{#1}{#2+\heightHost}{#3};
    \virus{#1+\shiftViruses}{#2+\heightHost}{#4};
    \virus{#1+2*\shiftViruses}{#2+\heightHost}{#5};
    \virus{#1+3*\shiftViruses}{#2+\heightHost}{#6};
    \virus{#1+4*\shiftViruses}{#2+\heightHost}{#7};
    
    \foreach \i in {0,1,2,3,4}{
        \draw (#1 + \widthVirus/6 + \i * \shiftViruses,#2 + \heightVirus) -- ++(0,\heightHost-\heightVirus);
        \draw (#1 + \widthVirus/2 + \i * \shiftViruses,#2 + \heightVirus) -- ++(0,\heightHost-\heightVirus);
        \draw (#1 + \widthVirus *5/6 + \i * \shiftViruses,#2 + \heightVirus) -- ++(0,\heightHost-\heightVirus);
    }
    
    \virus{#1}{#2}{#3};
    \virus{#1+\shiftViruses}{#2}{#4};
    \virus{#1+2*\shiftViruses}{#2}{#5};
    \virus{#1+3*\shiftViruses}{#2}{#6};
    \virus{#1+4*\shiftViruses}{#2}{#7};
}

\def\reproduction#1#2{
    \fill[white] (#1 + 2 * \shiftViruses,#2+\heightVirus + \heightHost/2) rectangle ++ (\widthVirus, -\gapRandomMaps);
    \draw (#1 + 2 * \shiftViruses,#2+\heightVirus + \heightHost/2) -- ++ (\widthVirus, 0);

    \ancestralLine{#1}{#2}{2}{1}{#1}{#2}{3}{1}
    \ancestralLine{#1}{#2}{2}{2}{#1}{#2}{3}{2}
    \ancestralLine{#1}{#2}{2}{3}{#1}{#2}{3}{3}

    \virus{#1 + \shiftViruses*2}{#2}{dark-gray,dark-gray,dark-gray}
}

\def\hostreplacement#1#2#3#4#5{
    \fill[white] (#1,#2+\heightVirus + \heightHost/2) rectangle ++ (4*\shiftViruses + \widthVirus, -\gapRandomMaps);
    \draw (#1,#2+\heightVirus + \heightHost/2) rectangle ++ (4*\shiftViruses + \widthVirus, 0);

    \foreach \particle in {1,...,5}{
        \foreach \locus in {1,...,3}{
            \ancestralLine{#3}{#4}{#5}{\locus}{#1}{#2}{\particle}{\locus}
        }
    }

    \foreach \i in {0,1,2,3,4}{
        \virus{#1 + \i*\shiftViruses}{#2}{light-gray, dark-gray, light-gray}
    }
}

\def\recombination#1#2{

    \fill[white] (#1 + 2 * \shiftViruses,#2+\heightVirus + \heightHost/2) rectangle ++ (\widthVirus, -\gapRandomMaps);
    \draw (#1 + 2 * \shiftViruses,#2+\heightVirus + \heightHost/2) -- ++ (\widthVirus, 0);

    \ancestralLine{#1}{#2}{5}{3}{#1}{#2}{3}{3}
    \ancestralLine{#1}{#2}{1}{1}{#1}{#2}{3}{1}
    \ancestralLine{#1}{#2}{1}{2}{#1}{#2}{3}{2}

    \virus{#1+\shiftViruses*2}{#2}{dark-gray,dark-gray,light-gray}
}

\def\reinfection#1#2#3#4#5#6{
    \fill[white] (#4 + \shiftViruses*3,#5+\heightVirus + \heightHost/2) rectangle ++ (\widthVirus, -\gapRandomMaps);
    \draw (#4 + \shiftViruses*3,#5+\heightVirus + \heightHost/2) -- ++ (\widthVirus, 0);

    \ancestralLine{#1}{#2}{#3}{1}{#4}{#5}{#6}{1}
    \ancestralLine{#1}{#2}{#3}{2}{#4}{#5}{#6}{2}
    \ancestralLine{#1}{#2}{#3}{3}{#4}{#5}{#6}{3}

    \virus{#4 + \shiftViruses*3}{#5}{light-gray, dark-gray, light-gray}
}

\def\mutation#1#2{

    \fill[white] (#1,#2+\heightVirus + \heightHost/2) rectangle ++ (\widthVirus, -\gapRandomMaps);
    \draw (#1,#2+\heightVirus + \heightHost/2) -- ++ (\widthVirus, 0);

    \draw[mark=x, mark size=4pt, line width=\widthThickLines] plot coordinates {(#1 + \widthVirus/6,#2+\heightVirus + \heightHost/2- \gapRandomMaps)};

    \draw[line width=\widthThickLines] (#1 + \widthVirus/6,#2+\heightVirus + \heightHost/2- \gapRandomMaps) -- ++(0,-\heightHost/2 + \gapRandomMaps);

    \draw[line width=\widthThickLines] (#1 + \widthVirus/2,#2+\heightVirus + \heightHost/2- \gapRandomMaps) -- ++(0,-\heightHost/2 + \gapRandomMaps);

    \draw[mark=x, mark size=4pt, line width=\widthThickLines] plot coordinates {(#1 + \widthVirus/2,#2+\heightVirus + \heightHost/2- \gapRandomMaps)};

    \draw[line width=\widthThickLines] (#1 + \widthVirus*5/6,#2+\heightVirus + \heightHost/2- \gapRandomMaps) -- ++(0,-\heightHost/2 + \gapRandomMaps);

    \draw[mark=x, mark size=4pt, line width=\widthThickLines] plot coordinates {(#1 + \widthVirus*5/6,#2+\heightVirus + \heightHost/2- \gapRandomMaps)};

    \virus{#1}{#2}{light-gray, dark-gray, dark-gray}
}

\def\ancestralLine#1#2#3#4#5#6#7#8{
    \draw[line width = \widthThickLines] (#1 + #3*\shiftViruses-\shiftViruses + #4*\widthVirus/3 - \widthVirus/6 , #2 + \heightHost) -- ++(0, -\heightHost / 2 - \gapRandomMaps + \heightVirus) -- (#5 + \shiftViruses*#7-\shiftViruses + \widthVirus*#8/3 - \widthVirus/6, #2 + \heightHost-\heightHost / 2 - \gapRandomMaps + \heightVirus) -- ++ (0, -\heightHost / 2 +\gapRandomMaps);
}

\definecolor{inactive}{gray}{0.9}
\definecolor{light-gray}{gray}{1}
\definecolor{dark-gray}{gray}{0.5}

\usepackage{tikz}
\usetikzlibrary{patterns}
\usetikzlibrary{decorations.pathreplacing}

\newcounter{zmycounter}

\usepackage{hyperref}

\usepackage{graphicx}%
\usepackage{amsmath,amssymb,amsfonts,amsbsy}%
\usepackage{amsthm}%
\usepackage{setspace}
\usepackage{mathtools}
\usepackage[normalem]{ulem} 
\usepackage{mathrsfs}%
\usepackage{xcolor}%
\usepackage{dsfont}%
\usepackage[numbers]{natbib}
\usepackage{comment}
\usepackage{bm}
\usepackage{tikz}
\usepackage{nicefrac}

\usepackage{scalerel,stackengine}
\newcommand\pig[1]{\scalerel*[5pt]{\big#1}{%
  \ensurestackMath{\addstackgap[1.5pt]{\big#1}}}}

\newcommand{\unlabel}[1]{}

\newcommand{\defeq}{\vcentcolon=}
\newcommand{\eqdef}{=\vcentcolon}
\newcommand{\dequal}{\stackrel{d}{=}}

\newcommand{\ttaud}[0]{\widetilde{\tau}^\dagger}
\newcommand{\ttau}[0]{\widetilde{\tau}}

\newcommand{\mybreak}{\\[0.7em]}

\theoremstyle{plain}
\newtheorem{theorem}{Theorem}[section]%
\newtheorem{assumptions}[theorem]{Assumptions}%
\newtheorem{assumption}[theorem]{Assumption}%
\newtheorem{lemma}[theorem]{Lemma}

\newtheorem{proposition}[theorem]{Proposition}%

\theoremstyle{definition}
\newtheorem{example}[theorem]{Example}%
\newtheorem{remark}[theorem]{Remark}%
\newtheorem{definition}[theorem]{Definition}%

\newcommand{\cadlag}[0]{c\`{a}dl\`{a}g}

\newcommand{\Dir}[0]{\textnormal{Dir}}
\newcommand{\Geo}[0]{\textnormal{Geo}}
\newcommand{\Exp}[0]{\textnormal{Exp}}

\newcommand{\rep}[0]{{\tt rep}}
\newcommand{\reinf}[0]{{\tt reinf}}
\newcommand{\mut}[0]{{\tt mut}}
\newcommand{\death}[0]{{\tt death}}
\newcommand{\recomb}[0]{{\tt recomb}}
\newcommand{\m}[0]{{\tt m}}

\newcommand{\map}[0]{{\tt map}}

\newcommand{\leb}[0]{\boldsymbol{\lambda}}

\newcommand{\IN}[0]{\mathbb{N}}
\newcommand{\IE}[0]{\mathbb{E}}
\newcommand{\IV}[0]{\mathbb{V}}
\newcommand{\IP}[0]{\mathbb{P}}
\newcommand{\IR}[0]{\mathbb{R}}
\newcommand{\1}[0]{\mathds{1}}

\newcommand{\bfY}[0]{\mathbf{Y}}
\newcommand{\bfX}[0]{\mathbf{X}}
\newcommand{\bfZ}[0]{\mathbf{Z}}
\newcommand{\bfk}[0]{\mathbf{k}}
\newcommand{\bfr}[0]{\mathbf{r}}
\newcommand*\bfell{\ensuremath{\boldsymbol\ell}}
\newcommand*\bfm{\ensuremath{\boldsymbol m}}
\newcommand{\0}[0]{\mathbf{0}}

\newcommand{\jN}[0]{j_N}

\newcommand{\cP}[0]{\mathcal{P}}
\newcommand{\cL}[0]{\mathcal{L}}
\newcommand{\cM}[0]{\mathcal{M}}
\newcommand{\cN}[0]{\mathcal{N}}
\newcommand{\cG}[0]{\mathcal{G}}
\newcommand{\cV}[0]{\mathcal{V}}

\newcommand{\ctVs}[0]{\widetilde{\mathcal{V}}^{\downarrow}}
\newcommand{\cVs}[0]{\cV^{\downarrow}}
\newcommand{\cR}[0]{\mathcal{R}}
\newcommand{\cI}[0]{\mathcal{I}}
\newcommand{\cB}[0]{\mathcal{B}}

\newcommand{\cZ}[0]{\mathcal{Z}}

\newcommand{\tm}[0]{\tilde{m}}
\newcommand{\tn}[0]{\tilde{n}}

\newcommand{\bftZ}[0]{\widetilde{\mathbf{Z}}}

\newcommand{\fDir}[0]{f^\textnormal{Dir}}

\newcommand{\WY}[0]{W_\bfY}
\newcommand{\mset}[1]{\{\!\vert #1 \vert\!\}}
\newcommand{\bigmset}[1]{\big\{\!\big\vert #1 \big\vert\!\big\}}

\newcommand{\svec}[3]{
\begin{smallmatrix}
#1\\[0.3em]
#2\\[0.3em]
#3
\end{smallmatrix}
}

\title{On a Neutral Host-Virus Model with Recombination}
\author{Raphael Eichhorn\footnote{University of Lübeck, Ratzeburger Allee 160, 23562 Lübeck, Germany, \texttt{eichhorn.raphael@gmail.com}} , Cornelia Pokalyuk\footnote{University of Lübeck, Ratzeburger Allee 160, 23562 Lübeck, Germany, \texttt{cornelia.pokalyuk@uni-luebeck.de}}}

\date{\today}

\begin{document}

\maketitle

\begin{abstract}
Motivated by observations in sequence data of herpesviruses, we introduce a multi-locus model for the joint evolution of different genotypes in a virus population that is distributed across a population of hosts.
In the model, virus particles replicate, recombine, and mutate within their hosts at rates that act on different time scales.
Furthermore, virus particles are exchanged between hosts at reinfection events and hosts are replaced by primary infected hosts when they die.
We determine the asymptotic type distribution observed in a single host in the limit of large host and virus populations under asymptotic rate assumptions by tracing back the ancestry of the sample.
The proposed model may serve as a null model for the evolution of  virus populations that are capable of persistence and can be used to estimate the strengths of different evolutionary forces driving genetic diversity, see also~\cite{EichhornGoerzerEtAl}.
\end{abstract}

\section{Introduction}\label{sec1}

A main topic in population genetics is to explain the observed diversity within populations.
Next to selective mechanisms that act on specific genetic types also neutral evolutionary forces leave intricate patterns of diversity within genomic data.
Here, we are interested in the neutral diversity that arises in viral populations.
Differentiating between neutral and selective patterns of diversity is, in the case of viral populations, often of clinical relevance because, e.g., the speed at which advantageous variants spread in a population is faster than the speed of neutral ones.

Our main motivation for analyzing the genetic diversity within viral populations is to explain the diversity patterns observed in \emph{Human Cytomegalovirus}.
This virus persistently infects its hosts, i.e.\ once a host is infected it remains infected until the end of the host's life.
Furthermore, a previous infection does not protect the host from another infection, in particular at different infection events diversity present in the viral population can be imported into single hosts. 

Viral populations are typically spread across a host population. In population genetics terms, this corresponds to a population evolving according to an island model, where the number of islands is equal to the number of hosts. 
Reinfection events are migration events between islands and the event that a host dies and a so far uninfected host is primary infected, meaning infected for the first time, corresponds to an extinction-recolonisation event.

We will consider the evolution of multiple genetic loci of the viral population. Mutations occur at a constant rate. In this manner genetic diversity is maintained in the viral population.

We are interested in the sequence type frequency distribution within hosts of a set of $L$ loci with $K$ different possible types at each locus.
Linkage between loci can be broken up due to a general recombination mechanism.
Our main results show that the sequence type distribution is asymptotically given by a Dirichlet distribution with (potentially random) parameters if virus reproduction and reinfection are acting on the same scale, virus reproduction is faster than host replacement, mutation, and recombination, as well as if recombination is faster than mutation.  

A Dirichlet distribution also arises as the stationary distribution in the multi-type Wright-Fisher diffusion with parent-independent mutation (see, e.g.\,~\cite{Ewens2004}, Section 5.10).
For a proof one can study the asymptotic genealogy of the stationary multi-type Moran model with parent-independent mutation forwards in time and relate it to a Hoppe urn (see~\cite{Hoppe1984} for the urn model,~\cite{Tavare1987} for a construction thereof using Yule processes with immigration,~\cite{DonnellyTavare1987} for the Dirichlet result which exploits the construction) and apply Theorem 9.12 in~\cite{Ethier1986}.
Also the proof of our main result relies on couplings with Hoppe urns and the mentioned construction.

In~\cite{Harris2017Hubbell}, Harris et al.\ consider an islands-mainland model originally going back to Hubbell \cite{Hubbell} and MacArthur and Wilson~\cite{MacArthurWilson}.
They show convergence to a hierarchical Dirichlet process for the species type distribution across the islands when the migration rate and the rate of occurrence of new species are appropriately scaled with the population sizes of the islands and (the larger) mainland, and migration between islands is negligible.
A main ingredient of the proof in \cite{Harris2017Hubbell} (which is based on an investigation of the corresponding generator) is a separation of the time scales on which the dynamics within islands and within the mainland are happening.
Here, we also consider a scaling limit, though in our model only islands and no mainland exist (and migration between islands is not negligible). We take an genealogical approach to analyze our model.
The genealogy of a sample of virions taken from a randomly chosen host can be considered on three different time scales.
This separation of time scales eventually  allows us to proof the claimed convergence to a Dirichlet distribution.

We apply our results in~\cite{EichhornGoerzerEtAl} to fit our model to genetic data sampled from patients infected with Human Cytomegalovirus.

\section{Model and Main Results}\label{sec:model_main_results}
We consider a population of $M \in \IN$ hosts that are infected by a particular persistent virus.
We assume that the number of virus particles (or ``virions'') per host is constantly equal to $N \in \IN$.
We are interested in the allele frequencies at $L \in \IN$ loci. At locus $\ell \in [L]$, the number of possible genotypes (or alleles) is $K_\ell \in \IN$, where $[L] \defeq \{1, ..., L \}$.
For simplicity of notation we assume that $K_\ell = K$ is the same for all loci but our results are easily extended to the case of varying $K_\ell$.
The type of a virus particle is represented by a vector in $[K]^L$.
We model the dynamics of the host-parasite system by a Markov process $\bfX^{M, N, L, K}$ taking values in $S^{[M] \times [N] \times [L]}$ where $S \defeq [K]$.
We keep $L$ and $K$ fixed throughout the article and later consider large viral and host populations, more precisely $N$ and $M= M_N$ will converge to infinity jointly.
Therefore, we simply write $\bfX^N$ and $S^N$ in the following.
A state $x \in S^N$ of the process gives full information about the genetic types of each of the $M \cdot N$ virus particles at all loci.
We use $x_{m,n} \in [K]^L$ to access the type vector of the $n$-th virus particle in the $m$-th host and $x_{m,n,\ell} \in [K]$ to access the type of this particle at the $\ell$-th locus.
Hence, $x_{m,n} = (x_{m,n,1},...,x_{m,n,L})$ can be interpreted as the ``sequence type'' of particle $n$ in host $m$.

We use a random mapping representation to define our model. This means that the dynamics of the model over time is given by a random sequence $(\m_i, t_i)_{i \in \mathbb{Z}}$ of events $\m_i$ and time points $t_i$ at which they happen.
Events change the state of the process and are given by maps $S^N \rightarrow S^N$.
The events we consider are virus replication (at rate $N \gamma_N$), reinfection (at $\lambda_N$), host replacement (at rate $1$), mutation (at rate $\mu_N$) and recombination (at rate $\rho_N$).
All mentioned rates are ``per host'' rates, which will become clear from the definitions below.
We let $m, m_1, m_2, m' \in [M], n, n_1, n_2, n' \in [N]$ and $\ell \in [L]$ and define the following maps:\\

\noindent \textbf{Virus Replication}
The virus particle with index $n_1$ in host $m$ has an offspring which replaces the particle with index $n_2$ in the same host:
\begin{align*}
    \big(\rep_{m, n_1, n_2} (x)\big)_{m',n'} \defeq
    \begin{cases}
        x_{m',n'}   &\text{if $(m',n') \neq (m, n_2)$,}\mybreak
        x_{m, n_1}  &\text{if $(m',n') = (m, n_2)$.}
    \end{cases}
\end{align*}
For a single host this can be interpreted as a reproduction event in a Moran model.
We denote the set of all replication maps by
\begin{equation}\label{eq:g_rep}
    \cG_{\rep}^{N} \defeq \cG_{\rep}^{M,N} \defeq \big\{\rep_{m, n_1, n_2}: m \in [M], n_1, n_2 \in [N] \big\}.
\end{equation}
We suppress the dependence on $M$ of this and of other quantities in the rest of the paper since in the main theorems $M$ will be chosen dependent on $N$.\\

\noindent \textbf{Reinfection}
Virion $n_1$ in host $m_1$ reinfects host $m_2$ by replacing particle $n_2$ in host $m_2$ with a copy of particle $n_1$:
\begin{align*}
    \big(\reinf_{m_1, n_1, m_2, n_2} (x)\big)_{m', n'} \defeq
    \begin{cases}
        x_{m', n'}      &\text{if $(m', n') \neq (m_2, n_2)$,}\mybreak
        x_{m_1, n_1}    &\text{if $(m', n') = (m_2, n_2)$.}
    \end{cases}
\end{align*}

\noindent \textbf{Host Replacement}
Host $m_2$ dies and is instantly replaced by a primary infected host which is infected by virion $n_1$ from host $m_1$. We assume that at this event all virus particles in host $m_2$ die and are replaced by copies of particle $n_1$ from host $m_1$:
\begin{align*}
    \big(\death_{m_1, n_1, m_2} (x)\big)_{m', n'} \defeq
    \begin{cases}
        x_{m', n'}      &\text{if $m' \neq m_2$,}\mybreak
        x_{m_1, n_1}    &\text{if $m' = m_2$.}
    \end{cases}
\end{align*}

\noindent \textbf{Mutation to Type $\bfk$}
We model parent independent mutations.
A virus particle with index $n$ in host $m$ mutates to type $\bfk \in [K]^L$, independent of the type before the mutation:
\begin{align*}
    \big(\mut^{\bfk}_{m, n}(x)\big)_{m', n'} \defeq
    \begin{cases}
        x_{m', n'}      &\text{if $(m', n') \neq (m, n)$,}\mybreak
        \bfk            &\text{if $(m', n') = (m, n)$.}
    \end{cases}
\end{align*}

\noindent \textbf{Recombination}
Let $\bfr$ be a set of loci, i.e.\ $\bfr \subseteq [L] $.
In host $m$, the particle with index $n$ dies and is replaced by a new recombinant offspring from particles $n_1$ and $n_2$. The recombinant carries the types of parent $n_1$ at all loci in $\bfr$ and the types of parent $n_2$ at the loci not in $\bfr$:
\begin{align*}
    \big(\recomb_{m, n, n_1, n_2}^\bfr (x)\big)_{m', n', \ell} \defeq
    \begin{cases}
        x_{m', n', \ell}      &\text{if $(m', n') \neq (m,n)$,}\mybreak
        x_{m, n_1, \ell}       &\text{if $(m', n') = (m, n)$ and $\ell' \in \bfr$,}\mybreak
        x_{m, n_2, \ell}       &\text{if $(m', n') = (m, n)$ and $\ell' \notin \bfr$.}
    \end{cases}
\end{align*}
We define the sets $\cG^{N}_{\reinf}$,  $\cG^{N}_{\death}$, $\cG^{N}_{\mut}$ and $\cG^{N}_{\recomb}$ analogously to (\ref{eq:g_rep}).

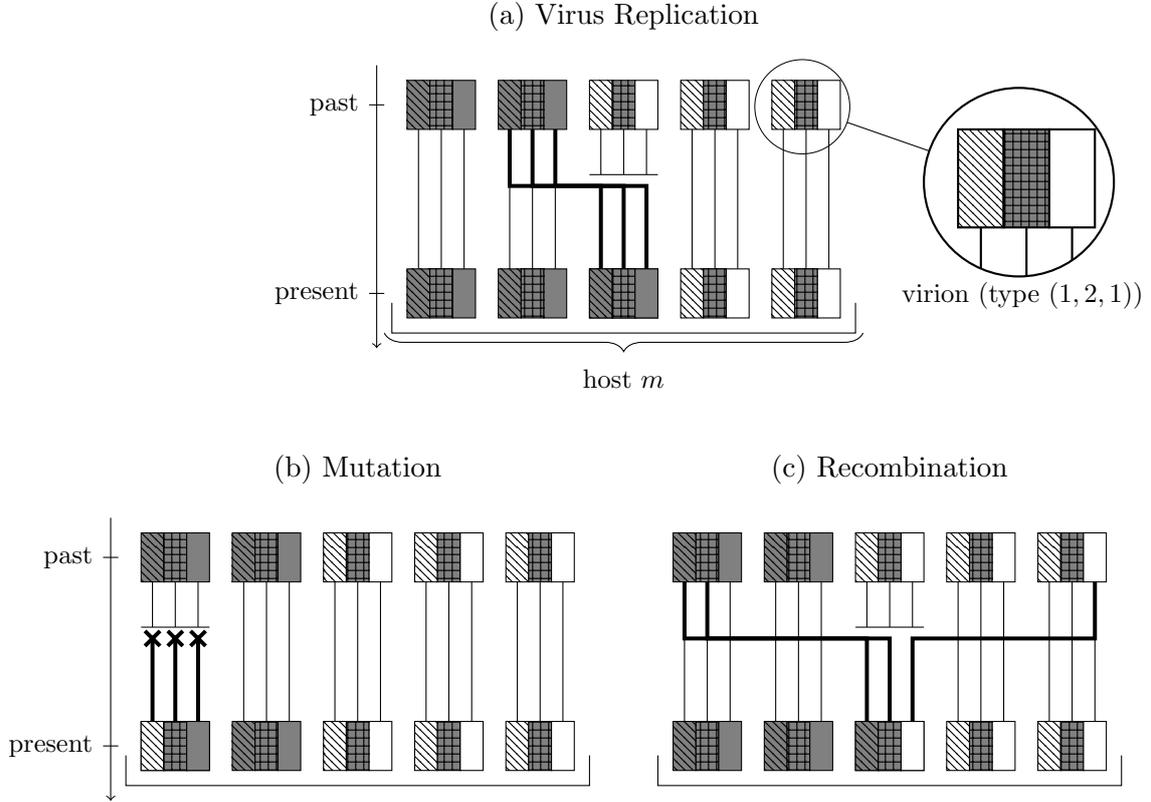
\begin{figure}[!htb]
    \centering
    \begin{tikzpicture}[spy using outlines={circle, magnification=2, size=2.5cm, connect spies}]

    \node at (3.5 + 2*\shiftViruses +\widthVirus/2, 10) {(a) Virus Replication};
    \draw[decorate,decoration={brace,amplitude=7pt,mirror}] (3.5-\hostMargin * 1.5,6 - \hostMargin) -- ++(\shiftViruses * 4 + \widthVirus + \hostMargin * 3,0) node[midway,yshift=-0.6cm,font=\small] {host $m$};

    \node at (2*\shiftViruses +\widthVirus/2, 4) {(b) Mutation};

    \node at (7 + 2*\shiftViruses +\widthVirus/2, 4) {(c) Recombination};

    \host{3.5}{6}{dark-gray,dark-gray,dark-gray}{dark-gray,dark-gray,dark-gray}{light-gray,dark-gray,light-gray}{light-gray,dark-gray,light-gray}{light-gray,dark-gray,light-gray}{1}{1};

    \host{0}{0}{dark-gray,dark-gray,dark-gray}{dark-gray,dark-gray,dark-gray}{light-gray,dark-gray,light-gray}{light-gray,dark-gray,light-gray}{light-gray,dark-gray,light-gray}{1}{1};

    \host{7}{0}{dark-gray,dark-gray,dark-gray}{dark-gray,dark-gray,dark-gray}{light-gray,dark-gray,light-gray}{light-gray,dark-gray,light-gray}{light-gray,dark-gray,light-gray}{0}{0};

    \recombination{7}{0};
    \mutation{0}{0}
    \reproduction{3.5}{6}

    \spy on (8.7, 8.8) in node [left] at (12.8,7.8);
    \node[font=\small] at (11.6, 6.3) {virion (type $(1,2,1)$)};

\end{tikzpicture}
    \caption{Illustration of the maps which involve only a single host. In the pictures, the number of hosts $M$ is arbitrary (because only a single host of the total population is shown for every type of event). The number of virus particles per host is $N=5$ and the number of loci is $L=3$.
    At each locus there are two possible types, light ($1$) or dark ($2$), so $K=2$. In picture (a), the map $\rep_{m, 2, 3}$ is shown. That is, in the depicted host $m$, the virus particle at index $2$ splits into two. The offspring particle replaces the old virus particle at index $3$. In picture (b) the map $\mut^{(1,2,2)}_{m, 1}$ is shown. In picture (c) the map $\recomb_{m, 3, 1, 5}^{\{1,2\}}$ is shown: the particle at index 3 dies and is replaced by a recombinant offspring of particles $1$ and $5$ that carries the types of particle $1$ at loci $1$ and $2$ and the type of particle $5$ at locus $3$.}
    \label{fig:within_host}
\end{figure}

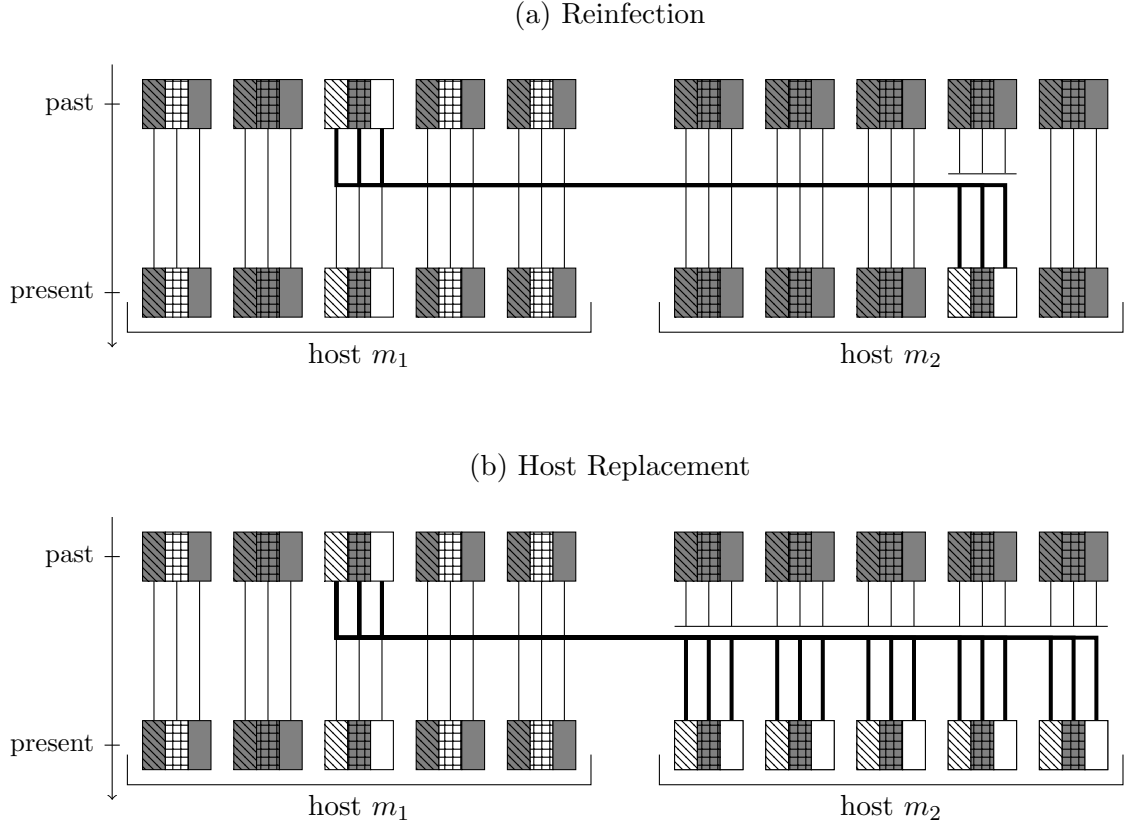
\begin{figure}[!htb]
    \centering
    \begin{tikzpicture}

    \host{7}{6}{dark-gray,dark-gray,dark-gray}{dark-gray,dark-gray,dark-gray}{dark-gray,dark-gray,dark-gray}{dark-gray,dark-gray,dark-gray}{dark-gray,dark-gray,dark-gray}{0}{0};

    \host{0}{6}{dark-gray,light-gray,dark-gray}{dark-gray,dark-gray,dark-gray}{light-gray,dark-gray,light-gray}{dark-gray,light-gray,dark-gray}{dark-gray,light-gray,dark-gray}{1}{1};

    \host{7}{0}{dark-gray,dark-gray,dark-gray}{dark-gray,dark-gray,dark-gray}{dark-gray,dark-gray,dark-gray}{dark-gray,dark-gray,dark-gray}{dark-gray,dark-gray,dark-gray}{0}{0};

    \host{0}{0}{dark-gray,light-gray,dark-gray}{dark-gray,dark-gray,dark-gray}{light-gray,dark-gray,light-gray}{dark-gray,light-gray,dark-gray}{dark-gray,light-gray,dark-gray}{1}{1};

    \hostreplacement{7}{0}{0}{0}{3};
    \reinfection{0}{6}{3}{7}{6}{4}

   \node at (3.3 + 2*\shiftViruses +\widthVirus/2, 4) {(b) Host Replacement};
   \node at (3.3 + 2*\shiftViruses +\widthVirus/2, 10) {(a) Reinfection};

   \node at (0 + \widthVirus/2 + 2*\shiftViruses, -0.5) {host $m_1$};
   \node at (7 + \widthVirus/2 + 2*\shiftViruses, -0.5) {host $m_2$};
   
   \node at (0 + \widthVirus/2 + 2*\shiftViruses, 5.5) {host $m_1$};
   \node at (7 + \widthVirus/2 + 2*\shiftViruses, 5.5) {host $m_2$};

\end{tikzpicture}
    \caption{Illustration of the maps which involve two hosts. In Picture (a) the map $\reinf_{m_1, 3, m_2, 4}$ is shown. Here, the particle with index $3$ in host $m_1$ splits into two and the offspring particle replaces the particle with index $4$ in host $m_2$. In Picture (b) the map $\death_{m_1, 3, m_2}$ is shown. All particles in host $m_2$ are replaced by offspring of particle $3$ in host $m_1$.}
    \label{fig:between_host}
\end{figure}

The maps introduced above can be interpreted visually in a graphical representation of branching and coalescing lines (for reproduction, reinfection and recombination), blocking symbols (for death of a particle or host) and stars (for mutations). We give illustrations for all effects in Figures \ref{fig:within_host} and \ref{fig:between_host}.

The generator of the main process studied in this work is defined to be the linear operator acting on functions $f:S^N \rightarrow \IR$ given by
\begin{align*}%
    G^{N}f(x)    \defeq  &\sum_{m = 1}^M \sum_{n_1 = 1}^N \sum_{n_2 =1 }^N                              \frac{\gamma_N}{N}                  \Big(  f \big(\rep_{m, n_1, n_2}             (x) \big) - f\big(x\big)\Big)\\
                        + &\sum_{m_1=1}^M \sum_{n_1=1}^N \sum_{m_2=1}^M \sum_{n_2=1}^N                  \frac{\lambda_N}{N^2M}              \Big(  f \big(\reinf_{m_1, n_1, m_2, n_2}    (x) \big) - f\big(x\big)\Big)\\
                        + &\sum_{m=1}^M \sum_{n=1}^N \sum_{n_1=1}^N \sum_{n_2=1}^N \sum_{\bfr \in \cR}  \frac{\rho_N p^\bfr_\recomb}{N^3}   \Big(  f \big( \recomb_{m, n, n_1, n_2}^\bfr (x) \big) - f\big(x\big)\Big)\\
                        + &\sum_{m = 1}^M \sum_{n = 1}^N \sum_{\bfk \in [K]^L}                          \frac{\mu_N \nu (\{\bfk\})}{N}      \Big(  f \big(\mut^{\bfk}_{m, n}          (x) \big) - f\big(x\big)\Big)\\
                        + &\sum_{m_1=1}^M \sum_{n_1=1}^N \sum_{m_2=1}^M                                 \frac{1}{MN}                        \Big(  f \big( \death_{m_1, n_1, m_2}        (x) \big) - f\big(x\big)\Big),
\end{align*}
where the set $\cR \subseteq \cP \left([L]\right)$ denotes the possible recombinations and $p^\bfr_\recomb > 0$ is the probability that a recombination is an $\bfr$-recombination. The probability that a mutation results in type $\bfk$ is given by $\nu(\{\bfk\})$.%

We need to make some assumptions on the rates of the different events and on the recombination mechanism.
In particular, recombination should be strong enough to break linkage between loci.
This condition is formalized in the following:

\begin{assumption}[Possible Recombinations]\label{assumptions:recombination}
    For all loci $\ell_1 \neq \ell_2 \in [L]$ there exists $\bfr \in \cR$ such that $\big\lvert \{\ell_1, \ell_2\} \cap \bfr \big\rvert = 1$. %
\end{assumption}

\begin{example}[Recombinations]
    An example for a set of recombinations $\cR$ satisfying Assumptions \ref{assumptions:recombination} is the set of single crossover recombinations given by
    \begin{equation*}
        \cR = \big\{ \{1\}, \{1,2\}, \{1,2,3\}, ..., \{1,2, ..., L -1\} \big\}.
    \end{equation*}
    If $L>k\geq 1$, then another set of recombinations satisfying the assumption is the one in which exactly $k$ loci are inherited from one parent:
    \begin{equation*}
        \cR = \big\{ \bfr \in \cP \big( [L] \big) : |\bfr| = k \big\}.
    \end{equation*}
\end{example}

To analyze the sequence type frequencies we will assume that the different evolutionary forces like reproduction and recombination act on different time scales.

\begin{assumptions}[Asymptotic Rates]\label{assumptions:main_rates}
We assume that $N \to \infty$ and $M_N \to \infty$ while satisfying the following asymptotics:
\begin{enumerate}
    \item \label{ass:rep_reinf} Virus reproduction and reinfection act on the same time scale: we assume that there exists a constant $\theta>0$ such that%
    \begin{equation*}
        \lim_{N\rightarrow \infty} \frac{\lambda_N}{\gamma_N} = \theta.
    \end{equation*}
    \item \unlabel{ass:timescales} Virus reproduction and reinfection act on a faster time scale than host replacement, mutation and recombination, more precisely
    \begin{equation*}
        N, \mu_N, \rho_N \in o(\gamma_N).
    \end{equation*}
    \item \unlabel{ass:recomb} Recombination acts on a faster times scale than mutation, more precisely %
    \begin{equation*}
        \mu_N \in o(\rho_N).
    \end{equation*}
    \item \unlabel{ass:mut} The mutation rate is large enough such that some genetic diversity can be observed, more precisely 
    \begin{equation}\label{assumptions:main_rates_case_one}
         \lambda_N \in o(M \mu_N) \tag{i}
    \end{equation}
    or
    \begin{equation}\label{assumptions:main_rates_case_two}
        \lim_{N \to \infty} \frac{\lambda_N}{M \mu_N} = C \tag{ii}
    \end{equation}
    for some $C>0$.
\end{enumerate}
\end{assumptions}

We denote the set of probability measures on a space $[K]^L$ and $[K]$ by $\cM_1 ([K]^L)$ and $\cM_1 ([K])$, respectively. 
\begin{assumption}[Mutation Distribution]\label{assumptions:mutations}
We assume $\nu \in \cM_1 \big([K]^L\big)$ to be a probability measure on the sequence type space $[K]^L$ such that each of its $L$ marginals $\nu_\ell$ has full support, i.e.
\begin{equation*}%
    \nu_\ell \big( \{k\}\big) > 0
\end{equation*}
for all $\ell \in [L]$ and $k \in [K]$.
\end{assumption}
Note that we do not make any assumptions on the joint mutation distribution, in particular we do \emph{not} assume that mutations at different loci  arise independently.

\begin{proposition}\label{prop:stationary}
    Under Assumptions \ref{assumptions:recombination} and \ref{assumptions:mutations} and if $\mu_N > 0$, a Markov process with generator $G^N$ %
    has a unique stationary distribution $\pi^{N}$.
    The stationary distribution has support on the whole state space $S^N$ if $N$ is large enough.%
\end{proposition}
A proof will be provided in Section \ref{sec:proofs}.

\begin{definition}[Host-Virus Process]\label{def:mainprocess}
Denote by
\begin{equation*}
{\bf X}^{N} = \big({\bf X}^{N}(t) \big)_{t\in \mathbb{R}} = \big(X^N_{m,n, \ell}(t) \big)_{(m,n, \ell)\in [M] \times [N] \times [L], t \in \mathbb{R}}
\end{equation*}
a stationary process with generator $G^N$ started at time $-\infty$ (which can be obtained by using a graphical construction, see Section \ref{sec:proofs}).
\end{definition}

\begin{definition}[Sequence Type Frequency in a Random Sample from a Single Host]\label{def:sample}
    For given $N, M_N\in \IN, j \in [N]$ and $T \in \IR$, let $\tilde{m}=\tm_N$ be uniformly distributed on the host indices $[M_N]$ and independent of everything else. Let
    \begin{equation*}
    \widetilde{\cI}^{N,j} := \{\tn_1, ... ,\tn_{j}\} \subset [N]    
    \end{equation*}
    be virus indices drawn independently without replacement from $[N]$. Then we define
    \begin{equation*}
        W^{N,j}=W^{N,j} (T) \defeq \frac{1}{j} \sum_{\tn \in \widetilde{\cI}^{N,j}} \delta_{X^N_{\tm, \tn} (T)} \in \cM_1 \big([K]^L \big)
    \end{equation*}
    to be the sequence type frequencies in a random sample of size $j$ taken from a single random host at time $T$.
    
    Additionally, we abbreviate the \emph{exact} sequence type frequencies observed when sampling all $N$ particles from a randomly chosen host by setting $$W^{N}= W^{N} (T) \defeq W^{N,N} (T).$$
\end{definition}
In the following, we also need an estimate of the marginal type frequencies in the whole population.
As an approximation for these frequencies we use the frequencies observed when taking a sample 
$\mathfrak{s}=\{(m_1, n_1), ..., (m_{h_N}, n_{h_N})\}$
 of virus particles of size $h_N$ at time $T$.
\begin{definition}[Marginal Type Frequencies]
For the process $\bfX^{N}$ and a locus $\ell \in [L]$ the type  frequencies at locus $\ell$ in the whole population  at time $T$ are
\begin{equation*}
    \bar{X}_\ell^{N} (T) \defeq \frac{1}{MN} \sum_{m \in [M]} \sum_{n \in [N]}\delta_{X^N_{m, n, \ell}(T)} \in \cM_1 ([K])
\end{equation*}
and the type frequencies at locus $\ell$ in the sample  $\mathfrak{s}$ are
\begin{equation*}
    \bar{X}_\ell^{N, \mathfrak{s}} (T) \defeq \frac{1}{h_N} \sum_{(m,n)\in \mathfrak{s}} \delta_{X^N_{m, n, \ell}(T)} \in \cM_1 ([K]).
\end{equation*}
We denote the vector of the marginal type frequencies in the whole population and in the sample $\mathfrak{s}$, respectively, at time $T$ by
\begin{equation*}
\bar{X}^N(T) \defeq \big(\bar{X}_1^N (T), ..., \bar{X}_L^N (T)\big) \in \big(\cM_1 ([K])\big)^L
\end{equation*}
and 
\begin{equation*}
\bar{X}^{N,\mathfrak{s}}(T) \defeq \big( \bar{X}_1^{N, \mathfrak{s}} (T) , ...,  \bar{X}_L^{N, \mathfrak{s}} (T) \big) \in \big(\cM_1 ([K])\big)^L.
\end{equation*}
\end{definition}

As a first result, we give the limiting distribution of $\bar{X}^N (T)$.

\begin{theorem}\label{thm:MainResult1}
    Fix $T>0$. 
    Under Assumptions \ref{assumptions:recombination}, \ref{assumptions:main_rates} and \ref{assumptions:mutations} the marginal type frequencies $\bar{X}^N (T)$ converge in distribution to some limiting random variable $\Bar{X}$ taking values in $\big(\cM_1 ([K])\big)^L$ as $N \to \infty$.
    If
    \begin{equation}\label{eq:main_thm_cas_one}
        \lim_{N \to \infty} \frac{\lambda_N}{M \mu_N} = 0, \tag{i}
    \end{equation}
    then $\Bar{X}$ is deterministic and almost surely equals $(\nu_\ell)_{\ell \in [L]}$. If
    \begin{equation}\label{eq:main_thm_case_two}
        \lim_{N \to \infty} \frac{\lambda_N}{M \mu_N} = C, \tag{ii}
    \end{equation}
    then $\Bar{X} = \big(\Bar{X}_\ell \big)_{\ell \in [L]}$, where $\Bar{X}_\ell$ follows a $\Dir \big(\frac{1+\theta}{C} \nu_\ell \big)$ distribution and the $L$ components are independent.
\end{theorem}
\begin{remark}
    In other words, in  
    Case \eqref{eq:main_thm_case_two}
    the marginal type frequencies $\bar{X}^N (T)$ converge in distribution as $N\rightarrow \infty$ to 
    the stationary distribution of a vector of $L$ independent multitype Wright-Fisher diffusions with parent independent mutation, and the mutation coefficient to type $k$ in the $\ell$-th diffusion is given by $\frac{1}{2}\frac{C}{1+\theta} \nu_\ell (\{k\})$.
    
\end{remark}

Now we can state our main result about the sequence type frequencies in a randomly chosen host.

\begin{theorem}\label{thm:MainResult2}
Consider the setting of Theorem \ref{thm:MainResult1}.
In Case \eqref{eq:main_thm_cas_one}, the sequence type frequencies in a randomly sampled host $W^{N} (T)$ converge in distribution to a $\Dir (\theta \nu^\otimes)$ distributed random variable where $\nu^\otimes$ denotes the product measure of the marginals of $\nu$.
    
In Case \eqref{eq:main_thm_case_two}, the sequence type frequencies converge in distribution to a Dirichlet distribution with random parameter $\theta \psi^\otimes$ where $\psi$ has the distribution of $\Bar{X}$, see Theorem \ref{thm:MainResult1}, and $\psi^\otimes$ again denotes the corresponding product measure of the marginals.
\end{theorem}
\begin{remark} %
    The density of the Dirichlet distribution with random parameter can be found in Subsection \ref{subsec:compound_dirichlet}.
\end{remark}

\begin{theorem}\label{thm:MainResult3} Consider the setting of Theorem \ref{thm:MainResult1}.
    If $\big\lvert \bar{X}^{N,\mathfrak{s}}(T) -  \bar{X}^{N}(T) \big\rvert \rightarrow 0 $ almost surely, then %
    $\big(W^{N} (T),\bar{X}^{N,\mathfrak{s}}(T)\big) \to \big(\Dir \big(\theta\Bar{X}^\otimes \big), \bar{X})$ in distribution.
\end{theorem}
\begin{remark}
By Skorokhod's Representation Theorem, one can construct $(\bfX^N)_{N \in \IN}$ on a joint probability space such that the almost sure convergence of $\Bar{X}^{N} (T) \to \Bar{X}$ is guaranteed.
The almost sure convergence of $|\bar{X}^{N,\mathfrak{s}}- \bar{X}^{N}| \rightarrow  0 $ is then not only fulfilled if the sample $\mathfrak{s}$ is generated by sampling virus particles uniformly at random from the whole population, but for example also when one takes a sample from asymptotically infinitely many hosts and the proportion of viral particles drawn within the single hosts is negligible in all hosts.
Furthermore, it is possible that the sample $\widetilde{\mathcal{I}}$ is a subset of $\mathfrak{s}$. The latter sampling scheme is considered in~\cite{EichhornGoerzerEtAl}.
\end{remark}

\begin{remark}
From the proofs of Theorems \ref{thm:MainResult2} and \ref{thm:MainResult3} it follows that also the sequence type frequencies observed when taking a sample of less than $N$ (but still asymptotically infinitely many) virus particles from a single host converge in distribution to the claimed Dirichlet distributions.
\end{remark}

Proofs of the three theorems are given in Section \ref{sec:proofs}.
Essentially, they all rely on an analysis of the ancestry of a \emph{small} sample of virus particles taken from a single host.
That is, we determine $W^{N,j} (T)$ for $j = \jN \to \infty$ sufficiently slowly by studying the genealogy of the sampled particles.
We first provide a sketch of this genealogy and then use it to sketch the proofs of the three theorems.

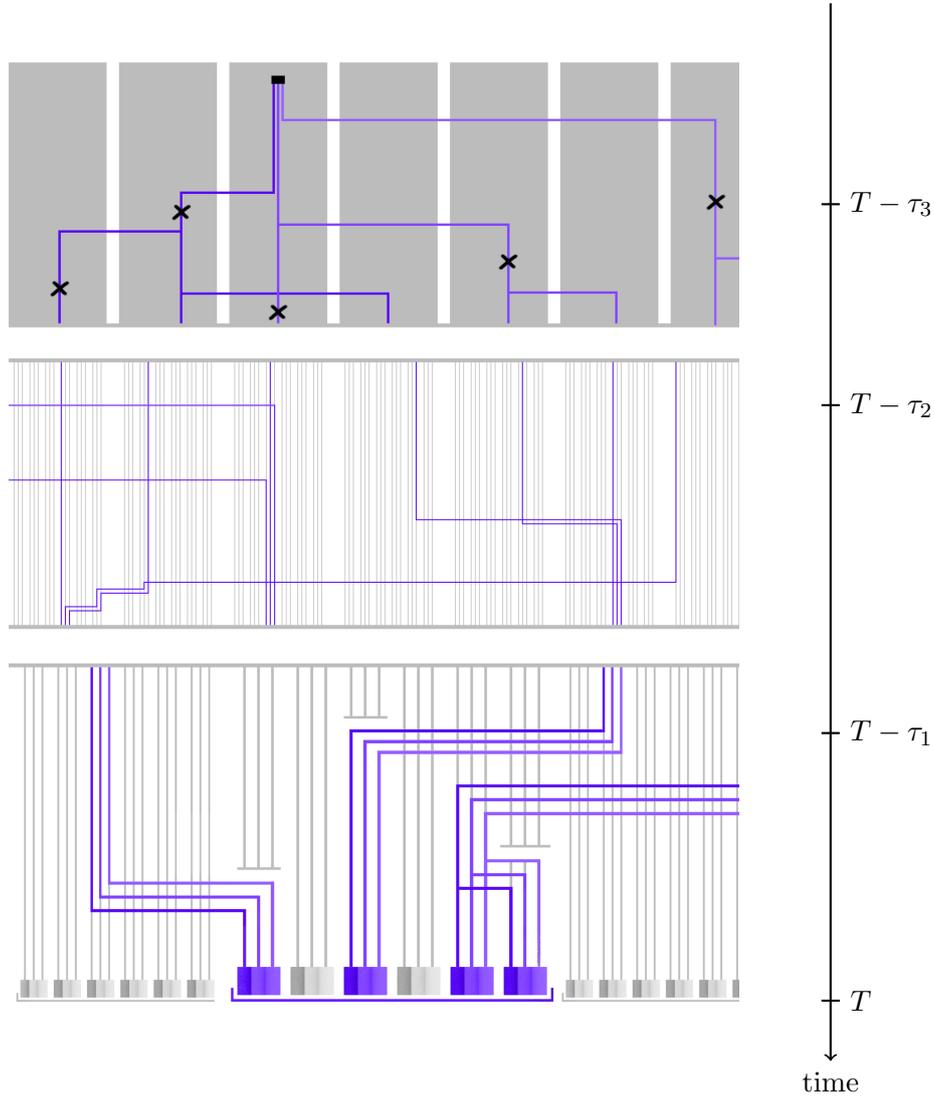
\begin{figure}
    \centering

    \begin{tikzpicture}
        \node[anchor=south west, inner sep=0] (img) {\includegraphics[width=0.8\linewidth]{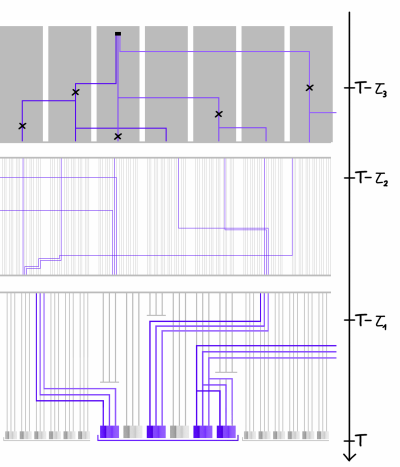}};
        \begin{scope}[x={(img.south east)},y={(img.north west)}]

            \fill[white] (0.8,0) rectangle (1,1);
            \draw[->, thick] (0.9,1) -- ++(0,-1) node[below] {time};
            \draw[-, thick] (0.89,0.057) -- ++(0.02,0) node[right] {$T$};

            \draw[-, thick] (0.89,0.31) -- ++(0.02,0) node[right] {$T-\tau_1$};
            \draw[-, thick] (0.89,0.62) -- ++(0.02,0) node[right] {$T-\tau_2$};
            \draw[-, thick] (0.89,0.81) -- ++(0.02,0) node[right] {$T-\tau_3$};
            
        \end{scope}
    \end{tikzpicture}
    
    \caption{Simplified sketch of the three phases of the genealogy when $N=6$ and $L=3$.
    At time $T$, a sample of $\jN = 4$ virions is taken from a single host.
    The sampled virions (and their loci) are depicted in shades of purple.
    During Phase 1, which is depicted at the bottom, ancestral lines are affected by reinfection and viral reproduction.
    They either merge while still in the sampled host or they are carried into different, separate hosts by reinfection. They are not yet affected by recombination, hence loci are still linked.
    Therefore, ancestral lines are always depicted in groups of $L=3$ locus-wise ancestral lines during this phase.
    The three lines leaving the plot to the right  simply leave w.h.p. the sampled host for a host that is not depicted.
    The end of Phase 1 is reached at time $T-\tau_1$ when all ancestral lines have left the sampled host.
    In the second phase, depicted in the center, ancestral lines undergo recombination.
    This splits up the groups of locus-wise lines from the first phase.
    Lines (locus-wise or not) may still experience reinfection, carrying them into different hosts. But, once two lines are in separate hosts, they will not end up w.h.p. in a common hosts again during this phase.
    The end of Phase 2 is reached at time $T-\tau_2$ when all loci are unlinked and all locus-wise ancestral lines are in separate hosts.
    During Phase 3, locus-wise ancestral lines are affected by mutation, and they may merge again.
    Note that locus-wise lines are hit by mutations while they are single.
    That is, no mutation determines the type of a sampled virus particle at more than one locus.
    }
    \label{fig:sketch_proof}
\end{figure}

\begin{proof}[Sketch of the genealogy.]
To prove our main results we take a backwards approach and determine the asymptotic genealogy of samples of sizes $(\jN)_{N \in \IN}$ from a random host, where $\jN \to \infty$ sufficiently slowly.
The choice of $\jN$ is discussed in Proposition \ref{assumptions:rates}.
Our rate assumptions guarantee that, asymptotically, the genealogy can be decomposed into three phases which are illustrated in Figure \ref{fig:sketch_proof}.
 
In the first phase, which is the shortest, the ancestral lines of the sample are hit by viral reproduction events, which yield a coalescence of lines within the sampled host, and reinfection events, which transport the lines backwards in time into different hosts. With high probability (w.h.p.), no other events affect the genealogy of the sample in this phase. In classical population genetics terms reinfection events carrying lines away from the sampled hosts correspond to ancestral lines being hit by a mutation in the infinite alleles model.
Since viral reproduction and reinfection act on the same time scale, the genealogy of the sample in the first phase is asymptotically given by a Hoppe urn followed backwards in time, which corresponds to the coalescent in the infinite alleles model.
The end of Phase 1 is defined to be $T-\tau_1$ which is the first time at which no ancestral line is in the \emph{sampled} host anymore.
At the end of Phase 1, w.h.p.\ all ancestral lines are located in \emph{different} hosts.
Due to coalescences, the number of ancestral lines has decreased w.h.p. (since $j_N \rightarrow \infty$) by time $T-\tau_1$. We denote the number of lines that are left at time $T-\tau_1$ by $\xi^N$. 
During the first phase, it is appropriate to think of an ancestral line as a virus particle before time $T$ that is an ancestor to a sampled particle.
One does not have to worry about single loci during this phase, since recombination is not at play yet.
This is reflected in the depiction of the first phase in Figure \ref{fig:sketch_proof}, where ancestral lines come in groups of $L$ ``locus wise ancestral lines''.

We follow the genealogy of the sample further in Phase 2. 
Due to the rate assumptions, recombination is the main driver of this phase.
If a recombination event falls on a group of locus wise ancestral lines, then the group is split in two with positive probability.
This is exactly the case, when the recombination event is an $\bfr$-recombination, the group of lines contains two distinct loci $\ell_1$ and $\ell_2$, and $|\bfr \cap \{\ell_1, \ell_2\}| = 1$.
After such an event, for a short time the host in which the event happened contains two groups of lines that are ancestral to the same sampled particles at different loci.
Since virus reproduction and reinfection act on a faster time scale, shortly after that either the two ancestral lines will be located in different hosts or the lines coalesce again with high probability.
Again, w.h.p.\ no other events have an effect on the genealogy of the sample.
We denote the end of this phase by $T-\tau_2$ and define it to be the first time point at which each host is occupied with at most one locus wise ancestral line.
In particular, a total of $L \xi^N$ hosts each carry exactly one locus wise ancestral line at this time point. 

In Phase 3, locus wise ancestral lines of the same locus may coalesce, and they may be hit by mutations.
This way, the genealogy of each single locus is described by the coalescent in the infinite alleles model which is a Hoppe urn followed backwards in time.
As a consequence, asymptotically, the types of the lines at the time $T-\tau_2$ are drawn from a realization of the stationary distribution of $L$ multitype Wright-Fisher diffusions.
The genealogies of the all single loci in the third phase are, asymptotically, independent: 
at the beginning of Phase 3, all lines are w.h.p.\ located in separate hosts.
During Phase 3, not only locus wise ancestral lines of the same locus may merge but also locus wise ancestral lines of different loci.
This introduces some dependence between loci.
However, since recombination and reinfection happen on a faster time scale, this dependence is immediately resolved.
\end{proof}

Now we can use the just described asymptotic genealogy to sketch the proofs of our main results.

\begin{proof}[Sketch of proof of Theorem \ref{thm:MainResult1}]
The proof of this result essentially relies on an analysis of Phase 3 of the genealogy.
The marginal type frequencies $\bar{X}^N (T)$ can be approximated as follows:
    for $\xi \in \IN$, let
    \begin{equation*}
        (m_1, n_1), ... , (m_{\xi \cdot L}, n_{\xi \cdot L})
    \end{equation*}
    be i.i.d.\ uniformly distributed on $[M] \times [N]$.
    Then, for each $\ell \in [L]$, the entry $\Bar{X}^N_\ell (T)$ can be approximated by 
    \begin{equation*}
        \Hat{X}_\ell^N (T) \defeq \frac{1}{\xi} \sum_{i=(\ell-1) \xi}^{\ell \xi} \delta_{X_{m_i, n_i, \ell}^N (T)}.
    \end{equation*}
    If $\xi = \xi_N \to \infty$ slow enough as $N \to \infty$, then w.h.p.\ $m_1, ..., m_{\ell \xi}$ are pairwise different and hence can be interpreted as the lines at the beginning of Phase 3 in Figure \ref{fig:sketch_proof}.
    If the mutation rate is sufficiently high as defined in Case \eqref{eq:main_thm_cas_one}, then mutations fall on all lines in Phase 3 before any coalescence happens and, by the law of large numbers, $\hat{X}_\ell^N$ converges to a constant.
    If mutation and coalescence act on the same time scale %
    as defined in Case \eqref{eq:main_thm_case_two}, then the genealogy of the third phase is at each locus given by an independent Hoppe's urn and, by classical results, $\hat{X}^N$ asymptotically equals a vector of independent Dirichlet distributions.
\end{proof}

\begin{proof}[Sketch of proof of Theorem \ref{thm:MainResult2}]
    The proof of Theorem \ref{thm:MainResult2} relies on the analysis of all three phases of the genealogy.
    As we will see, each of the three phases contributes one piece to the claimed limiting distribution.

    Similar to the preceding sketch of proof of the first result, we can approximate $W^N (T)$ by $W^{N, \jN} (T)$, where $\jN \to \infty$ sufficiently slowly.
    This insight allows us to study $W^{N, \jN}(T)$ instead of $W^N (T)$ for the rest of this sketch.

    Using the analysis of Phase 1, one can conclude that the sampled particles can be grouped into ``families of the first Phase''. For a better understanding, see also the bottom of Figure \ref{fig:sketch_proof} where we observe two families of size one and one family of size two.
    These families are represented by their unique ancestors at at time $T-\tau_1$, and all members of a family carry their unique ancestor's type.
    Hence, for computing $W^{N,\jN}(T)$, only the family sizes and the ancestor's type of each family are relevant.
    By the relationship of the first phase with the infinite alleles model, one can conclude that the family sizes in the sample are precisely the family sizes arising in Hoppe's urn after $\jN$ draws.
    The family sizes in Hoppe's urn have been studied extensively. In particular, it is well known that the type frequencies observed when letting $\jN \to \infty$ and coloring each family with an i.i.d.\ type, follow a Dirichlet distribution.

    By arguing that the families' ancestors' types are almost i.i.d.\ draws from a (deterministic or random) distribution, Phase 1 yields the fact that the limiting distribution is a Dirichlet distribution (with deterministic or random parameter).
    
    The other two phases determine the parameter of this Dirichlet distribution:
    Phase 2, which is driven by recombination, is responsible for the fact that the parameter of the Dirichlet distribution is a ``product'' ($\otimes$) of the single locus type frequencies.
    Phase 3 determines these single-locus type frequencies as given in Theorem \eqref{thm:MainResult1}: in case \eqref{eq:main_thm_cas_one}, they approach the deterministic parameter $\nu$, in case \eqref{eq:main_thm_case_two}, they are again given by independent Dirichlet distributions.
\end{proof}

\begin{proof}[Sketch of proof of Theorem \ref{thm:MainResult3}]
The key idea here is - as in the last sketch of proof - to regard the coloring of the
Hoppe's urn from the first phase as a coloring with the product of the marginal
type frequencies at $T-\tau_2$, given by $\bar{X}^N (T-\tau_2)$.

Since Phases 1 and 2 are asymptotically instantaneous on the time scale of Phase 3, we can approximate the state of the realization of the stationary distribution at the end of Phase 2 by the frequencies at the sampling time $T$ if almost sure convergence of $|\bar{X}^{N, \mathfrak{s}}- \bar{X}^N|_\infty$ to 0 is assumed.
\end{proof}

\begin{remark}
The main results should also hold under weaker assumptions: 
One can relax the assumption that viral replication within hosts follows Moran dynamics in favor of more general (discrete or continuous time) Cannings dynamics, as long as replication events asymptotically only cause pairwise (Kingman-like) coalescences in the backwards process, see also \cite{MoehleSagitov2001} for conditions that yield convergence of ancestral processes arising from Cannings models to a Kingman coalescent.
This possible generalization is particularly relevant for viral populations, in which a single virion may infect a cell and be replicated multiple times.
Under the specific rate assumptions studied in this paper, one can also relax the assumption that at primary infection only a single virion from a single host gives offspring to all virions in the primary infected host.
This is due to the fact that host replacements/primary infections are asymptotically invisible in the ancestral process of a sample of size $\jN$.

It should be possible to obtain similar results when modeling even more general neutral replication dynamics within host:
A potential generalization would be accounting for highly skewed offspring distributions.
When embedded in the same separation-of-scales setting, this kind of offspring distributions can change the genealogies within hosts leading to, e.g., multiple instead of binary mergers in the ancestral process.
In this case, the ``outer'' Dirichlet distribution in Theorem \ref{thm:MainResult2} would be replaced by another distribution.
\end{remark}

\section{Proofs}\label{sec:proofs}

This section is subdivided into five subsections. In the first subsection, we establish some preliminaries including a construction of the process ${\bf X}^{N}$, the proof of Proposition \ref{prop:existence}, and the definition of the sample sizes $(\jN)_{N \in \IN}$ mentioned earlier.
Here, we also define a reversed Hoppe urn which will be an important tool in the subsequent subsections.
Subsections \ref{sec:first_phase} to \ref{sec:third_phase}
are devoted to a rigorous treatment of the Phases 1 to 3 described in the sketch of the genealogy above.
Finally, in the last subsection, we give the proofs of the main results. 

\subsection{Preliminaries}

The existence of the Markov process ${\bf X}^{N}$ with generator $G^{N}$ is covered by standard theory.
Nevertheless, we provide a pathwise construction here. It will be useful for the definition of dual processes and couplings thereof.
We follow the random mapping construction described in~\cite{Swart22}, Section 2.3.

First of all, we rewrite the generator as
\begin{equation*}
    G^N f(x) = \sum_{\m \in \cG^N} r_{N, \m} \big( f(\m(x))-f(x) \big)
\end{equation*}
where
\begin{equation*}
    \cG^N \defeq \cG^N_{\rep} \cup \;
                        \cG^N_{\reinf} \cup \;
                        \cG^N_{\recomb} \cup \;
                        \cG^N_{\mut} \cup \;
                        \cG^N_{\death}
\end{equation*}
and
\begin{equation*}
    r_{N, \m} \defeq
    \begin{cases}
        \frac{\gamma_N}{N}                          &\text{if } \m \in \cG^N_\rep,\\[0.4em]
        \frac{\lambda_N}{N^2 M}                     &\text{if } \m \in \cG^N_\reinf,\\[0.4em]
        \frac{\rho_N p_{\recomb}^\textbf{r}}{N^3}   &\text{if } \m = \recomb_{m, n, n_1, n_2}^\bfr,\\[0.4em]
        \frac{\mu_N \nu (\{\bfk\})}{N}              &\text{if } \m = \mut^{\bfk}_{m, n},\\[0.4em]
        \frac{1}{MN}                                &\text{if } \m \in \cG^N_\death.\\[0.4em]
    \end{cases}
\end{equation*}
Then, the total rate at which random maps arrive is defined by
\begin{equation*}
    r_N \defeq \sum_{\m \in \cG^N} r_{N, \m} = MN \gamma_N + M \lambda_N + M \mu_N + M \rho_N + M.
\end{equation*}
Denote the Lebesgue measure by $\leb$.
Let $$\mathfrak{t} = \mathfrak{t}^N \sim \text{PPP}(r_N d\leb)$$  be a Poisson point process   on $\IR$ with intensity $r_N d \leb$ and let $(\map_t)_{t \in \IR} = (\map_t^N)_{t \in \IR}$ be an independent family of i.i.d.\ random variables with
\begin{equation*}
    \IP \big(  \map_t = \m\big) = \frac{r_{N, \m}}{r_N}.
\end{equation*}
for all $\m \in \cG^N$.
Then, by the Poisson Coloring Theorem,
\begin{equation*}
  \omega(A)=  \omega^N (A) \defeq \int \1_A \big(\map_t, t\big) \mathfrak{t} (dt), \hspace{2em}  A \in  \cP \big(\cG^N \big) \otimes \cB,
\end{equation*}
where $\cB$ denotes the Borel-$\sigma$-algebra on $\IR$, defines a Poisson point process $\omega^N$ on $\cG^N \times \IR$ with intensity $r_N d \leb \otimes \sum_{\m \in \cG^N} \frac{r_{N,\m}}{r_N} \delta_{\m}$.
The PPP $\omega^N$ will have the interpretation of a (random) collection of maps and times at which they are applied.
That is, $(\m, t) \in \omega^N$ means that at time $t$ map $\m$ is happening.
As described in~\cite{Swart22}, one can now unambiguously define a random function $\IR \ni t \mapsto \mathfrak{m}_t^{\omega}$ by setting
\begin{equation*}
    \mathfrak{m}_t^{\omega^N} \defeq
    \begin{cases}
    \m &\text{if } (\m,t) \in \omega^{N},\mybreak
    1 &\text{otherwise}
    \end{cases}
\end{equation*}
where $1$ denotes the identity map, since for each $t \in \IR$ there is at most one $\m \in \cG^N$ such that $(\m,t) \in \omega^N$.
The aim now is to define piecewise constant right continuous functions $[s,\infty) \ni t \mapsto \bfX^{N}(t)$ that solve 
\begin{equation}\label{eqn:problem}
    \bfX^{N} (s) = x \text{  and  } \bfX^{N} (t) = \mathfrak{m}_t^{\omega} \big( \bfX^{N} (t^-) \big).
\end{equation}
We write
\begin{equation*}
    \omega_{s, u} = \big\{(\m, t) \in \omega : t \in (s,u]\big\}.
\end{equation*}
This set contains only finitely many elements which we can order as
\begin{equation*}
    \omega_{s, u} = \big\{ (\m_n, t_n),... (\m_1, t_1) \big\}
\end{equation*}
with $t_n < t_{n-1} < ... < t_1$. The unique solution of \eqref{eqn:problem} is then given by:
\begin{align*}
    &\bfX^{N} (t) = x            &&\text{ for } t \in [s, t_n),\\
    &\bfX^{N} (t) = \m_n(x)       &&\text{ for } t \in [t_n, t_{n-1}),\\
    &\bfX^{N} (t) = \m_{n-1} \circ \m_n(x)       &&\text{ for } t \in [t_{n-1}, t_{n-2}),\\
    & ...\ .
\end{align*}
As described in~\cite{Swart22}, one can now define a stochastic flow $(\mathbb{X}_{s,u})_{s \leq u}$ which is right-continuous in $s$ and $t$ and has independent increments by
\begin{equation*}
    \mathbb{X}_{s,u} \defeq \m_1 \circ \m_2 \circ ... \circ \m_n .
\end{equation*}
The unique solution of (\ref{eqn:problem}) can now be rewritten as
\begin{equation*}
    \bfX^{N} (t) = \mathbb{X}_{s,t} (x) .
\end{equation*}

\begin{proposition}\label{prop:existence}
    Let  $\omega$ be defined as above and let $X^{N}_0$ be a $S^N$ valued random variable independent of $\omega$.
    Set $\bfX^{N} (t) = \mathbb{X}_{0,t}\big(X^N_0\big)$. Then $\big(\bfX^N (t)\big)_{t \geq 0}$ is a Markov process with generator $G^N$.
\end{proposition}
\begin{proof}
    This is Proposition 2.7 in~\cite{Swart22}.
\end{proof}

\begin{proof}[Proof of Proposition \ref{prop:stationary}]

We prove the whole statement for $N$ large enough by showing irreducibility on $S^N$.
For small $N$ (as can be seen trivially for $N=1$), irreducibility on $S^N$ is lost but the existence of a unique stationary distribution still holds.
This can be shown by studying the paths of potential influence as done in \cite{Swart22}.
We leave out the argument here to avoid excessive notation.

Irreducibility on $S^N$ for $N$ large enough can be seen as follows:
Under Assumptions~\ref{assumptions:mutations}, mutations can produce any of the $K$ possible types at any locus but mutations alone may not be able to produce all sequence types.
However, any arbitrary sequence type $(k_1, ..., k_L) \in [K]^L$ can be created in a single host by first creating sequences with types $$(k_1, \cdot, ... , \cdot),(\cdot, k_2, \cdot, ... , \cdot),..., (\cdot, ..., \cdot, k_L)$$ from mutations and then recombining them appropriately.
Because of Assumptions \ref{assumptions:recombination}, there is a finite sequence of recombination maps that produce exactly the sequence type $(k_1, ..., k_L)$.
Indeed, one can first create sequences $(k_1, k_2, \cdot, ... \cdot)$, $(\cdot, k_2, k_3, \cdot, ... \cdot)$, ... .
These sequences can be further recombined to sequences $(k_1, k_2, k_3, \cdot, ... , \cdot), (\cdot, k_2, k_3, k_4 \cdot, ... , \cdot),$ ... .
To arrive at a sequence of type $(k_1, k_2, k_3, \cdot, ... , \cdot)$ one applies a recombination map $\recomb^\bfr_{m,n,n_1,n_2}$ with $1 \in \bfr$ and $3 \notin \bfr$, which exists by Assumption \ref{assumptions:recombination}, as well as $x_{m,n_1}=(k_1,k_2,\cdot, ..., \cdot), x_{m,n_2}=(\cdot,k_2, k_3,\cdot, ..., \cdot)$.
Continuing this procedure yields the desired sequence type $(k_1, ..., k_L)$.
For $N$ large enough, it is possible to generate each possible sequence type in one single host.
The types can then be arbitrarily distributed across the host population by virus reproduction and reinfection.
Since all the described events happen with positive rate and and do not depend on the initial condition, the chain is irreducible on $S^N$.
\end{proof}

Next we state a weaker kind of exchangeability, namely that the distribution of the stationary process at any time is invariant under ``reasonable relabeling'' of hosts and virus particles.
A relabeling is a permutation
\begin{align*}
    &\sigma : [M] \times [N] \rightarrow [M] \times [N],\\
    &(m,n) \mapsto (\sigma (m,n)_1, \sigma (m,n)_2),
\end{align*} satisfying $\sigma (m,n')_1 = \sigma (m,n'')_1$ for all $n', n'' \in [N]$ and $m \in [M]$.
That is, $\sigma$ can change indices of hosts and virions but virions that belong to a common host before the relabeling still have to belong to a common host after the relabeling. For any state $x \in S^N$ and any relabeling $\sigma$, we define the relabeled state $x^\sigma \in S^N$ by
\begin{equation*}
    x^\sigma_{m,n} = x_{\sigma^{-1}((m,n))}.
\end{equation*}

\begin{proposition}\label{prop:exchangeability}
    Let $\pi^N$ be the unique stationary distribution and $\sigma : [M] \times [N] \to [M] \times [N]$ a relabeling.
    Then,
    \begin{equation}\label{eq:exchangeable}
        \pi^N (x) = \pi^N (x^\sigma)
    \end{equation}
    for all states $x \in S^N$.
\end{proposition}

\begin{proof}
    Start the process $\bfX^N$ from some initial distribution $\mu$ which is invariant under relabeling in the sense of Equation \eqref{eq:exchangeable}.
    Then, at any time $t$, the law of $\bfX^N(t)$ is still invariant under relabeling since the generator is invariant under relabeling, i.e.\ $G^N f(x) = G^N f(x^\sigma)$.
    Therefore also the limiting law
    \begin{equation*}
        \lim_{t \to \infty} \cL \big( \bfX^N(t) \big)
    \end{equation*}
    is invariant under relabeling.
    But this limiting law is precisely the unique stationary distribution $\pi^N$.
\end{proof}

As outlined in Section \ref{sec:model_main_results}, the proof of the main results relies on an analysis of the asymptotic genealogy of samples of $\jN$ virions taken from a single host.
We have not specified $\jN$ yet, but we do so now.

\begin{lemma}[Sample Size]\label{assumptions:rates}
Under the Assumptions \ref{assumptions:main_rates}, there is a sequence of sample sizes $(\jN)_{N \in \IN}$ satisfying the following asymptotics:
\begin{enumerate}
    \item \label{ass:timescales} \begin{equation*}
        \max \big\{\jN N, \jN^{2} \mu_N, \jN^{2} \rho_N \big\} \in o(\gamma_N).
    \end{equation*}
    \item \label{ass:sample_size}
    \begin{equation*}
          \jN \rightarrow \infty \quad \text{with} \quad \jN^3 \in o(M_N).
    \end{equation*}
    \item %
    \begin{equation*}
         \jN \log (\jN) \mu_N \in o(\rho_N).\end{equation*}
    \item \label{assumptions:rates_mutation_cases} %
    Either
    \begin{equation*}
        \lim_{N \to \infty} \frac{\log(\jN) \big)^3 \lambda_N}{M \mu_N} = 0
    \end{equation*}
    or
    \begin{equation*}
        \lim_{N \to \infty} \frac{\lambda_N}{M \mu_N} = C
    \end{equation*}
    for the value $C>0$ given in Assumptions \ref{assumptions:main_rates}.
\end{enumerate}
\end{lemma}
\begin{proof}
One can verify that in Case \eqref{assumptions:main_rates_case_two} of Assumptions \ref{assumptions:main_rates},
\begin{equation*}
\jN \defeq \min \bigg\{
{\Big(\frac{\gamma_N}{N}\Big)}^{\frac{1}{2}},
{\Big(\frac{\gamma_N}{\mu_N}\Big)}^{\frac{1}{4}},
{\Big(\frac{\gamma_N}{\rho_N}\Big)}^{\frac{1}{4}},
{\big(M_N\big)}^{\frac{1}{4}},
{\Big(\frac{\rho_N}{\mu_N}\Big)}^{\frac{1}{2}}
\bigg\}    
\end{equation*}
satisfies the desired conditions.
In Case \eqref{assumptions:main_rates_case_one}, 
\begin{equation*}
    j'_N \defeq \min \bigg\{ \jN,  { \Big( \frac{M_N \mu_N}{\lambda_N} \Big)}^{\frac{1}{6}} \bigg\}
\end{equation*}
satisfies the conditions.
\end{proof}

We conclude this subsection with the definition of an ordered, reversed Hoppe urn that we will need in our proofs.
For this, we need to introduce some more notation.
Let
\begin{equation*}
\cV \defeq \big\{v \in \IN^n : n \in \IN_0 \big\}    
\end{equation*}
be the set of all integer-valued vectors. Here we use the convention that $\IN^0$ contains exactly one element which is the empty vector denoted by empty brackets, $\IN^0 = \big\{() \big\}$.
For a vector $v = (v_1, ..., v_{n_v}) \in \cV$, denote by $$\bfell(v) = n_v \in \IN_0$$ its length.
Denote by $$\cVs = \big\{ (v_1,...v_n) \in \cV : v_1 \geq v_2 \geq ... \geq v_n \big\} \cup \IN^0$$ the set of all descending integer-valued vectors.
Let $$\bm{s} : \cV \, \cup \, \cM (\IN) \rightarrow \cVs$$ be a sorting function which sorts its input in descending order.
That is, for any input \textit{vector} $v = (v_1, ... , v_n)$ with arbitrary $n \in \IN$, $\bm{s}(v) = (v_{\sigma (1)}, ..., v_{\sigma (n)})$ such that $\sigma : [n] \rightarrow [n]$ is a permutation and $v_{\sigma (1)} \geq v_{\sigma (2)} \geq ... \geq v_{\sigma (n)}$. Similarly, if the input is a finite \textit{multiset of integers} (with set brackets denoted by $\{\vert$ and $\vert\}$) $$\mset{m_1, m_2, ..., m_n},$$ then $\bm{s} (\mset{m_1, m_2, ..., m_n }) = (m_{\sigma (1)}, ..., m_{\sigma (n)})$ such that $\sigma$ is a permutation and $m_{\sigma (1)} \geq m_{\sigma (2)} \geq ... \geq m_{\sigma (n)}$.
We define 
\begin{gather}\label{eq:def_plus}
    \oplus : \cVs \times \IN \rightarrow \cVs\\
    \big((v_1, \dots, v_n), u\big) \mapsto (v_1, \dots, v_n) \; \oplus \; u \defeq \bm{s} \big( (v_1, \dots, v_n, u) \big) \nonumber
\end{gather}
to be an operator that inserts $u$ (at a right place) into the sorted vector $(v_1, ..., v_n)$.
We define 
\begin{gather}\label{eq:def_minus}
    \ominus : \cVs \times \IN \rightarrow \cVs\\
    \big((v_1, \dots, v_n), u\big) \mapsto (v_1, \dots, v_n) \; \ominus \; u \defeq (v_1, \dots, v_{i-1}, v_{i+1}, \dots v_n) \nonumber
\end{gather}
with $i = \inf \{k \in [n]: v_k= u\}$ to be an operator that removes $u$ from the sorted vector $(v_1,...,v_n)$.
    
For a vector $v = (v_1, ..., v_n)\in \cV$ we define the multiplicity of value $k \in \IN$  in $v$ as $\bfm:\cV \times \IN \rightarrow \IN$ by
\begin{equation*}
\bfm (v, k) = \big\vert \{i \in [n] : v_i = k \} \big\vert.
\end{equation*}

The process we define next is similar to the coalescent of the Infinite Alleles Model as described by Watterson \cite{Watterson1984}.
On different time scales  asymptotic genealogies obtained in our model can be appropriately described by this process. It can also be interpreted as a Hoppe urn followed backwards in time with a specific ordering, which is why we chose the name ``reversed Hoppe's urn'' for this process.

\begin{definition}[Reversed Hoppe's Urn]\label{def:backwards_hoppes_urn}
We define the (ordered) reversed Hoppe urn with parameter $\theta>0$, speed $c>0$ and $d\in \IN$ initial particles to be a continuous time Markov chain $\big(\bfZ (t)\big)_{t \geq 0}$ taking values in $\cVs \times \cVs$ with initial state $\smash{\bfZ(0) = \big(( \underbrace{1,\dots,1}_{d} ),()\big)}$, jumping from $z = (z^{(1)}, z^{(2)})$ to
\begin{align*}
&\big(z^{(1)} \; \ominus \; k, z^{(2)} \; \oplus \; k\big) & &\text{at rate}  & &c \cdot \theta \cdot \bfm (z^{(1)} ,k)\\
\intertext{and to}
&\big(z^{(1)} \; \ominus \; i \; \ominus \; j \; \oplus \; (i+j),\; z^{(2)}\big) & &\text{at rate} & &c \cdot \bfm(z^{(1)}, i) \big(\bfm(z^{(1)}, j)-  \1_{i=j}\big).
\end{align*}
for any $i,j,k \in \IN$ with $i \neq j$.
\end{definition}
\begin{remark}\label{rem:hoppe}
    Note that the reversed Hoppe's urn eventually reaches an absorbing state $\bfZ_\infty$ with $\bfZ_\infty^{(1)} = ()$. At every jump, the number of entries of the first component decreases by one, so an absorbing state is reached after $d$ many jumps.
    Note also that the distribution of the final state only depends on $\theta$ but not on $c$.
    We call $\bfZ_\infty^{(2)}$ the \emph{family sizes at absorption}.
\end{remark}

Now we are ready for the rigorous study of the genealogy.

\subsection{First Phase}\label{sec:first_phase}
In this section we formally show that by the end of the first phase (which is formally defined in Equation \eqref{eq:tau1} below), all ancestral lines of the sample have left with high probability  the sampled host and the genealogy of the sample until this time is given by a reversed Hoppe urn, see Proposition \ref{lem:g2Y}.

Recall that $\tm$ is the index of the randomly sampled host and $\widetilde{\cI}^N$ are the indices of the sampled virus particles within that host.
By assumption, $\tm$ is uniformly distributed on $[M]$ and the elements of $\cI^N$ arise as $\jN$ drawings without replacement from $[N]$.
Let \mbox{$[\jN]_0 \defeq [\jN] \cup \{0\}$}.
We will now define a Markov process
\begin{equation*}
\bfY^{N, 1} = \big(\bfY^{N, 1} (s) \big)_{s \geq 0} = \big( Y^N_{m,n} (s)\big)_{m \in [M], n \in [N], s \geq 0}
\end{equation*}
taking values in $S_\bfY^{N,1} \defeq [\jN]_0^{[M] \times [N]} \cup \{\dagger\}$ that tracks the ancestral particles of the sample.
We will refer to $\bfY^{N,1}$ as the ``dual of the first phase'' since its role will be similar to that of a pathwise dual of $\bfX^{N}$:
if $\bfY^{N,1} (s) \neq \dagger$ at time $s>0$, then the value $Y_{m, n}^{N,1 } (s) = k$ will have the interpretation that at time $T-s$, the virus particle $n$ in host $m$ is ancestral to $k$ sampled virus particles and has propagated its type to the sampled particle (i.e., no mutation or recombination has occurred along the way).
On the other hand, $\bfY^{N,1} (s) = \dagger$ will have the interpretation that an ``unwanted'' event has happened in the ancestry of the sample between times $T$ and $T-s$.
Unwanted events are events that break the relationship of $\bfX^{N}$ and $\bfY^{N,1}$ or that would make the dual process harder to analyze. 
Specifically, these include:
\begin{itemize}
    \item mutation or recombination events falling on ancestral lines
    \item host replacements or reinfection events causing ancestral lines to coalesce
    \item host replacements or reinfection events placing ancestral particles that were in separate hosts into a common host.
\end{itemize}
We will show later in Lemma \ref{lem:g2Y} that due to our rate assumptions these unwanted events are very unlikely to happen as $N\rightarrow \infty$.

We proceed to define $\bfY^{N,1}$ and couple it with $\bfX^N$ by first defining a dual map $\m^{-1,1}$ for each map $\m \in \cG^N$.
Then we consider the same PPP $\omega^N$ used to construct $\bfX^N$ but replace maps with their duals to obtain a dual PPP $\omega^{N,-1,1}$.
Fixing some initial state for $\bfY^{N,1}$ and reversing time, we apply the same type of random mapping construction used for $\bfX^N$ to define $\bfY^{N,1}$.
This procedure will be used throughout the paper to construct dual processes for the other phases, too.

\begin{definition}[Dual Maps of the First Phase]
We call $(m,n) \in [M] \times [N]$ a site.
For a state $y \in S_{\bfY}^{N,1} \setminus \{\dagger\}$ we say that there is an (ancestral) line at site $(m,n)$ if $y_{m,n} > 0$ and we say that $y_{m,n}$ is the number of progeny of the line at $(m,n)$.
We write
\begin{equation*}
y_m^{\#} \defeq \sum_{n \in [N]} \1_{y_{m,n}\neq0}    
\end{equation*}
for the number of lines in host $m \in [M]$ and 
\begin{equation*}
y_m^{\Sigma} \defeq \sum_{n \in [N]} y_{m,n}
\end{equation*}
for the total number of their progeny.
We abuse notation by writing $\dagger$ and $(\dagger)^{M \times N}$ for the same coffin state.
We define for any $m, m_1, m_2 \in [M]$, $n, n_1, n_2 \in [N]$ and $y \in S_\bfY^{N,1} \setminus \{\dagger\}$:
\begin{align*}
    &\big(\rep^{-1,1}_{m, n_1, n_2} (y)\big)_{m',n'} & & \defeq \begin{cases}
        y_{m, n_1} + y_{m, n_2}     &\text{if $(m',n') = (m, n_1)$,}\\
        0                           &\text{if } (m',n') = (m, n_2),\\
        y_{m',n'}                   &\text{otherwise.}
    \end{cases}\\
    \\
    &\big(\reinf^{-1,1}_{m_1, n_1, m_2, n_2} (y)\big)_{m', n'} & & \defeq \begin{cases}
        \dagger                         &\text{if } y_{m_1}^{\#} > 0,\\
        y_{m_1, n_1} + y_{m_2, n_2}     &\text{if } y_{m_1}^{\#} = 0 \text{ and } (m',n') = (m_1, n_1),\\
        0                               &\text{if } y_{m_1}^{\#} = 0 \text{ and } (m',n') = (m_2, n_2),\\
        y_{m',n'}                       &\text{otherwise.}
    \end{cases}\\
    \\
    &\mut^{\bfk, -1,1}_{m, n}(y) & & \defeq \begin{cases}
        y           &\text{if $y_{m, n} = 0$,}\\
        \dagger     &\text{otherwise.}
    \end{cases}\\
    \\
    &\big(\death_{m_1, n_1, m_2}^{-1,1} (y)\big)_{m', n'} & & \defeq \begin{cases}
        \dagger                 &\text{if } y_{m_1}^{\#} > 0 \text{ or } y_{m_2}^{\#} > 1,\\
        y_{m_2}^{\Sigma}        &\text{if } y_{m_1}^{\#} = 0, y_{m_2}^{\#} = 1 \text{ and } (m', n') = (m_1, n_1),\\
        0                       &\text{if } y_{m_1}^{\#} = 0, y_{m_2}^{\#} = 1 \text{ and } m' = m_2,\\
        y_{m',n'}               &\text{otherwise.}
    \end{cases}\\
    \\
    &\recomb_{m, n, n_1, n_2}^{\bfr, -1,1} (y) & & \defeq \begin{cases}
        y           &\text{if $y_{m, n} = 0$,}\\
        \dagger     &\text{otherwise.}
    \end{cases}
\end{align*}
Furthermore, we set the image of $\dagger$ under any of the dual maps to be $\dagger$ itself.
In the following we will write $\cG^{N,-1,1}$ for the set of dual maps for the first phase. That is,
\begin{equation*}
    \cG^{N,-1,1} \defeq \big\{\m^{-1, 1} : \m \in \cG^{N}\big\}.
\end{equation*}
\end{definition}
We can now define the dual PPP for the first phase by
\begin{equation*}
    (\m, t) \in \omega^N \iff (\m^{-1,1}, t) \in \omega^{N,-1,1}
\end{equation*}
almost surely.
This allows us to formally define the dual process of the first phase in analogy to \eqref{eqn:problem}.
\begin{definition}[Dual Process of the First Phase]\label{def:dual_1}
Let 
\begin{equation*}
\omega^{N,-1,1}_{(-\infty, T]} := \big\{(\m^{-1,1}, t) \in \omega^{N,-1} : t \leq T\big\} =  \big\{(\m^{-1,1}_1, t_1), \big(\m^{-1,1}_2, t_2\big), ... \big\}   
\end{equation*}
such that $T > t_1 > t_2 > ...$ .
We define the dual process of the first phase to be a piecewise constant stochastic process
$(\bfY^{N,1} (s))_{s \geq 0}$ with \cadlag{} paths taking values in \mbox{$S_\bfY^{N,1} = [\jN]_0^{[M] \times [N]} \cup \{\dagger\}$} by setting
\begin{align*}%
    Y^{N,1}_{m,n} (0) =
    \begin{cases}
        1 &\text{if } m = \tm \text{ and } n \in \cI^N,\mybreak
        0 &\text{otherwise.}
    \end{cases}
\end{align*}
and
\begin{align*}
    &\bfY^{N} (s) = \bfY^{N} (0)                            &&\text{ for } s \in [0, T-t_1),\\
    &\bfY^{N} (s) = \m_1^{-1,1}(\bfY^{N} (0))               &&\text{ for } s \in [T-t_1, T-t_2),\\
    &\bfY^{N} (s) = \m_{2}^{-1,1} \circ \m_1^{-1,1} (\bfY^{N} (0))       &&\text{ for } s \in [T-t_2, T-t_3),\\
    & ...\ .
\end{align*}

\end{definition}
As long as the $\dagger$-state is not reached, the just defined reversed maps yield the following pathwise duality, see also Section 6.5 of~\cite{Swart22}.
\begin{proposition}\label{prop:maps_and_their_duals_phase_1}
    Let $\m \in \cG^N$ be a map and let $\m^{-1,1} \in \cG^{N,-1,1}$ be its corresponding dual.
    For $y \in S_\bfY^{N,1} \setminus \{\dagger\}$ such that $ \m^{-1,1} (y) \neq \dagger$ and arbitrary $x \in S^N$ we have
    \begin{equation*}
        \sum_{(m, n) \in [M] \times [N]} y_{m,n} \delta_{(\m (x))_{m,n}} = \sum_{(m, n) \in [M] \times [N]} \big(\m^{-1,1} (y)\big)_{m,n} \delta_{x_{m,n}} .
    \end{equation*}
\end{proposition}
\begin{proof}
    For any replication map $\rep_{m', n'_1, n'_2}$ we have by the definition of the replication map and its inverse, and by closely looking at which entries of $x$ remain unchanged
    \begin{align*}
        &\sum_{\mathclap{\substack{(m, n)\\ \in [M] \times [N]}}} y_{m,n}                   \delta_{(\rep_{m', n'_1, n'_2} (x))_{m,n}}\\
        &= \sum_{\mathclap{\substack{(m, n)\\ \notin \{(m',n'_1),\\ (m',n'_2)\}}}} y_{m,n}  \delta_{(\rep_{m', n'_1, n'_2} (x))_{m,n}}
                + y_{m',n'_1}\delta_{(\rep_{m', n'_1, n'_2} (x))_{m',n'_1}}         + y_{m',n'_2}\delta_{(\rep_{m', n'_1, n'_2} (x))_{m',n'_2}}\\
        &= \sum_{\mathclap{\substack{(m, n)\\ \notin \{(m',n'_1),\\ (m',n'_2)\}}}}          \big(\rep_{m', n'_1, n'_2}^{-1,1} (y)\big)_{m,n} \delta_{x_{m,n}}
                + y_{m',n'_1}\delta_{x_{m',n'_1}}                                   +  y_{m',n'_2}\delta_{x_{m',n'_1}}\\
        &= \sum_{\mathclap{\substack{(m, n)\\ \notin \{(m',n'_1),\\ (m',n'_2)\}}}}          \big(\rep_{m', n'_1, n'_2}^{-1,1} (y)\big)_{m,n} \delta_{x_{m,n}}
                + \big( y_{m',n'_1} +  y_{m',n'_2}\big)\delta_{x_{m',n'_1}}         + 0 \cdot \delta_{x_{m',n'_2}}\\
        &= \sum_{\mathclap{\substack{(m, n)\\ \in [M] \times [N]}}} \big(\rep_{m', n'_1, n'_2}^{-1,1} (y)\big)_{m,n} \delta_{x_{m,n}} .
    \end{align*}
    Similar calculations apply for the other types of maps.
\end{proof}

Recall that the sequence type frequency in the small random sample from a single, randomly chosen host was defined as
\begin{equation*}
    W^{N,\jN} (T) \defeq \frac{1}{\jN} \sum_{\tn \in \widetilde{\cI}^{N,\jN}} \delta_{X^N_{\tm, \tn} (T)}
\end{equation*}
where $\jN$ is the number of the sampled virus particles, $\tm$ is the index of the sampled host  and $\widetilde{\cI}^N = \{\tn_1, ..., \tn_{\jN}\}$ are the indices of the virus particles which arise as independent draws without replacement from $[N]$.

During the first phase we can express $W^{N,\jN}(T)$ in terms of the dual process of the first phase $\bfY^{N,1}$.
To this end, we define for any $s \geq 0$ such that ${\bf Y}^{N,1}(s) \neq \dagger$,
\begin{equation*}\label{eq:WY}
W_\bfY^{N, 1} (T, s) \defeq \frac{1}{\jN} \sum_{(m, n) \in [M] \times [N]} Y_{m, n}^{N,1} (s) \delta_{X^N_{m, n} ((T-s)^-)}.
\end{equation*}
This implies for $T-s \notin \mathfrak{t}^N$  that
\begin{equation*}%
    W_\bfY^{N,1} (T, s) = \frac{1}{\jN} \sum_{(m, n) \in [M] \times [N]} Y_{m, n}^{N,1} (s) \delta_{X^{N}_{m, n} (T-s)}.
\end{equation*}
Note that, as long as ${\bf Y}^{N,1}$ is not in the $\dagger$-state, $W_{\bf Y}^{N,1}$ is constant in $s$.
\begin{proposition}\label{prop:WY_equals_W}
For all times $s>0$ it holds almost surely that either $\bfY^{N,1} (s) = \dagger$ or
\begin{equation}\label{eq:WY_equals_W}
    W^{N,\jN} (T) = \WY^{N,1} (T,s).
\end{equation}
\end{proposition}

\begin{proof}
For the proof we work inductively along the jumps made by $\bfX^{N}$ and $\bfY^{N,1}$.
As outlined above, $\bfX^N$ is constructed from $\omega^N$ and $\bfY^{N,1}$  is constructed from
\begin{equation*}
    \omega^{N,-1,1} = \big\{ (\m^{-1,1}, t) : (\m, t) \in \omega^N \big\}.
\end{equation*}
For any $s>0$, there are almost surely only finitely many elements in
\begin{equation*}
    \omega_{[T-s, T]}^{N, -1, 1} \defeq \omega^{N,-1,1} \; \cap \; \big( \cG^{N,-1,1} \times [T-s, T] \big).
\end{equation*}
Denote the number of elements by $n_\omega = |\omega_{[T-s, T]}^{-1}|$. We can then order the set by the second component:
\begin{equation*}
\omega_{[T-s, T]}^{-1} = \big\{ (\m_{n_\omega}^{-1}, t_{n_\omega},), (\m_{{n_\omega}-1}^{-1}, t_{{n_\omega}-1}),..., (\m_1^{-1}, t_1) \big\}
\end{equation*}
such that $T-s < t_{n_\omega} < t_{{n_\omega}-1} < ... < t_1 < t_0 \defeq T$. Define $s_i \defeq T-t_i$ for all $i \in \{0,...,{n_\omega}\}$.

Now we can start with the induction.
For $s_0=0$ the statement holds, since 
\begin{align*}
 W_\bfY^{N, 1} (T, 0)&= \frac{1}{\jN} \sum_{(m, n) \in [M] \times [N]} Y_{m, n}^{N,1} (0) \delta_{X^N_{m, n} ((T-0)-)}\mybreak
 &=\frac{1}{\jN} \sum_{n \in \cI^N} \delta_{X^N_{m, n} (T)} = W^{N,\jN} (T),
\end{align*}
where we have used the definition of $\bfY^{N,1}(0)$ and the fact that $\bfX^N$ does not jump at time $T$ almost surely.

For the induction step, assume that \eqref{eq:WY_equals_W} holds for all $s' \in [0, s_i]$ for some $i \in \{1, ... ,{n_\omega}-1\}$.
If $s' \in (s_i, s_{i+1})$ then,
according to the induction hypothesis,
\begin{equation*}
W_\bfY^{N, 1} (T, s')=W^{N, \jN}(T)    
\end{equation*}
since ${\bf Y}^{N,1}$ and ${\bf X}^N$ are constant on the time interval $[s_i, s_{i+1})$. 
Now let $s'= s_{i+1}$ and assume that $\bfY^{N,1} (s_{i+1}) \neq \dagger$.
By the definition of $\bfY^{N,1}$ this directly implies that also for any $0 \leq s'' \leq s'$ it holds that $\bfY^{N,1} (s'') \neq \dagger$ and in particular for $\bfY^N (s_i)$.
Therefore, given $\bfY^{N,1} (s_{i+1}) \neq \dagger$, we have by the induction hypothesis and by using the coupled construction of $\bfX^{N}$ and $\bfY^{N,1}$ that

\begin{align*}
 W_\bfY^{N, 1} \big(T, s_{i+1}\big)&\stackrel{\text{Def. } W_\bfY^{N, 1}}{=} \frac{1}{\jN} \sum_{(m, n) \in [M] \times [N]} Y_{m, n}^{N,1} (s_{i+1}) \delta_{X^{N}_{m, n} ((T-s_{i+1})^-)}\mybreak
 & \stackrel{\text{Def.\ } \bfY^{N, 1}}{=} \frac{1}{\jN} 
 \sum_{(m, n) \in [M] \times [N]} \pig( \m^{-1,1}_{i+1} \big({\bf Y}^{N,1}(s_i)\big) \pig)_{m,n} \delta_{X_{m,n}^N((T-s_{i+1})^-)} \mybreak
 & \stackrel{\text{Prop.\ } \ref{prop:maps_and_their_duals_phase_1}}{=} \frac{1}{\jN} 
 \sum_{(m, n) \in [M] \times [N]} Y_{m, n}^{N,1} (s_i) \delta_{(\m_{i+1}(\bfX^N((T-s_{i+1})^-)))_{m,n}}
 \mybreak
 & \stackrel{\text{Def. } \bfX^N}{=} \frac{1}{\jN} 
 \sum_{(m, n) \in [M] \times [N]} Y_{m, n}^{N,1} (s_i) \delta_{X_{m,n}^N(T-s_{i})} \mybreak
 & \stackrel{\text{ind.}}{=} W^{N,\jN}(T),
\end{align*}
 where we used Proposition \ref{prop:maps_and_their_duals_phase_1} in the third equation and the induction hypothesis in the last equation.

\end{proof}

Proposition \ref{prop:WY_equals_W} already gave a hint on how the dual process $\bfY^{N,1}$ may help computing $W^{N,\jN}$.
However, not much is known about $\bfY^{N,1}$ so far.
Therefore, we explore some properties of this process in the following paragraphs.
In particular, we will rigorously show that the genealogy of the sample in the first phase can be represented by a Hoppe urn.
But first we state another property of $\bfY^{N,1}$.
For simplicity we defined $\bfY^{N,1}$ with state space $S_\bfY^{N,1} = [\jN]_0^{[M] \times [N]} \cup \{\dagger\}$ but in fact much less states are reached.

\begin{proposition}
For all times $s \geq 0$ it holds almost surely that either $\bfY^{N,1} (s) = \dagger$ or
\begin{equation}\label{eq:lines_are_single}
\sum_{n \in [N]} \1_{Y^{N,1}_{m,n} (s) } \leq 1 \text{ for all } m \neq \tm.
\end{equation}
\end{proposition}
\begin{proof}
 Condition \eqref{eq:lines_are_single} is satisfied for $s=0$, since by definition $Y^{N,1}_{m,n} (0) = 0$ whenever $m \neq \tm$.
 To see that the condition holds for arbitrary $s > 0$, one can work inductively along the random maps arriving until time $s$ and realize that all of the maps from Definition \ref{def:dual_1} map
 \begin{itemize}
  \item $\dagger$ to $\dagger$ and
  \item states satisfying equation \eqref{eq:lines_are_single} to $\dagger$ or to states satisfying equation \eqref{eq:lines_are_single}.
 \end{itemize}
\end{proof}

Now we want to establish the relationship of $\bfY^{N,1}$ with Hoppe's urn.
To this end, we transform $\bfY^{N,1}$ with a function
\begin{equation*}
    f:S_\bfY^{N,1} \rightarrow \cVs  \times \cP \big( [\jN] \times [M] \times [N] \big) \cup \big\{(\dagger, \dagger)\big\}
\end{equation*}
where
\begin{equation*}
    f(y) =
    \begin{cases}
        (\dagger, \dagger), &\text{if } y = \dagger,\\[0.5em]
        \pig(\bm{s} \big(\mset{y_{\tm, n} : n \in [N], y_{\tm, n} \neq 0}\big),\\
         \phantom{xx} \big\{(y_{m, n}, m, n) : n \in [N], m \neq \tm, y_{\tm, n} \neq 0\big\}\pig), &\text{otherwise.}
    \end{cases}
\end{equation*}
That is, the first component of
\begin{equation*}
f\big(\bfY^{N,1}\big) = \big(f^{(1)}(\bfY^{N,1}),f^{(2)}(\bfY^{N,1})\big)    
\end{equation*}
tracks the ordered family sizes within the sampled host and the second component tracks the family sizes outside of the sampled host together with their virus and host index.
This way, the type frequencies can be rewritten another way as long as $\bfY^{N,1} (s) \neq \dagger$:
\begin{equation}\label{eq:typefrequencies}
    W_f^{N} (T,s) = \frac{1}{\jN} \Bigg(\sum_{n \in [N]} Y^{N,1}_{\tm, n} (s) \delta_{X^N_{\tm,n} (T-s)} + \sum_{\substack{(y_{m,n}, m, n) \\ \in f^{(2)}(\bfY^{N,1}(s))}} y_{m,n} \delta_{X^{N}_{m,n} (T-s)} \Bigg).
\end{equation}
When not keeping track of the indices of the host and virus of ancestral lines not located in the sampled host we arrive at the following function  $g:S_\bfY^{N,1} \rightarrow \cV \times \cV$
\begin{equation*}
    g   (y) =
    \begin{cases}
        (\dagger, \dagger), &\text{if } y = \dagger\\[0.5em]
        \pig( \bm{s}\big(\mset{y_{\tm, n} : n \in [N], y_{\tm, n} \neq 0}\big),\\
        \phantom{xxx} \bm{s}\big(\mset{   y_{m, n} : n \in [N], m \neq \tm, y_{\tm, n} \neq 0}\big)\pig), &\text{otherwise.}
    \end{cases}
\end{equation*}

\begin{definition}
    We define the first time at which all ancestral lines have left the sampled host $\widetilde{m}$ (or $\dagger$ was reached) by
    \begin{equation}\label{eq:tau1}
        \tau_1^N \defeq \inf \Big\{ s \geq 0: f^{(1)} \big(\bfY^{N,1} (s)\big) \in \big\{\dagger, () \big\} \Big\} .
\end{equation}
\end{definition}

\begin{remark}
Note that from the perspective of $\bfY^{N,1}$, the role of $\tau_1^N$ is that of a stopping time.
From the perspective of $\bfX^N$ it is an ``independent time'', since  $\bfX^N (T-t)$ only depends on $\omega_{(-\infty,T-t)}$ while $\tau_1^N$ only depends on $\omega_{(T-t, \infty)}$.
The two random sets are independent by the definition of Poisson point processes.
More precisely, for any  $A \in \cP \big(S^N \big)$ it holds that
\begin{align*}
\IP \pig( \bfX^N \big(T-\tau_1^N\big) \in A\pig) &= \int \IP \pig( \bfX^N (T-t) \in A \; | \; \{ \tau_1^N = t \}\pig) d F_{\tau_1^N}(t)\\
&= \int \IP \pig(\bfX^N (T-t) \in A\pig) d F_{\tau_1^N}(t)\\
&= \pi^N (A).
\end{align*}
Here, we have used that $\big\{\bfX^N (T-t) \in A\big\}$ is independent of $\{ \tau_1^N = t \}$ and the law of total probability where $F_{\tau_1^N}$ is the CDF of $\tau_1^N$.
\end{remark}
Now we define the number of ancestral lines at the end of the first phase by 
\begin{equation*}
\xi^N \defeq \big\lvert g^{(2)} \big(\bfY^{N,1} (\tau_1^N)\big) \big\rvert
\end{equation*}
and
\begin{equation}\label{eq:w_g}
W_g^{N} \defeq \frac{1}{\jN} \sum_{i=1}^{\xi^N} g^{(2)}_i \big(\bfY^{N,1} (\tau_1^N)\big) \delta_{X_{i,1}^N (T-\tau_1^N)}.
\end{equation}
This quantity corresponds to a coloring of the family sizes $g^{(2)}_i \big(\bfY^{N,1} (\tau_1^N)\big)$ of the sample in the first phase with the types of the particles with index $1$ in the first $\xi^N$ hosts, given by $\big( X_{i,1}^N\big)_{i \in [\xi^N]}$. We can assume this ordering without loss of generality due to exchangeability.

\begin{proposition}\label{prop:phase1_something}
For any $s>0$ it holds that
\begin{equation}\label{eq:Wf_equals_W}
    \IP \pig( W^{N, \jN} (T) = W_f^N (T,s) \; \big\vert \; \bfY^{N,1} (s) \neq \dagger \pig) = 1.
\end{equation}
This in particular implies that
\begin{equation}
 \IP \Big( W^{N,\jN} (T) = \sum_{\substack{(y_{m,n}, m, n)\\ \in f^{(2)}(\bfY^{N,1}(\tau_1^N))}} y_{m,n} \delta_{X^N_{m,n} (T-\tau_1^N)}  \; \big\vert \; \bfY^{N,1} \big(\tau^N_1 \big) \neq \dagger \Big) = 1.
\end{equation}
Furthermore, conditioned on $Y^{N,1}(\tau_1)\neq \dagger$
\begin{equation*}
 W_g^N \dequal W_f^N (T, \tau_1).
\end{equation*}
\end{proposition}
\begin{proof}
 The first two equations 
 follow by construction. 
 The last equation is a consequence of the exchangeability of the sampled lines %
 and that at time $\tau_1$ %
 there is at most one line per host %
 if $Y^{N,1}(\tau_1)\neq \dagger$.
\end{proof}

\begin{lemma}[Distribution of the Family Sizes]\label{lem:g2Y}%
There exists a random variable $Z^N$ taking values in $\cVs$ such that
\begin{equation*}
     \IP \Big( g^{(2)} \pig(\bfY^{N,1} \big(\tau^N_1 \big) \pig) = Z^N \; \big\vert \; \bfY^{N,1} \big(\tau_1^N \big) \neq \dagger \Big) = 1
\end{equation*}
and $Z^N$ has the distribution of the family sizes at absorption of a reversed Hoppe urn with $\jN$ initial particles and parameter $\theta_N = \frac{\lambda_N}{\gamma_N}$.
Furthermore, it holds that
\begin{equation*}\label{eq:phase_1_whp}
 \lim_{N \to \infty}\IP \pig( \bfY^{N,1} (\tau_1^N) = \dagger\pig) = 0 .
\end{equation*}
\end{lemma}

\begin{proof}
To prove the first statement, we couple $g \big(\bfY^{N,1}\big)$ with a process $\bftZ$.
We first define $\bftZ$ and then verify that it is indeed a Hoppe urn.
After that, we prove the second statement.
For simpler notation we will suppress the dependencies on $N$ in this proof and additionally write $\bfY = \bfY^{N,1}$.

We define a stochastic process $\big(\bftZ(s)\big)_{s \geq 0}$ taking values in $\cVs \times \cVs$ with initial state
\begin{equation*}
    \bftZ (0) = g\big(\bfY (0)\big) = \big( \underbrace{(1,...,1)}_{\jN} ,()\big).
\end{equation*}
As before, denote by
\begin{equation*}
\big\{ (\m_1^{-1}, s_1,), (\m_{2}^{-1}, s_{2}),...\big\}
\end{equation*}
with $0\eqdef s_0 < s_1 < s_2 < ... $ the random maps and the jump times used to construct $\bfY$, where $s_i = T - t_i$. \\
Now set
\begin{equation*}
 \bftZ (s) = g\big(\bfY (0)\big) \text{ for } s \in (0,s_1)
\end{equation*}
and for any $i \in \IN$, $s \in [s_i, s_{i+1})$ set
\begin{equation*}
\bftZ (s) =
\begin{cases}
    g\pig(\m_i \big(\bfY (s_i^-)\big)\pig)      &\text{if } \bfY (s_i^-) \neq \dagger,\; \m_i \big(\bfY (s_i^-)\big) \neq \dagger,\\
                                                &\phantom{\text{, }}\\
    \bftZ (s_i^-)                               &\text{if } \bfY (s_i^-) \neq \dagger,\; \m_i^{-1} \big(\bfY (s_i^-)\big) = \dagger \text{ and }\\
                                                &\phantom{\text{, }}\m_i \neq \reinf_{m, n, \tm, n_2} \text{ for all } m,n,n_2,\\
                                                &\phantom{\text{, }}\\                                    
    \pig( \bftZ^{(1)} (s_i^-) \;\ominus \; Y_{\tm, n_2} (s_i^-),
                                                &\text{if } \bfY (s_i^-) \neq \dagger,\;  \m_i^{-1} \big(\bfY (s_i^-)\big) = \dagger \text{ and }\\
    \phantom{xxx}\bftZ^{(2)} (s_i^-) \;\oplus \; Y_{\tm, n_2} (s_i^-) \pig)   
                                                &\phantom{\text{, }}\m_i = \reinf_{m, n, \tm, n_2} \text{ for some } m,n,n_2,\\
                                                &\phantom{\text{, }}\\
    \pig( \bftZ^{(1)} (s_i^-) \;\ominus Z^{(1)}_{n_2} (s_i^-),
                                                &\text{if } \bfY (s_i^-) = \dagger, \; \m_i^{-1} \big(\bfY (s_i^-)\big) = \dagger \text{ and }\\
    \phantom{xxx}\bftZ^{(2)} (s_i^-) \;\oplus \; Z^{(1)}_{n_2} (s_i^-) \pig)
                                                &\phantom{\text{, }}\m_i = \reinf_{m, n, \tm, n_2} \text{ with } n_2 \leq \bfell \big( \bftZ^{(1)} (s_i^-) \big) ,\\
                                                &\phantom{\text{, }}\\
    \pig( \bftZ^{(1)} (s_i^-) \;\ominus Z^{(1)}_{n_1} (s_i^-) \;\ominus Z^{(1)}_{n_2} (s_i^-)
                                                &\text{if } \bfY (s_i^-) = \dagger, \; \m_i^{-1} \big(\bfY (s_i^-)\big) = \dagger \text{ and }\\
    \phantom{xxx}\oplus  \big( Z^{(1)}_{n_1} (s_i^-) + Z^{(1)}_{n_2} (s_i^-)\big), \; \bftZ^{(2)} (s_i^-)\pig)
                                                &\phantom{\text{, }}\m_i = \rep_{\tm,n_1, n_2} \text{ with } n_1, n_2 \leq \bfell \big( \bftZ^{ (1)} (s_i^-) \big) \\
                                                &\phantom{\text{, }}\text{with } n_1 \neq n_2,\\
                                                &\phantom{\text{, }}\\
    \bftZ (s_i^-)                               &\text{otherwise,}
\end{cases}
\end{equation*}
where $\oplus$ and $\ominus$ are defined as in Equations \eqref{eq:def_plus} and \eqref{eq:def_minus} of the last section.
Next, we prove that $\bftZ$ is indeed a reversed Hoppe urn.
To this end, note that for  $s, \delta > 0$, $z =(z^{(1)}, z^{(2)}) \in \cVs \times \cVs$   and $k \in \IN$
\begin{align*}
&\IP \pig( \bftZ (s+\delta) = (z^{(1)} \; \ominus \; k , z^{(2)} \; \oplus \; k ) \; | \; \bftZ (s) = z\pig)\\[0.5em]
&= \underbrace{\IP \pig( \bftZ (s+\delta) = (z^{(1)} \; \ominus \; k , z^{(2)} \; \oplus \; k ) \; | \; \bftZ (s) = z, \bfY (s) = \dagger \pig)}_{(*)} \IP \pig( \bfY (s) = \dagger \; | \; \bftZ (s) = z \pig)\\[0.5em]
&\phantom{===} + \underbrace{\IP \pig( \bftZ (s+\delta) = (z^{(1)} \; \ominus \; k , z^{(2)} \; \oplus \; k ) \; | \; \bftZ (s) = z, \bfY (s) \neq \dagger \pig)}_{(**)}\IP \pig( \bfY (s) \neq \dagger \; | \; \bftZ (s) = z\pig)
\end{align*}
by the law of total probability.
For the first term we get by definition of the PPP and of $\bftZ$ that
\begin{align*}
(*) = \IP \pig( \big\{&\exists! (\m^{-1}, s') \in \omega^{-1} \; \cap \cG^{-1} \times (s, s+\delta) \big\}\\ \cap \big\{ & \m^{-1} = \reinf_{m,n,\tm,i}^{-1} \text{ for some } m \in [M], n \in [N] \text{ and } i \in [\bfell (z^{(1)})] \text{ s.t. } z^{(1)}_i = k \big\} \pig)\\
 +o(\delta&) .
\end{align*}
Since the two components of $\omega^{-1}$ are independent and the random maps are i.i.d., this equals
\begin{align*}
    &\IP \pig( \big\{\exists! (\m^{-1}, s') \in \omega^{-1} \; \cap \cG^{-1} \times (s, s+\delta) \big\}\pig)\\
    &\cdot \IP\pig( \big\{\m^{-1} = \reinf_{m,n,\tm,i}^{-1} \text{ for some } m \in [M], n \in [N] \text{ and } i \in [\bfell (z^{(1)})] \text{ s.t. } z^{(1)}_i = k \big\} \pig) + o(\delta).
\end{align*}
There are $\bfm (z^{(1)}, k) \cdot M \cdot N$ many reinfection maps satisfying the condition in the latter probability and each of the maps has probability $\frac{\lambda_N}{N^2 M r_N}$.
Therefore we get
\begin{align*}
    (*) &= \big(\delta + o(\delta)\big) r_N \frac{\lambda_N}{N^2 M r_N} \bfm (z^{(1)}, k) \cdot M \cdot N\\
        &= \big(\delta + o(\delta)\big) \frac{\lambda_N}{N} \bfm (z^{(1)}, k).
\end{align*}
For the second term we get by the law of total probability that
\begin{align*}
    (**) &= \IP  \pig( \bftZ (s+\delta) = (z^{(1)} \; \ominus \; k , z^{(2)} \; \oplus \; k ) \; | \; \bftZ (s) = z, \bfY (s) \neq \dagger \pig)\\[.5em]
    &=\sum_{y \in S_\bfY^{N,1} \setminus \{\dagger\}} \IP \pig( \bfY (s) =y \; | \; \bftZ (s) = z, \bfY (s) \neq \dagger \pig)\\[-0.8em]
    &\phantom{=\sum_{y \in S_\bfY^{N,1} \setminus \{\dagger\}}} \cdot \underbrace{\IP \pig( \bftZ (s+\delta) = (z^{(1)} \; \ominus \; k , z^{(2)} \; \oplus \; k ) \; | \; \bftZ (s) = z, \bfY (s) = y \pig)}_{(***)}
\end{align*}
where we use the convention that $\IP(A|B)=0$ if $\IP(B)=0$.
Note in particular that
\begin{equation*}
    \IP \big(\bftZ (s) = z, \bfY (s) = y \big)=0
\end{equation*}
and
\begin{equation*}
    \quad \IP \big( \bfY (s) =y \; | \; \bftZ (s) = z, \bfY (s) \neq \dagger \big) = 0
\end{equation*}
whenever $z \neq g(y)$ and $y \neq \dagger$.
Therefore, $(***)$ equals either zero or
\begin{align*}
     \IP \pig(\bftZ &(s+\delta) = (z^{(1)} \; \ominus \; k , z^{(2)} \; \oplus \; k ) \; \big\vert \; \bftZ (s) = g(y), \bfY (s) = y \pig)\\[0.5em]
     = \IP \pig( \big\{&\exists! (\m^{-1}, s') \in \omega^{-1} \; \cap \cG^{-1} \times (s, s+\delta)\; \big\}\\ \cap \; &\big\{\m^{-1} = \reinf_{m,n,\tm,i}^{-1} \text{ for some } m \in [M], n \in [N] \text{ and } i \in [N] \text{ s.t. } y_{\tm,i} = k \big\} \pig)
 +o(\delta)
\end{align*}
by definition of $\omega$ and $\bftZ$.
Since
\begin{equation*}
z^{(1)} = g^{(1)}(y) = \bm{s} \big(\mset{y_{\tm, n} : n \in [N], y_{\tm, n} \neq 0} \big),    
\end{equation*}
there are exactly $\bfm (z^{(1)}, k)$ many entries of $y$ that are equal to $k$.
Thus, again $\bfm (z^{(1)}, k) \cdot M \cdot N$ reinfection maps satisfy the condition in the last probability.
This yields
\begin{align*}
    \IP \pig(\bftZ (s+\delta) = (z^{(1)} \; \ominus \; k&, z^{(2)} \; \oplus \; k ) \; | \; \bftZ (s) = g(y), \bfY (s) = y \pig)\\
    = \big(&\delta + o(\delta)\big) r_N \frac{\lambda_N}{N^2 M  } \bfm (z^{(1)}, k) \cdot M \cdot N r_N
\end{align*}
Now it is straightforward to check that also
\begin{equation*}
    \IP \pig( \bftZ (s+\delta) = (z^{(1)} \; \ominus \; k , z^{(2)} \; \oplus \; k ) \; | \; \bftZ (s) = z\pig) = \big(\delta + o(\delta)\big) r_N \frac{\lambda_N}{Nr_N} \bfm (z^{ (1)}, k).
\end{equation*}
Thereby we have derived an expression for Part $(i)$ of the definition of the reversed Hoppe urn.
Now we turn to Part $(ii)$: for $s, \delta > 0$, $z =(z^{(1)}, z^{(2)}) = \big((z_1^{(1)},... z_p^{(1)}),(z_1^{(2)},... z_q^{(2)}) \big) \in \cVs \times \cVs$ and $i,j \in \IN$ with $i \neq j$,
\begin{align*}
    \IP &\pig( \bftZ (s+\delta) = \big(z^{(1)} \; \ominus \; i \; \ominus \; j \; \oplus \, (i+j), z^{(2)} \big) \; | \; \bftZ (s) = z\pig)\\[0.5em]
    =&\underbrace{\IP\pig( \bftZ (s+\delta) = \big(z^{(1)} \; \ominus \; i \; \ominus \; j \; \oplus \, (i+j), z^{(2)} \big) \; | \; \bfY(s) = \dagger , \bftZ (s) = z\pig)}_{(\diamondsuit)} \IP \big( \bfY(s) = \dagger \; | \; \bftZ (s) = z\big)\\[0.5em]
    +&\underbrace{\IP \pig( \bftZ (s+\delta) = \big(z^{(1)} \; \ominus \; i \; \ominus \; j \; \oplus \, (i+j), z^{(2)}\big) \; | \; \bfY(s) \neq \dagger , \bftZ (s) = z\pig)}_{(\diamondsuit \diamondsuit)} \IP \big( \bfY(s) \neq \dagger \; | \; \bftZ (s) = z\big)
\end{align*}
again by the law of total probability. By definition of the PPP and of $\bftZ$ we have that
\begin{align*}
    (\diamondsuit) = \IP \pig( \big\{&\exists! (\m^{-1}, s') \in \omega^{-1} \; \cap \cG^{-1} \times (s, s+\delta)\big\}\\ 
    \cap \; & \big\{\m^{-1} = \rep_{\tm,n_1, n_2}^{-1} : n_1 \neq n_2 \text{ and }\{z_{n_1}, z_{n_2}\} = \{i,j\} \big\} \pig)\\
 +o(\delta&). 
\end{align*}
There are exactly $\bfm(z^{(1)},i) \big(\bfm(z^{(1)},j)-1_{i=j} \big)$ reproduction maps satisfying the condition and each of the maps has probability $\frac{\gamma_N}{r_NN}$.
Therefore, and due to independence, similarly to the calculations above, $( \diamondsuit )$ equals
\begin{equation*}
    \big(\delta + o(\delta)\big) r_N \frac{\gamma_N}{r_N N} \bfm(z^{(1)},i) \big(\bfm(z^{(1)},j)-1_{i=j} \big)
\end{equation*}
For the term $(\diamondsuit \diamondsuit)$ we proceed as above and apply the law of total probability to arrive at
\begin{align*}
    (\diamondsuit \diamondsuit) =
    \sum_{y \in S_\bfY^{N,1} \setminus \{\dagger\}}& \underbrace{\IP \pig( \bftZ (s+\delta) = \big(z^{(1)} \; \ominus \; i \; \ominus \; j \; \oplus \, (i+j), z^{(2)}\big) \; | \; \bftZ (s) = z, \bfY (s) = y \pig)}_{(\diamondsuit \diamondsuit \diamondsuit)} \\
    &\cdot \IP \pig( \bfY (s) =y \; \big\vert \; \bftZ (s) = z, \bfY (s) \neq \dagger \pig).
\end{align*}
Again, $(\diamondsuit \diamondsuit \diamondsuit)$ equals either zero or
\begin{align*}
    \IP \big( \bftZ (s+\delta&) = (z^{(1)} \; \ominus \; i \; \ominus \; j \; \oplus \, (i+j), z^{(2)}) \; | \; \bftZ (s) = g(y) , \bfY (s) = y \big)\\
    = \IP \big(\{&\exists! (\m^{-1}, s') \in \omega^{-1} \; \cap \cG^{-1} \times (s, s+\delta)\; \}\\ \cap \; &\{\m^{-1} = \rep_{\tm,n_1,n_2}^{-1} \text{ for some } n_1 \neq n_2 \in [N] \text{ s.t. } \{y_{\tm, n_1}, y_{\tm, n_2}\} = \{i,j\} \} \big)
\end{align*}
Since $z^{(1)} = g^{(1)}(y) = \bm{s}(\mset{y_{\tm, n} : n \in [N], y_{\tm, n} \neq 0})$, there are exactly $\bfm (z^{(1)}, i)$ (resp. $\bfm (z^{(1)}, j)$) many entries of $y$ that are equal to $i$ (resp. $j$).
Therefore, there are $\bfm (z^{(1)}, i) (\bfm(z^{(1)},j)-\1_{i=j})$ many pairs of indices $n_1$ and $n_2$ with $\{y_{\tm, n_1}, y_{\tm, n_2}\} = \{i,j\}$.
By the same calculations as for Part $(i)$ we get that
\begin{align*}
    (\diamondsuit \diamondsuit) = \big(\delta + o(\delta)\big)r_N\frac{\gamma_N}{r_NN} \bfm(z^{(1)},i) \big(\bfm(z^{(1)},j)-1_{i=j} \big)
\end{align*}
Putting everything together, we get
\begin{equation*}
    \lim_{\delta \searrow 0} \frac{1}{\delta}\IP \big( \bftZ (s+\delta) = (z^{(1)} \; \ominus \; k , z^{(2)} \; \oplus \; k ) \; | \; \bftZ (s) = z\big) = \frac{\lambda_N}{N} \bfm (z^{(1)}, k)
\end{equation*}
and 
\begin{equation*}
\lim_{\delta \searrow 0} \frac{1}{\delta}\IP \pig( \bftZ (s+\delta) = (z^{(1)} \; \ominus \; i \; \ominus \; j \; \oplus \, (i+j), z^{(2)}) \; | \; \bftZ (s) = z\pig) =  \frac{\gamma_N}{N} \bfm(z^{(1)},i) \big(\bfm(z^{(1)},j)-1_{i=j} \big)
\end{equation*}
which coincide with the transition rates of a Hoppe's urn with speed $\frac{\gamma_N}{N}$ and parameter $\theta_N = \frac{\lambda_N}{\gamma_N}$.
By definition it is clear that $\bftZ(s) = g \big(\bfY(s) \big)$ whenever $\bfY(s) \neq \dagger$.
Therefore the absorption time of the reversed Hoppe urn,
\begin{equation*}
    \tau_{\text{abs}} = \inf \big\{ s \geq 0 : \bftZ^{(1)}(s) = () \big\}
\end{equation*}
coincides with $\tau_1$ on the event that $\bfY(\tau_1) \neq \dagger$, i.e.
\begin{equation*}
    \IP (\tau_1 = \tau_{\text{abs}} \; | \; \bfY(\tau_1) \neq \dagger) = 1
\end{equation*}
and therefore also
\begin{equation*}
    \IP \pig(\bftZ(\tau_{\text{abs}})= g\big( \bfY(\tau_1) \big) \; | \; \bfY(\tau_1) \neq \dagger \pig) = 1.
\end{equation*}
Hence, $Z^N \defeq \bftZ^{(2)}(\tau_{\text{abs}})$ is the desired random variable from the statement of the Lemma.

Now only the asymptotic bound for $\IP \big(\bfY(\tau_1)=\dagger \big)$ remains to be proven.
Recalling that \mbox{$\bfell : \cV \rightarrow \IN$} was defined to be the length of a vector, we use the convention $\bfell (\dagger) = \dagger$ from here on.
It is easy to verify that $g \big(\bfY(s)\big)_{s \geq 0}$ and
\begin{equation*}
    g^{\#} \big(\bfY (s) \big)_{s \geq 0} = \pig( g^{\#,(1)} \big(\bfY (s) \big), g^{\#,(2)} \big(\bfY (s) \big) \pig) \defeq  \Big( \bfell \pig(g^{(1)} \big(\bfY(s)\big)\pig), \bfell \pig(g^{(2)} \big(\bfY(s)\big)\pig) \Big)_{s \geq 0}
\end{equation*}
are both Markov processes with state spaces $\cV \times \cV \cup \big\{(\dagger, \dagger) \big\}$ and $\IN_0 \times \IN_0 \cup \big\{(\dagger, \dagger) \big\}$, respectively.
The process $g \big(\bfY(s)\big)_{s \geq 0}$ has the interpretation that it keeps track of the family sizes (that is, the number of progeny of each ancestral line) within the sampled host in its first component and outside of it in its second component.
The process $g^{\#} \big(\bfY (s) \big)_{s \geq 0}$ is more crude and only counts the numbers of ancestral lines within and outside of the sampled host. 
Let $s_i$ be the jump times of $g \big(\bfY(s)\big)_{s \geq 0}$. These coincide with the jump times of $g^{\#} \big(\bfY (s)\big)_{s \geq 0}$.
Then,
\begin{align*}
    \big\{\bfY &(\tau_1) \neq \dagger \big\} = \pig\{g\big(\bfY(\tau_1)\big) \neq \dagger \pig\}\\[0.5em]
    = \pig\{&g^\# \big(\bfY(\tau_1) \big) \neq \dagger \pig\}\\[0.6em]
    = \pig\{&g^\# \big(\bfY(s_1) \big) \neq \dagger, g^\# \big(\bfY(s_2)\big) \neq \dagger, \dots, g^\# \big(\bfY(s_{\jN}) \big) \neq \dagger \pig\}.
\end{align*}
Note that almost surely $g^{\#, (1)} \big(\bfY(s_i)\big) \in \{\jN - i, \dagger \}$ and $g^{\#, (2)} \big(\bfY(s_i) \big) \in \{0,...,i\} \; \cup \; \{\dagger \}$.
Therefore, it holds that
\begin{align*}
 \IP \big(\bfY (\tau_1) \neq \dagger\big) &= \IP \pig(g^\# \big(\bfY(s_1)\big) \neq \dagger, g^\#(\bfY(s_2)) \neq \dagger, \dots, g^\# \big(\bfY(s_{\jN}) \big) \neq \dagger \pig)\\[0.7em]
 &= \IP  \pig(g^{\#,(2)} \big(\bfY(s_1) \big) \neq \dagger, g^{\#,(2)}\big(\bfY(s_2) \big) \neq \dagger, \dots, g^{\#,(2)} \big(\bfY(s_{\jN}) \big) \neq \dagger \pig)\\
 &\geq \prod_{i=0}^{\jN - 1} \min_{k \in \{0,...,i\}}  \IP \pig( g^{\#,(2)}\big(\bfY(s_{i+1})\big) \neq \dagger \; \big\vert \; g^{\#,(2)} \big(\bfY(s_i) \big) = k \pig)\\[0.7em]
 &\geq \pig( 1 - \max_{k \in \{0,...,\jN - 1\}}  \IP \pig( g^{\#,(2)} \big(\bfY(s_{i+1})\big) = \dagger \; \big\vert \; g^{\#,(2)}\big(\bfY(s_i)\big) = k \pig) \pig)^{\jN}\\[0.7em]
 &\geq 1 - \jN \max_{k \in \{0,...,\jN-1\}}  \IP \pig( g^{\#,(2)} \big(\bfY(s_{i+1})\big) = \dagger \; \big\vert \; g^{\#,(2)}\big(\bfY(s_i) \big) = k \pig).
\end{align*}
We want to show that $\IP \big(\bfY (\tau_1) \neq \dagger \big) \to 1$ as $N \to \infty$ by proving that the maximum in the last line is asymptotically of smaller order than $\frac{1}{\jN}$.
The argument of the maximum can be upper bounded by dividing the rate at which $g^{\#,(2)}\big(\bfY( \; \cdot \;)\big)$ jumps from state $k$ to the $\dagger$-state by the rate at which it jumps from state $k$ to state $k+1$ (which is a lower bound for the total jump rate from state $k$):
\begin{align*}
    \IP &\pig( g^{\#,(2)}\big(\bfY(s_{i+1})\big) = \dagger \; \big\vert \; g^{\#,(2)}\big(\bfY(s_i) \big) = k \pig)\\[0.5em]
    &\leq \frac{\frac{\lambda_N}{N^2M}(\jN-i+k)(k+1)N+\frac{\mu_N}{N}(k-i+\jN)+\frac{\rho_N}{N}(k-i+\jN) + 1 + \frac{1}{M}k(k+1)}{\frac{\lambda_N}{N^2 M}(\jN-i)(M-k-1)N}\\[0.5em]
    &\leq \frac{\frac{\lambda_N}{M}\jN^2+\mu_N \jN+\rho_N\jN + 1 + \frac{N}{M}\jN^2}{\frac{\lambda_N}{2}} \ll \frac{1}{\jN}
\end{align*}
as $N \to \infty$.
The numerator in the second line has the following interpretation: the first term is the rate at which the process jumps to $\dagger$ due to an unwanted reinfection.
The second and third term are the rates at which an ancestral line is hit by a mutation or recombination.
The fourth term is the rate at which the sampled host is replaced and the last term is the rate at which other unwanted host replacement events happen.
The denominator is the rate at which ancestral lines in the sampled host are hit by reinfections.
To justify that the whole expression is of smaller order than $\frac{1}{\jN}$ in the last step, we have used the rate assumptions given in Assumption \ref{assumptions:main_rates} and properties of the sample size $\jN$ given in Lemma \ref{assumptions:rates}.
Specifically, we have used point \ref{ass:sample_size} of Lemma \ref{assumptions:rates} for the first term, point \ref{ass:rep_reinf} of Assumption \ref{assumptions:main_rates} and point \ref{ass:timescales} of Lemma \ref{assumptions:rates} for the second and third term, point \ref{ass:timescales} of Lemma \ref{assumptions:rates} for the fourth term and all mentioned points for the last term.

\end{proof}

Under the rate assumptions the limit $\theta \defeq \lim_{N \to \infty} \theta_N$ is well-defined.
The following lemma states that  the number of ancestral lines present at the end of the first phase, which we denote by 
\begin{equation*}
\xi^N \defeq \bfell \pig( g^{(2)} \big(\bfY^{N,1} (\tau_1^N)\big)\pig),
\end{equation*}
is concentrated around $\theta \log(\jN)$.

\begin{lemma}[Bound on $\xi^N$]\label{Lemma:ControllSizeOfXi}
Under the rate assumptions it holds that
\begin{equation*}
    \lim_{N \to \infty} \IP \pig( (1-\alpha)\theta \log (j_N) < \xi^N < (1+\alpha)\theta \log (j_N)\pig) = 1.
\end{equation*}
for any $\alpha > 0$.
\end{lemma}
\begin{proof}
    Lemma \ref{lem:g2Y} implies that
    \begin{align*}
        &\limsup_{N \to \infty} \IP \pig( \xi^N \notin \big( (1-\alpha) \theta \log (\jN), (1+\alpha) \theta \log (\jN)\big) \pig)\\
        & \leq \limsup_{N \to \infty} \IP \pig( \big\{ \xi^N \notin \big( (1-\alpha) \theta \log (\jN), (1+\alpha) \theta \log (\jN) \big) \big\} \; \cap \; \big\{ \bfY^{N,1} (\tau_1^N) \neq \dagger \big\} \pig)\\
        & \phantom{mm} + \limsup_{N \to \infty} \underbrace{\IP \big( \bfY^{N,1} (\tau_1^N) = \dagger \big)}_{\to 0}\\
        & \leq \limsup_{N \to \infty} \IP \pig( \big\{ H^N \notin \big( (1-\alpha) \theta \log (\jN), (1+\alpha) \theta \log (\jN) \big)\big\} \; \cap \; \big\{ \bfY^{N,1} (\tau_1^N) \neq \dagger \big\}\pig)\\
        & \leq \limsup_{N \to \infty} \IP \pig( H^N \notin \big( (1-\alpha) \theta \log (\jN), (1+\alpha) \theta \log (\jN) \big) \pig)
    \end{align*}
    where $H^N$ is the number of elements of the second component of the final state of a Hoppe's urn with $\jN$ initial particles and parameter $\theta_N$.
    Denote by $\mu_N$ and $\sigma^2_N$ the mean and variance of $H^N$  for which it holds that
    \begin{equation*}
        \mu_N \in [\big(1 - \tfrac{\alpha}{2} \big) \theta \log(\jN), \big(1 + \tfrac{\alpha}{2} \big) \theta \log(\jN) ]
        \text{ and }
        \sigma^2_N \leq 2 \theta \log (\jN)
    \end{equation*}
    for $N$ large enough, see %
    Lemma \ref{lem:hoppe_fam_sizes}.
    Then, for $N$ large enough and by Chebyshev's inequality
    \begin{align*}
        \IP \pig( H^N \notin \big( &(1-\alpha) \theta \log (\jN), (1+\alpha) \theta \log (\jN) \big) \pig)\\[0.7em]
        &\leq \IP \pig( \big\lvert H^N-\mu_N \big\rvert \geq \tfrac{\alpha}{2}\theta \log (\jN)\pig)\\[0.7em]
        &\leq \IP \pig(\big\lvert H^N -\mu_N \big\rvert \geq \tfrac{\alpha}{4} \sigma_N^2\pig)\\[0.7em]
        &\leq \frac{1}{\frac{\alpha^2}{8}\sigma_N^2}.
    \end{align*}
    Thus,
    \begin{equation*}
        \limsup_{N \to \infty} \IP \pig( \xi^N \notin \big( (1-\alpha) \theta \log (\jN), (1+\alpha) \theta \log (\jN)\big) \pig) \leq \lim_{N \to \infty} \frac{1}{\frac{\alpha^2}{8} \sigma^2_N} = 0
    \end{equation*}
    which proves the claim.
    
\end{proof}

\subsection{Second Phase}

At the beginning of the second phase the process ${\bf Y}^{N,1}$ has w.h.p.\ not jumped to the $\dagger$-state yet and the number of families resulting from the first phase $\xi^N$  is bounded by $(1\pm\alpha) \theta \log(\jN)$ w.h.p.\ for any $\alpha>0$, see Lemma \ref{Lemma:ControllSizeOfXi}.
In the following we condition on these two events (for some fixed $\alpha>0$).~Recall that the overall goal is to determine the distribution of type frequencies in a random sample denoted by $W^{N,\jN} (T)$.
In Section \ref{sec:first_phase}, a first step towards this goal was made by essentially rewriting it as $W_g^N$ given in Equation~\eqref{eq:w_g}.
Now we will further determine $W_g^N$ by studying \mbox{$\big(X_{i,1}^N (T-\tau_1^N)\big)_{i \in [\xi^N]}$}.

Due to the assumptions on the asymptotics of the rates, no relevant recombination events happened in the first phase of the genealogy.
Therefore, all ancestral loci of the sampled particles remained completely linked until time $\tau_1^N$ and it sufficed to consider $S_\bfY^{N,1} \defeq [\jN]_0^{[M] \times [N]} \cup \{\dagger\}$ as the state space for the dual process of the first phase.
This resulted in tracking ancestral particles of the sample and the number of their descendants in the sample.
More precisely, $$Y_{m,n}^{N,1} (s) = y_{m,n} \in [\jN]_0 \cup \{\dagger\}$$ had the interpretation that the virus particle at $(m,n)$ at time $T-s$ is ancestral to $y_{m,n}$ particles in the sample at time $T$, if $y_{m,n}\neq \dagger$.

In the second phase, the asymptotic rate assumptions cause particles to undergo recombination.
Therefore, the dual process will not only have to track ancestral \textit{particles} but also ancestral \textit{loci}, resulting in a bigger state space.
To this end, we define a dual process $\bfY^{N,2}$ of the second phase with state space $S_{\bfY}^{N,2}  \defeq\left([\jN]_0^L\right)^{M \times N} \; \cup \; \{ \dagger\}$.
For simplicity we denote generic states in the second phase again by $y$, though in this subsection they are elements of $S_{\bfY}^{N,2}$ (and not of $S_{\bfY}^{N,1}$).
Just like in the first phase, $\dagger$ is an artificially introduced state reached whenever unlikely events happen that complicate the analysis. At the beginning of Phase 2 all $\xi^N$ particles are located in different hosts. By a relabeling of hosts and viral particles, we can assume that at time $T- \tau^N_1$ the $\xi^N$ ancestral lines of the sampled viral particles are located in hosts 1, \dots, $\xi^N$, each at the position of the viral particle with label 1. Since in the phase we still do not consider the types of the viral particles at the different loci, we can associate to the $i$-th viral particle the vector $(i,...,i)$ of length $L$. More precisely, we will consider a process ${\bf Y}^{N,2}$ with initial state
\begin{equation*}
    Y^{N,2}_{m,n} (0) = 
    \begin{cases}
        (m,...,m)   &\text{if } n=1 \text{ and } m \in [\xi^N],\\[0.5em]
        \0          &\text{otherwise,}
    \end{cases}
\end{equation*}
where $\0 \defeq (0,...,0)$. Also at later times points 
the value $Y_{m,n,\ell}^{N,2} (s) = i$ will have the interpretation that the virus particle with index $n$ in host $m$ at time $T-\tau_1^N-s$ is ancestral to the virus particle $1$ in host $i$ at time $T-\tau_1^N$ at locus $\ell$. This information will allow us to distinguish pairs of ancestral lines located in the same host, that were at the beginning of Phase 2 still linked, from those pairs that were already at the beginning of Phase 2 located in two different hosts. While the former event is likely to occur in Phase 2, namely when an recombination event separates two lines that have been linked before, the latter event is unlikely (and unwanted) to happen.

Just like in the first phase, we proceed by defining duals to each map. We use $2$ as an upper index to distinguish the dual maps from those of the first phase.
For $y \in S_\bfY^{N,2} \; \setminus \; \{ \dagger \}$ and $m, m', m_1, m_2 \in[M]$, $n, n', n_1,n_2 \in [N]$ and $\ell, \ell' \in [L]$, define
\begin{align*}
    &\big(\rep^{-1,2}_{m, n_1, n_2} (y) \big)_{m',n'}\defeq 
    \begin{cases}
        y_{m, n_1} + y_{m, n_2},     &\text{if $(m',n') = (m, n_1)$,}\\[0.5em]
        \0,                                      &\text{if } (m',n') = (m, n_2),\\[0.5em]
        y_{m',n'},                               &\text{otherwise.}
    \end{cases}
\end{align*}
In Phase 2 at each locus at most one ancestral  line of a sampled virus particle is in each host. Hence for each $\ell\in \{1, ..., L\}$ at most one summand of the sum $y_{m, n_1, \ell} + y_{m, n_2, \ell}$ is not equal to 0 and therefore $y_{m, n_1, \ell} + y_{m, n_2, \ell} = \max \big\{y_{m, n_1,\ell}, y_{m, n_2,\ell} \big\}$.
Similarly, we define 
\begin{align*}
    &\big(\reinf^{-1,2}_{m_1, n_1, m_2, n_2} (y)\big)_{m', n'}\defeq \begin{cases}
        \dagger, &\text{if } y^{(2)\#}_{m_1}> 0,\\[0.5em]
        y_{m_1, n_1} + y_{m_2, n_2}, &\text{if } y^{(2)\#}_{m_1} = 0 \text{ and } (m',n') = (m_1, n_1),\\[0.5em]
        \0, &\text{if } y^{(2)\#}_{m_1} = 0 \text{ and } (m',n') = (m_2, n_2),\\[0.5em]
        y_{m',n'}, &\text{otherwise,}
    \end{cases}
    \end{align*}
    where $y^{(2)\#}_{m_1} \defeq \sum_{n \in [N]} \1_{y_{{m_1},n} \neq \0}$ is the number of ancestral lines in host $m_1$.
    \begin{align*}
    &\mut^{\bfk, -1,2}_{m, n}(y)\defeq 
    \begin{cases}
        y, &\text{if $y_{m, n} = \0$,}\\[0.5em]
        \dagger, &\text{otherwise,}
    \end{cases}\\
    \\
    &\big(\death_{m_1, n_1, m_2}^{-1,2} (y)\big)_{m', n'}\defeq \begin{cases}
        \dagger,                 &\text{if } y^{(2)\#}_{m_1} > 0 \text{ or } y^{(2)\#}_{m_2} > 1,\\[0.5em]
        y^{(2)\Sigma}_{m_2},     &\text{if } y^{(2)\#}_{m_1} = 0, y^{(2)\#}_{m_2} = 1 \text{ and } (m', n') = (m_1, n_1),\\[0.5em]
        0,                       &\text{if } y^{(2)\#}_{m_1} = 0, y^{(2)\#}_{m_2} = 1 \text{ and } m' = m_2,\\[0.5em]
        y_{m',n'},         &\text{otherwise,}
    \end{cases}
    \end{align*}
    where $y^{(2)\Sigma}_m \defeq \sum_{n \in [N]} y_{m,n}$, which  is the label of the virus that caused the infection event at host replacement (forward in time).
    \begin{align*}
    &\big(\recomb_{m, n, n_1, n_2}^{\bfr, -1, 2} (y)\big)_{m', n', \ell'} \defeq 
    \begin{cases}
        y_{m', n', \ell'}    &\text{if } m' \neq m\text{ or }n' \notin \{n, n_1, n_2\},\\[0.5em]
        y_{m, n, \ell'}      &\text{if } m' = m, n'=n_1 \text{ and } \ell' \in \bfr,\\[0.5em]
        y_{m, n, \ell'}      &\text{if } m' = m, n'=n_2 \text{ and } \ell' \notin \bfr,\\[0.5em]
        0                       &\text{otherwise.}
    \end{cases}\\
\end{align*}
As before, we denote by $\omega_{(-\infty, T]}$ the PPP used to define $\bfX^N$ until time $T$.
We write $\omega_{(-\infty, T-\tau_1^N]}$ for the PPP until time $T-\tau_1^N$.
We define
\begin{align*}
    \omega^{(2)} =\; &\omega_{(-\infty, T-\tau_1^N]} - \big(T-\tau_1^N\big)\\
    \defeq \;& \big\{ \big(\m, t - (T-\tau_1^N)\big): (\m,t) \in \omega_{(-\infty, T-\tau_1^N]} \big\}.
\end{align*}
Then, $\omega^{(2)}$ is a PPP on $\mathcal{G} \times (-\infty, 0]$, it is independent of $\omega_{(T-\tau_1^N, \infty)}$ and it has the same intensity as $\omega$.
One can order $\omega^{(2)}$ by the second component, i.e.
\begin{equation*}
    \omega^{(2)}= \{(\m_1, t_1), (\m_2, t_2), ... \}
\end{equation*}
with $0>t_1 > t_2 > \dots$\ .
Now $\big(\bfY^{N,2}(s)\big)_{s > 0}$ can be defined by
\begin{align*}
    \bfY^{N,2}&(s) = \bfY^{N,2}(0)  & &\text{ for } s \in (0, -t_1]\\
    \bfY^{N,2}&(s)= \m^{-1,2}_i \big( \bfY^{N,2} (t_i) \big) & &\text{ for } s \in (-t_i, -t_{i+1}], i \in \IN\\
    ... &
\end{align*}
Heuristically, we define the end of the second phase to be the first time point at which all loci of the ancestral lines have become ``completely unlinked'' or the state $\dagger $ is reached.
Formally, this is
\begin{equation*}
    \tau_2^{N} = \inf \Big\{ s \geq 0 : \bfY^{N,2} (s) = \dagger \text{ or } {\textstyle \sum_{l=1}^L } \1_{Y^{N,2}_{m,n,\ell} (s) \neq 0} \leq 1 \text{ for all } m \in [M], n \in [N] \Big\},
\end{equation*}
i.e.\ the first time at which in every host there is at most one ancestral particle and at most one locus of this particle is ancestral to the sample (or $\dagger$ is reached).

\begin{proposition}\label{prop:properties_phase_2}
    At all times $s\geq0$, almost surely either $\bfY^{N,2}(s) = \dagger$ or the following statements hold:
    \begin{itemize}
        \item[(i)] For all $i \in [\xi^N]$ and $\ell \in [L]$ there is exactly one particle $(m,n) \in [M] \times [N]$ such that $Y^{N,2}_{m,n,\ell} (s) = i$.
        \item[(ii)] There is no $m \in [M]$ such that there are $n_1 \neq n_2 \in [N]$ and $\ell_1, \ell_2 \in [L]$ with
        $$0 \notin \pig\{Y_{m,n_1,\ell_1}^{N,2} (s), Y_{m,n_2,\ell_2}^{N,2} (s)\pig\}$$
        such that $Y_{m,n_1,\ell_1}^{N,2} (s) \neq Y_{m,n_2,\ell_2}^{N,2} (s)$.
    \end{itemize}
\end{proposition}

\begin{remark}
    The first property says that for each sampled particle and each locus,  there is exactly one particle that is ancestral to it at time $T-\tau_1^N-s$.
    The second property says that no host contains two particles that are ancestral to \textit{different} sampled particles.
\end{remark}

\begin{proof}[Proof of Proposition \ref{prop:properties_phase_2}]
    As before one can argue inductively along the jump chain of $\bfY^{N,2}$ to verify the lemma.
    The key idea is that the dual random maps of the second phase are constructed  such, that the properties in the lemma are fulfilled.
\end{proof}

The next proposition, which follows by construction,  formalizes the duality relationship of the second phase.
It says that if $\bfY^{N,2} (s) \neq \dagger$, then the types of the particles from the definition of $W^{N}_g$ can be traced backwards in time.

\begin{proposition}\label{prop:dual_second_phase}
    For all times $s \geq 0$ it holds almost surely that either $\bfY^{N,2} (s) = \dagger$ or
    \begin{equation*}
        X_{i,1,\ell}^{N}(T-\tau_1^N) = \sum_{(m,n,\ell) \in [M]\times[N]\times[L]} \1_{Y^{N,2}_{m,n,\ell} (s) = i} X^{N}_{m,n,\ell} (T-\tau_1^N-s)
    \end{equation*}
    for all $i \in [\xi^N]$.
\end{proposition}

By Proposition \ref{prop:properties_phase_2}, there is only one $(m,n,\ell) \in [M]\times[N]\times[L] $ such that $Y_{m,n,\ell}^{N,2} (s) = i$, so Proposition~\ref{prop:dual_second_phase} is equivalent to
\begin{equation*}
    X_{i,1,\ell}^N (T-\tau_1^N) = X_{m,n,\ell}^N (T- \tau_1^N-s)
\end{equation*}
where $(m,n,\ell) \in [M] \times [N] \times [L]$ is the unique site s.t.\ $Y_{m,n,\ell}^{N,2} (s) = i$.

Proposition \ref{prop:dual_second_phase} only gives a  duality relationship as long as $\bfY^{N,2}(s) \neq \dagger$.
The next lemma states that during all times $s$ of the second phase, with high probability $\bfY^{N,2}(s) \neq \dagger$.

\begin{lemma}\label{lem:phase_2_whp}
    It holds that
    \begin{equation}\label{eq:phase_2_whp}
        \lim_{N \to \infty} \IP \pig( \bfY^{N,2} (\tau_2^{N}) = \dagger\pig) = 0.
    \end{equation}
\end{lemma}

\begin{remark}\label{rem:phase_2_final_positions}
    Lemma \ref{lem:phase_2_whp} implies that at the end of Phase 2 each of the $\xi^N$ ancestral lines of the sampled particles that were left at the end of Phase 1 is split up into $L$ locus-wise ancestral lines and all $L\xi^N$ locus-wise ancestral lines are located in $L \xi^N$  different hosts.
    By another relabeling we can assume without loss of generality that at the end of Phase 2 the first $L \xi^N$ hosts contain the ancestral lines of the sample, and more precisely the first $\xi^N$ hosts contain the ancestral lines of the first locus, hosts $\xi^N+1$ to $2\xi^N$ contain the ancestral lines of the second locus, etc.
    This is formalised in equation \eqref{eq:W_ell} below.
\end{remark}

\begin{proof}
    First, we prove the statement for the process ${\bf Y}^{N,2}$ restricted to two loci, i.e. for a subset $\mathcal{L}'\subset \mathcal{L} \defeq [L]$ with  $|\mathcal{L}'|=2$. Denote the probability of a recombination event to fall between the two loci of $\mathcal{L}'$ by $p_{\mathcal{L}'}$, which is positive by Assumption \ref{assumptions:recombination}.  Note that the model is consistent in the number of loci in the following sense:
    if $\mathbf{X}^N$ models loci $\cL$ and
    \begin{equation*}
        \bfX^N \vert_{\cL'}
    \end{equation*}
    is the restriction of $\bfX^N$ to loci $\cL'$, then $\bfX^N \vert_{\cL'}$ follows the model dynamics of the host-virus model with two loci (with appropriate parameters).
    Also, if Assumptions \ref{assumptions:recombination}, \ref{assumptions:main_rates} and \ref{assumptions:mutations} are satisfied for $(\bfX^N)_{N \in \IN}$, then they are also satisfied for $(\bfX^N \vert_{\cL'} )_{N \in \IN}$.
    A similar relationship holds also for the backwards processes.
    To be precise, if $\bfY^{N,2}$ is the dual process of the second phase tracking all $L = |\cL|$ loci, then
    the restriction
    \begin{equation*}
        \bfY^{N,2} \vert_{\cL'}
    \end{equation*}
    has the same distribution  as the corresponding two-locus dual process that considers only the two loci in $\mathcal{L}'$, for any $\cL'$ with $|\cL'| = 2$.
    In the following, we denote the restriction of ${\bf Y}^{N,2}$ to $\mathcal{L}'$ by  ${\bf Y}^{N,2, \mathcal{L}'}$ and  the corresponding state space by $S_{\bfY}^{N,2, \mathcal{L}'}$  . 
    
    Consider the function $h:S_{\bfY}^{N,2, \mathcal{L}'} \rightarrow \IN_0^3 \cup \{\dagger\}$,
    \begin{equation*}
        h(y) = \bigg(
    \begin{smallmatrix}
    h_1 (y)\\
    h_2 (y)\\
    h_3(y)
    \end{smallmatrix} \bigg)^{\textrm{T}}
    \end{equation*}
     where for a vector $v$ we denote by $v^{\textrm{T}}$ its transpose and for $y \in S_{\bfY}^{N,2,\mathcal{L}'} \setminus \{\dagger\}$,
    \begin{align*}
        h_1 \big( y \big) = \sum_{\substack{(m,n)\\ \in [M] \times [N]}} \sum_{i=1}^{\xi^N} \1_{y_{m,n} = (i,i)}
    \end{align*}
    is the \textit{number of unrecombined ancestral particles} in the total population,
    \begin{equation*}
        h_2 \big(y \big) = \sum_{i=1}^{\xi^N} \sum_{m \in [M]} \1_{(\sum_{n \in [N]} y_{m,n})=(i,i)} - h_1 \big( y \big)
    \end{equation*}
    is the \textit{number of hosts containing both ancestors of a recombined particle} and
    \begin{equation*}
        h_3 \big( y \big) = \sum_{m=1}^M \1_{\big(\sum_{n=1}^N \sum_{i=1}^{\xi^N} \1_{y_{m,n}=(i,0)}+\1_{y_{m,n}=(0,i)} \big)=1}
    \end{equation*}
    is the \text{number of hosts containing exactly one recombined particle}.
    Additionally, abusing notation we let $h(\dagger)  = \dagger$.

    By Proposition \ref{prop:properties_phase_2}, at all times $s \geq 0$ it holds almost surely that either \mbox{$\bfY^{N,2} (s) = \dagger$} or
    \begin{equation*}
        h_1 \big(\bfY^{N,2} (s) \big)+ h_2 \big(\bfY^{N,2} (s) \big) +\frac{1}{2} h_3 \big(\bfY^{N,2} (s)\big) = \xi^N .
    \end{equation*}
    Also note that
    \begin{equation*}
        \tau_2^{N, \mathcal{L}'} = \inf \pig\{ s \geq 0: h_3 \big(\bfY^{N,2, \mathcal{L}'} (s) \big) \in \big\{\dagger, 2 \xi^N \big\} \pig\}
    \end{equation*}
    and
    \begin{equation*}
        \pig\{h_3\big(\bfY^{N,2, \mathcal{L}'} (\tau_2^{N, \mathcal{L}'})\big)=\dagger \pig\} = \pig\{ \bfY^{N,2, \mathcal{L}'} (\tau_2^{N, \mathcal{L}'}) = \dagger\pig\} .
    \end{equation*}
    Therefore, in order to verify equation \eqref{eq:phase_2_whp}, it is sufficient to study the transformed process 
    \begin{equation*}
        \big(H(s) \big)_{s \geq 0} =\big(h(\bfY^{N,2,\mathcal{L}'}(s))\big)_{s \geq 0}
    \end{equation*}
    which is a continuous time Markov chain with initial state
    \begin{equation*}
        H(0) = h \big(\bfY^{N,2,\mathcal{L}'}(0) \big) = \bigg(\svec{\xi^N}{0}{0} \bigg)^{\textrm{T}},
    \end{equation*}
     making jumps from $(z_1,z_2,z_3) \in \IN_0^3$  to
    \begin{align*}    
    \bigg(&\svec{z_1-1}{z_2+1}{z_3}\bigg)^{\textrm{T}}     & \text{a}&\text{t rate}& & z_1 p_{\mathcal{L}'} \frac{\rho_N}{N^2}(N-1) & &\text{A particle recombines.}\\\\[0.9em]
    \bigg(&\svec{z_1}{z_2-1}{z_3+2}\bigg)^{\textrm{T}}    & \text{a}&\text{t rate}& &2 z_2 \frac{\lambda_N}{N M}(M-z_1-z_2-z_3) & & \text{Two recombined ancestral particles}\\
    & & & & & & &\text{are carried to different hosts via reinf.}\\[0.9em]
    \bigg(&\svec{z_1+1}{z_2-1}{z_3}\bigg)^{\textrm{T}}    & \text{a}&\text{t rate}& &2 z_2 \frac{\gamma_N}{N} & & \text{Two recombined ancestral particles}\\
    & & & & & & & \phantom{\text{}}\text{coalesce due to reproduction.}\\[0.9em]
    &\phantom{\svec{1}{1}{1}} \dagger                & \text{a}&\text{t rate}& & & &\text{An unwanted event happens. These are:}\\
    &   &   &   &   &z_2 \phantom{\frac{\mu_N}{N}}&   & \text{- particles in a common host merge via hr. }\\
    &   &   &   &   &+(z_1+2z_2+z_3)\frac{\mu_N}{N} \phantom{\frac{(z_3)^2}{M}}  &   & \text{- mutation of an ancestral particle}\\
    &   &   &   &   &+\frac{(z_1 + z_2 +z_3)^2}{M}   &   & \text{- particles go into a common host via hr.}\\
    &   &   &   &   &+(z_1 + 2z_2 +z_3) \frac{\lambda_N}{N^2 M}   &   & \text{- particles go into a common host via reinf.}\\
    &   &   &   &   &\quad \cdot(z_1+z_2+z_3-1) N  &   & \\
    \end{align*}
   The transition rates and the Markov property can be verified by counting random maps and by realizing that for $H(s) = (z_1,z_2,z_3) \neq \dagger$, the number of hosts with at least one ancestral particle is $z_1 + z_2 + z_3$ and the number of ancestral particles is $z_1 + 2z_2 +z_3$. 
    Note that by construction $z_2 \in \{0,1, \dagger\}$.
    We denote the total jump rate of the process from state $(z_1, z_2, z_3)$ by
    \begin{align*}
        r(&z_1,z_2,z_3) = z_1 p_{\mathcal{L}'} \frac{\rho_N}{N^2}(N-1) + 2 z_2 \frac{\lambda_N}{N M}(M-z_1-z_2-z_3) + 2 z_2 \frac{\gamma_N}{N}+ z_2\\
         + (&z_1+2z_2+z_3)\frac{\mu_N}{N} + \frac{(z_1 + z_2 +z_3)^2}{M} + (z_1 + 2z_2 +z_3) \frac{\lambda_N }{N^2 M}  (z_1+z_2+z_3),
    \end{align*}
    which is of order $\Theta \big(\frac{\lambda_N}{N} \big)$ for $z_2=1$, and at least of order $\Theta \big( \frac{\rho_N}{N} \big)$ and at most of order $\Theta \big(\xi^N \frac{\rho_N}{N} \big)$  for $z_2=0$.        
    Since, by assumption,
    \begin{equation*}
    r(z_1, 0, z_3) - z_1 p_{\mathcal{L}'}\frac{\rho_N}{N^2}(N-1) \in o \pig(\frac{\rho_N}{N}\pig),
    \end{equation*}
    the process typically jumps from a state of the form $(z_1, 0, z_3)$ to state $(z_1-1, 1, z_3)$ (and not to $\dagger$) due to the occurrence of a recombination event when $N$ is large.
    In particular, right at the beginning of Phase 2 starting from $(\xi^N,0,0)$ most likely a particle recombines, changing the state to $(\xi^N-1,1,0)$.
     Then, since $\rho_N \ll \lambda_N \in \Theta(\gamma_N)$    asymptotically immediately for $N \to \infty$  the two ancestors of the recombined particle either coalesce again (resulting in a jump back to $(z_1,0, z_3)$ %
     or one of the two ancestors of the recombined lines is carried to a different, so far empty host (resulting in a jump to 
     $(z_1-1, 0, z_3+2)$. We call the former  event an \textit{unsuccessful recombination}. It occurs asymptotically with probability
     \begin{equation*}
     \lim_{N \to \infty} \frac{2\gamma_N}{2 \gamma_N + 2\lambda_N} = \frac{1}{1+\theta}.  
     \end{equation*} 
     The latter event happens asymptotically with probability $\frac{\theta}{\theta+1}$. In analogy to the former event we call it a \textit{successful recombination}.
Successful and unsuccessful recombinations happen until eventually the state $(0,0,2 \xi^N)$ or $\dagger$ is reached.

Now we formalize the just described behavior to prove Equation \eqref{eq:phase_2_whp}.
We have $\tau_2^{N,\mathcal{L}'} = \min \big\{\ttau, \ttaud \big\}$, where 
\begin{equation*}
\ttaud \defeq\inf \big\{ s: {\bf Y}^{N, 2, \mathcal{L}'}(s)=\dagger \big\}
\end{equation*}
and
\begin{equation*}
  \ttau\defeq \sum_{i=1}^{\xi^N} \ttau_i  
\end{equation*}
is the waiting time for $\xi^N$ many successful recombination events,
\begin{equation*}
  \ttau_i = \sum_{j=1}^{K_i} \ttau_{i,j}^{(1)}+ \ttau_{i,j}^{(2)} + \ttau_i^{(1)}+ \ttau_i^{(2)}.  
\end{equation*}
Here, $K_i \sim \text{Geo}\big(\beta_N(i)\big)$ is the number of unsuccessful recombination events during the waiting time for the $i$-th successful recombination event. The parameter \mbox{$\beta_N(i)=\beta_N(i, \xi^N)$} depends on $\xi^N$ and $i$ and can be lower bounded by
\begin{equation*}
\uline{\beta}= \uline{\beta}(N) \defeq \frac{1}{1+\theta} - \varepsilon_N    
\end{equation*}
for some sequence $(\varepsilon_N)_{N \in \IN}$ with $\varepsilon_N \rightarrow 0$, see Lemma \ref{lem:UnsucRec}.
The random variable $\ttau_{i,j}^{(1)}$ is the waiting time for $j$-th recombination event during the waiting time on the $i$-th successful recombination event  and $\ttau_i^{(1)}$ is the waiting time for the last recombination event which leads to the $i$-th successful recombination event.
The sequences
\begin{equation*}
\ttau_{i,j}^{(1)}, \ttau_i^{(1)}, j=1, ..., K_i    
\end{equation*}
are independent and exponentially distributed with parameter $\frac{\rho_N}{N}p_{\mathcal{L}'} $.
The random variable $\ttau_{i,j}^{(2)}$ is the waiting time for the $j$-th jump back to $\big(\xi^N-i, 0, 2i \big)$ from state $\big(\xi^N-i-1, 1, 2i\big)$ and, analogously, $\ttau_{i}^{(2)}$ is the waiting time for the jump  to $\big(\xi^N-i-1, 0, 2(i+1)\big)$ from state $\big(\xi^N-i-1, 1, 2i\big)$.
The random variables $\big(\ttau_{i,j}^{(2)}\big)_j$ are independent and exponentially distributed with rate $\frac{\gamma_N}{N}$ conditioned to be smaller than independent Exp($\frac{\lambda_N}{N}$) distributed random variables. Furthermore, $\ttau_i^{(2)}$ is exponentially distributed with rate $\frac{\lambda_N}{N}$ conditioned to be smaller than an independent Exp($\frac{\gamma_N}{N}$) distributed random variable.
Hence, we have
\begin{equation*}
\ttau_{i,j}^{(2)}\leq \ttau_{i,j}^{(3)}   
\end{equation*}
with $\ttau_{i,j}^{(3)} \sim \Exp \big( \frac{\gamma_N}{N} \big)$ and 
\begin{equation*}
\ttau_{i}^{(2)}\leq \ttau_{i}^{(3)}    
\end{equation*}
with $\ttau_{i}^{(3)} \sim \Exp \big(\frac{\lambda_N}{N} \big).$
        
We can couple $K_i$ with a $K_i'$ such that $K_i \leq K_i'$ and $K_i' \sim \Geo \big(\uline{\beta}\big)$. Then
\begin{equation*}
\sum_{j=1}^{K_i} \ttau_{i,j}^{(1)} + \ttau_i^{(1)} \leq \sum_{j=1}^{K_i'} \ttau_{i,j}^{(1)} + \ttau_i^{(1)} \sim \ttau_i^{(4)}
\end{equation*}
with $\ttau_{i}^{(4)} \sim \Exp \big(\frac{\rho_N}{N}p_{\mathcal{L}'} \uline{\beta} \big)$. 
Similarly,
\begin{equation*}
\sum_{j=1}^{K_i} \ttau_{i,j}^{(2)} + \ttau_i^{(2)} \leq \sum_{j=1}^{K_i'} \ttau_{i,j}^{(3)} + \ttau_i^{(3)} \leq  \ttau_i^{(5)} 
\end{equation*}
with $\ttau_i^{(5)} \sim \Exp \pig(\frac{\min\{\gamma_N, \lambda_N\}}{N} \uline{\beta} \pig).$
$\ttau_i^{(4)}$ and $\ttau_i^{(5)}$ depend on each other over $K_i'$, however $\ttau_i^{(4)}\sim \ttau_1^{(4)}$  and $\ttau_i^{(5)}\sim \ttau_1^{(5)}$, as well as $\ttau_i^{(4)}, \ttau_i^{(5)}$ are independent of $\xi^N.$
Hence,
\begin{align*}
\IE \big[\ttau \big] &= \IE\pig[\IE\big[\ttau \; | \; \xi^N\big]\pig] = \IE \pig[\xi^N \big(\ttau_1^{(4)} + \ttau_1^{(5)} \big) \pig] =\IE \big[\xi^N \big] \pig(\IE \big[\ttau_1^{(4)} \big] + \IE \big[\ttau_1^{(5)}\big]\pig)\\[0.7em]
&= \IE \big[\xi^N \big] \IE \big[\ttau_1^{(4)} \big] \big(1+o(1) \big)\\[0.7em]
&= O \Big(\log(\jN) \frac{N}{\rho_N} \Big) 
\end{align*}
according to the rate assumptions, in particular since $\rho_N \ll \min \{\gamma_N, \lambda_N\}$ and
\begin{align*}
\IV \big[\ttau \big]  &= \IV \pig[\IE \big[\ttau \;| \; \xi^N \big] \pig] + \IE \pig[\IV \big[\ttau \;| \;\xi^N \big] \pig]\\[0.7em]
        &= \IV \pig[\xi^N \big(\ttau_1^{(4)} + \ttau_1^{(5)} \big) \pig] + \IE \pig[\xi^N \IV \big[\ttau_1^{(4)}+ \ttau_1^{(5)} \big] \pig]  \\[0.7em]
        &= \IV \big[\xi^N \big] \IV \big[\ttau_1^{(4)} + \ttau_1^{(5)} \big] + {\pig(\IE \big[\xi^N \big] \pig)}^2 \, \IV \big[\ttau_1^{(4)} + \ttau_1^{(5)} \big] + {\pig(\IE \big[\ttau_1^{(4)} + \ttau_1^{(5)}\big] \pig)}^2 \, \IV \big[\xi^N \big] \\[0.2em] & \quad \quad + \IE \big[\xi^N \big] \IV \big[\ttau_1^{(4)} + \ttau_1^{(5)} \big] \\[0.7em]
        &= O \Big(\log(\jN)^2 \frac{N^2}{\rho_N^2} \Big)
\end{align*}
Hence, we can estimate 
\begin{align}\label{eq:S}
\IP\left(\ttau >  \log(\jN)^2 \frac{N}{\rho_N}\right) & = \IP\left(\ttau- \IE[\ttau] > \log(\jN)^2 \frac{N}{\rho_N}- \IE[\ttau]\right) \notag \\[0.5em]
&\leq \frac{\IV[\ttau]}{(\log(\jN)^2 \frac{N}{\rho_N}- \IE[\ttau])^2}\\[0.7em]
&= O\left( \frac{\log(\jN)^2 \frac{N^2}{\rho_N^2}}{ \log(\jN)^4 \frac{N^2}{\rho_N^2}  }\right) \rightarrow 0 \notag
\end{align}
Furthermore, $\ttaud$ is bounded from below by an exponentially distributed random variable with parameter  $(2 \theta \log(\jN)) \frac{\mu_N}{N}$, and hence we can estimate 
\begin{align*}
\IP\left(\ttaud\leq \log(\jN)^2 \frac{N}{\rho_N}\right) \leq 
1- \exp\left( \frac{2\log(\jN)\mu_N}{N} (\log(\jN))^2 \frac{N}{\rho_N}\right) \rightarrow 0
\end{align*}
since $\mu_N \log(\jN)^3 \in o(\rho_N)$.
With this we can estimate
\begin{align*}
\IP \pig( \bfY^{N,2,\cL'} \big(\tau_2^{N,\cL'} \big) \neq \dagger\pig) &\geq \IP\pig(\ttaud > \log(\jN)^2 \tfrac{N}{\rho_N}, \ttau \leq \log(\jN)^2 \tfrac{N}{\rho_N}\pig)\\[0.7em]
    &= 1 - \IP\pig( \big\{\ttaud \leq \log(\jN)^2 \tfrac{N}{\rho_N} \big\} \cup \big\{\ttau > \log(\jN)^2 \tfrac{N}{\rho_N}\big\}\pig)\\[0.7em]
    &\geq 1 - \IP\pig( \ttaud \leq \log(\jN)^2 \tfrac{N}{\rho_N} \pig) - \IP\pig( \ttau > \log(\jN)^2 \tfrac{N}{\rho_N} \pig)\\[0.7em]
    &\to 1,
\end{align*}
which proves the lemma in the case of $|L|=2.$

For $L > 2$ we argue as follows. %
Assume we study loci $\cL = \{1,...,L \}$.
Due to the consistency of the process ${\bf Y}^{N,2}$ according to the restriction on subsets $\mathcal{L}'$ we have that if for some $s>0$ it holds that
\begin{equation}\label{eq:phase_2_multi_locus}
    h \pig(\bfY^{N,2, \cL'} (s) \pig) = \bigg(\svec{0}{0}{2\xi^N}\bigg)^{\textrm{T}} \text{ for all } \cL' \subset \cL \text{ with } |\cL'| = 2
\end{equation}
then
\begin{equation*}
    \tau_2^N \leq s \text{ and } \bfY^{N,2} (\tau_2^N) \neq \dagger.
\end{equation*}
In words, this means that if there is some point in time $s>0$ at which the ancestral lines of all \textit{pairs} $\cL'$ of loci are in separate hosts, then at this time the ancestral lines of \textit{all} loci are in separate hosts.
Therefore, in order to prove the lemma, it suffices to prove the existence of such a time $s$.
To show this, denote by
\begin{equation*}
    \sigma_2^{N, \cL'} \defeq \inf \Big\{s \geq \tau_2^{N,\cL'} :  \bfY^{N,2,\cL'}  (s) = \dagger \Big\}
\end{equation*}
the first time after $\tau_2^{N,\cL'}$ at which an unwanted state is reached.
If
\begin{equation*}
    \bfY^{N,2,\cL'} \big(\tau_2^{N,\cL'} \big) = \dagger,
\end{equation*}
then $\tau_2^{N,\cL'} = \sigma_2^{N, \cL'}$. Otherwise, $\tau_2^{N,\cL'} < \sigma_2^{N, \cL'}$ holds and between times $\tau_2^{N,\cL'}$ and $\sigma_2^{N, \cL'}$ all ancestral lines of the loci in $\cL'$ stay in separate hosts, i.e.
\begin{equation*}
    h \big(\bfY^{N,2,\cL'}(s) \big) = \bigg(\svec{0}{0}{2\xi^N}\bigg)^{\textrm{T}} \text{ for all } s \in \big[\tau_2^{N,\cL'}, \sigma_2^{N, \cL'} \big) .
\end{equation*}
In order to prove \eqref{eq:phase_2_multi_locus}, it suffices to show that the time intervals $\big\{[\tau_2^{N,\cL'}, \sigma_2^{N, \cL'}), \mathcal{L}'\subset \mathcal{L}\big\}$ overlap with high probability, i.e.
\begin{equation}\label{eq:nonempty}
    \bigcap_{\substack{\cL' \subset \cL \\ |\cL'|=2}} \big[\tau_2^{N,\cL'}, \sigma_2^{N, \cL'} \big) \neq \varnothing.
\end{equation}
This can be achieved by applying the just derived  upper bound for $\tau_2^{N,\cL'}$ and a lower bound for $\sigma_2^{N, \cL'} - \tau_2^{N,\cL'}$.
The quantity $\sigma_2^{N, \cL'} - \tau_2^{N,\cL'}$ is the waiting time for a jump to $\dagger$ when all ancestral lines are in separate hosts.  Given $\xi^N$, this time is exponentially distributed with rate
\begin{equation*}
\frac{2\xi^N\mu_N}{N} + \frac{4(\xi^N)^2}{M} + \frac{4\xi^N(\xi^N-1) \lambda_N}{MN}.    
\end{equation*}
Therefore, by the CDF of the exponential distribution and by Lemma \ref{assumptions:rates}
\begin{align*}
&\IP \pig( \sigma_2^{N, \cL'} - \tau_2^{N,\cL'} > \tfrac{c (\xi^N)^2 N}{\theta^2 \rho_N} \; \big\vert \; \xi^N\pig)\\
= &\exp{\Big(-\pig(\tfrac{2\xi^N\mu_N}{N} + \tfrac{4(\xi^N)^2}{M} + \tfrac{4\xi^N(\xi^N-1) \lambda_N}{MN}\pig) \tfrac{c(\xi^N)^2 N}{\theta^2 \rho_N} \Big)} \stackrel{N \to \infty}{\to} 1,
\end{align*}
for any $c>0$ since we conditioned on the event that $\xi^N \in (1\pm \alpha) \theta \log(j_N)$.
Since  \eqref{eq:S} implies that 
$\tau_2^{N,\cL'}$ is  with high probability less than $\log(\jN)^2 \frac{N}{\rho_N}$,  due to the bound on  $\sigma_2^{N, \cL'} - \tau_2^{N,\cL'}$ and since there exist only finitely many different subsets $\mathcal{L}' \subset \mathcal{L}$ we have with high probability
\begin{equation*}
    \frac{\log(\jN)^2 N}{\rho_N} \in \big[\tau_2^{N,\cL'},\sigma_2^{N, \cL'}\big)
\end{equation*}
for all $\cL' \subset \cL$ with $|\cL'| = 2$.
So for the event in \eqref{eq:nonempty} it holds that
\begin{align*}
    \IP \Big( \Big\{ \bigcap_{\substack{\cL' \subset \cL \\ |\cL'|=2}} & \big(\tau_2^{N,\cL'}, \sigma_2^{N, \cL'} \big] \neq \varnothing \Big\}\Big)\\[0.7em]
    \geq \; &\IP \Big( \bigcap_{\substack{\cL' \subset \cL \\ |\cL'|=2}} \Big\{ %
    \tfrac{\log(\jN)^3 N}{\rho_N} \in \big(\tau_2^{N,\cL'}, \sigma_2^{N, \cL'} \big] \Big\} \Big)
    \stackrel{N \to \infty}{\to} 1 
\end{align*}

\end{proof}

\subsection{Third Phase}\label{sec:third_phase}
Just like in the first two phases, we need to define an appropriate dual process and a coupling with simpler processes.
We denote this dual process of the third phase by $\bfY^{N,3} = \big(\bfY^{N,3}(s)\big)_{s \geq 0}$.
Due to the rate assumptions, there were with high probability no mutations in the first two phases of the ancestry of the sample.
Therefore, the state spaces for the dual processes of the first two phases were chosen just big enough to track lines and family sizes but we did not track any types arising from mutations.
In the first phase, only particle-wise ancestral lines and the total number of their progeny in the sample were considered, resulting in the state space
\begin{equation*}
    S_{\bfY}^{N,1} = [\jN]_0^{[M] \times [N]} \cup \{ \dagger \} .
\end{equation*}
These particle-wise family sizes were the appropriate choice since w.h.p. no recombinations happen during the first phase and hence loci remained linked.
On the timescale of the second phase, recombination is the driving force.
Therefore, it was necessary to consider a finer state space that allowed to track locus-wise ancestral lines, resulting in the choice
\begin{equation*}
    S_{\bfY}^{N,2} = \big([\jN]_0^L\big)^{M \times N} \; \cup \; \{ \dagger\}.
\end{equation*}
In the third phase of the ancestry, mutations will be observed.
These ultimately determine the types of the sampled virus particles.
Therefore, it will be necessary to consider an even bigger state space for the dual process which can track locus-wise family sizes along with types of families resulting from mutations.

To this end, define (for each locus) by $\ctVs$ the set of possible ``typed family sizes'' ordered by their size.
Let $\ctVs_0 \defeq \big\{ () \big\}$,
\begin{equation*}
    \ctVs_i \defeq \pig\{ v = \big((n_1,k_1),...,(n_i,k_i) \big) \in \big( [\jN] \times [K] \big)^i  : n_1 \geq n_2 \geq ... \geq n_i \pig\}
\end{equation*}
for $i \in \IN$, and
\begin{equation*}
    \ctVs \defeq \bigcup_{i \in [j_N]_0} \ctVs_i.
\end{equation*}
For $v = \big((n_1,k_1),...,(n_i,k_i)\big) \in \ctVs_i$, the first components $n_1, ..., n_i$ have the interpretation of family sizes at a specific locus and the second components $k_1, ..., k_i$ correspond to their types.
Abusing notation, we write
\begin{equation*}
    v^{(1)} \defeq (n_1, ..., n_i)
\end{equation*}
for the sorted family sizes (without types) of $v$.
In analogy to the first phase, we define
\begin{gather}
    \oplus : \ctVs \times \big([\jN] \times [K] \big) \rightarrow \ctVs \nonumber \\[0.5em]
    \pig(\big((n_1,k_1),...,(n_i,k_i)\big), \big(n', k' \big)\pig) \mapsto \big((n_1,k_1),...,(n_i,k_i)\big) \oplus \big(n', k' \big) \nonumber \\[0.5em]
    \defeq \big((n_1,k_1),...,(n_j, k_j), (n',k'), (n_{j+1},k_{j+1}),...,(n_i,k_i)\big) \nonumber
\end{gather}
where $j$ is the largest index such that $n_j \geq n'$, or zero if there is no such index.
In words, ``$\oplus$'' inserts a typed family size into a sorted vector of typed family sizes at the right position.
We define the state space of the third phase by
\begin{equation*}
    S_{\bfY}^{N,3} \defeq [\jN]_0^{M \times N \times L} \times \big( \ctVs \big)^L \cup \{ \dagger \}.
\end{equation*}
For the rest of this section we denote generic elements of $S^{N,3}_\bfY$ by $y$.
A state
\begin{equation*}
y= \big(y_1, y_2, \dots, y_{L+1}\big) \in S_{\bfY}^{N,3} \setminus \{\dagger\}   
\end{equation*}
tracks locus-wise ancestral lines that have not yet been hit by a mutation in its first component \mbox{$y_1=(y_{1,m,n,\ell})_{(m,n,\ell)\in [M]\times[N]\times[L]} \in [\jN]_0^{M \times N \times L}$}.
The remaining $L$ components track the locus-wise ancestral lines that have been hit by a mutation.
More precisely, $(n,k) \in y_{\ell+1} \in \ctVs$ will have the interpretation that for locus $\ell \in [L]$, a total of $n$ locus-wise ancestral lines of the sample taken at $\tau_2^N$ have merged and have been hit by a mutation to type $k \in [K]$ at that locus.
In this phase it is possible to consider these family sizes for each locus independently because, as we will see, mutations will w.h.p.\ only affect one locus.

In line with Remark \ref{rem:phase_2_final_positions} we set as initial state for $\ell\in [L]$
\begin{equation*}
    \bfY^{N,3}_{1, m,n,\ell} (0) =
        \begin{cases}
            1       &\text{if } n=1 \text{ and } \lceil \frac{m}{\xi^N} \rceil = \ell ,\mybreak
            0       &\text{otherwise,}
        \end{cases}
\end{equation*}
and
\begin{equation*}
    \bfY^{N,3}_{1+\ell} (0) = () \hspace{1cm} \text{for all } \ell \in [L].
\end{equation*}
Next, we define the dual maps of the third phase.
For given $y \in S_\bfY^{N,3} \setminus \{\dagger\}$, only the duals of mutation maps may affect all components of $y$.
The other dual maps may only change the first component of $y$. We define

\begin{align*}
    &\big(\mut^{\bfk, -1, 3}_{m, n}(y)\big)_{1,m',n'} =
    \begin{cases}
        y_{1, m',n'}      &\text{if } (m,n) \neq (m',n') \text{ or } y_{1, m',n'} = \0,\\[0.3em]
        \0                      &\text{if } (m,n) = (m',n') \text{ and } \exists! \ell' \in [L]: y_{1, m',n',\ell'} \neq 0,\\[0.25em]
        \dagger                 &\text{otherwise}
    \end{cases}
\end{align*}
and
\begin{align*}
    &\big(\mut^{\bfk, -1, 3}_{m, n}(y)\big)_{\ell+1} =
    \begin{cases}
        y_{\ell+1}         &\text{if } y_{1, m,n} = \0,\\[0.3em]
        y_{\ell+1} \oplus \big(y_{1, m,n,\ell}, \bfk_\ell \big) &\text{if } \ell \text{ is the only locus in $L$ s.t.\ } y_{1, m,n,\ell} \neq 0,\\[0.25em]
        \dagger                 &\text{otherwise.}
    \end{cases}
\end{align*}
This definition has the following interpretation: if a mutation happens at some particle $(m,n)$ that is not ancestral to the sample, then it has no visible effect.
If a mutation happens at a particle that is ancestral to the sample, then one of the following happens:
\begin{itemize}
    \item If only one locus of the particle affected by the mutation is ancestral to the sample, then the mutation determines the type of all the descendant particles at that locus.
    Therefore, the number of descendants and the type they received by the mutation is kept track of in the corresponding component of $y$.
    The ancestral line need not be followed up further, so the corresponding particle in the system $y_1$ is set to $\0$.
    \item If more than one locus of the particle affected by the mutation is ancestral to the sample, then some dependence between to loci would arise. Since in Phase 2 all loci were separated by recombination and since this separation is - as we will see - maintained for most of the time of Phase 3 with high probability, this scenario is unlikely to happen.
    Hence, all such events can be aggregated in the $\dagger$-state.
\end{itemize}
The remaining dual maps only affect the first component of $y \in S_\bfY^{N,3} \setminus \{ \dagger \}$.
That is, for any dual map of the third phase $\m^{-1, 3}$ which is not a mutation map and for any $y \in S_\bfY^{N,3} \setminus \{ \dagger \}$ and $\ell \in [L]$ we set
\begin{equation*}
    \big(\m^{-1,3} (y)\big)_{\ell+1} = y_{\ell+1}. 
\end{equation*}
Furthermore, we set for any $\ell\in [L]$
\begin{align*}   
    &\big(\rep^{-1, 3}_{m, n_1, n_2} (y)\big)_{1, m',n', \ell} = 
    \begin{cases}
        y_{1, m, n_1,\ell} + y_{1, m, n_2,\ell}   &\text{if $(m',n') = (m, n_1)$,}\\[0.5em]
        \0                                          &\text{if $(m',n') = (m, n_2)$,}\\[0.5em]
        y_{1,m',n',\ell}                           &\text{otherwise.}
    \end{cases}\\
\end{align*}
Similar to the second phase we define the number of (ancestral) particles in host $m$ to be $y^{(3), \#}_{1, m} \defeq \sum_{n \in [N]} \1_{y_{1,m,n} \neq \0}$ where again $\0 \defeq (0,...,0) \in [\jN]_0^L$.
We say \textit{``host $m$ has $d$ lines''} if $y^{(3), \#}_{1, m}=d$.
This allows us to define the dual of reinfection maps for any $\ell\in [L]$ by
\begin{align*}
    &\big(\reinf^{-1, 3}_{m_1, n_1, m_2, n_2} (y)\big)_{1, m', n',\ell} =
    \begin{cases}
        y_{1, m',n',\ell}              &\text{if } y_{1,m_2,n_2} = \0\\[0.5em]
        \dagger                         &\text{if } y_{1,m_2,n_2} \neq \0 \text{ and}\\
                                        &\hspace{0.5cm} y_{1,m_1,n_1} \neq \0 \text{ or}\\
                                        &\hspace{0.5cm} m_1 \text{ is not empty and $\exists$ a host with $2$ lines}\\[0.5em]
                                        &\text{otherwise,}\\
        y_{1,m',n',\ell}               &\hspace{0.5cm} \text{if } (m',n') \notin \{(m_1,n_1),(m_2,n_2)\}\\
        y_{1,m_2,n_2,\ell}             &\hspace{0.5cm} \text{if } (m',n') = (m_1,n_1)\\
        \0                              &\hspace{0.5cm} \text{if } (m',n') = (m_2,n_2).
    \end{cases}
\end{align*}
In words, backwards in time reinfections are invisible if there is no ancestral line at the reinfect\textit{ed} particle.
Unlikely (and for the study of the process unwanted) reinfection events leading to the $\dagger$-state are
\begin{itemize}
    \item a reinfection event causing ancestral lines to directly coalesce (this is the case $y_{1,m_2,n_2} \neq \0$ and $y_{1,m_1,n_1} \neq \0$)
    \item  reinfection events that occur when there is already a host present with two ancestral lines (that have been separated by a recombination event just before) and involves a reinfecting hosts, which contain an ancestral line, and a reinfected particle at which there is an ancestral line
\end{itemize}
Otherwise, the ancestral particles are traced back just as before:
the duals of recombination maps are defined analogously to the ones from the second phase:
\begin{align*}
    &\big( \recomb_{m, n, n_1, n_2}^{\bfr, -1, 3} (y) \big)_{1, m', n', \ell'} = 
    \begin{cases}
        y_{1, m, n, \ell'}            &\text{if } m' = m, n'=n_1 \text{ and } \ell' \in \bfr,\\[0.5em]
        y_{1, m, n, \ell'}            &\text{if } m' = m, n'=n_2 \text{ and } \ell' \notin \bfr,\\[0.5em]
        0                                &\text{if } m' = m, n'=n \text{ and } \ell' \in [L],\\[0.5em]
        y_{1, m', n', \ell'}          &\text{if } m' \neq m\text{ or }n' \notin \{n, n_1, n_2\}.
    \end{cases}\\
\end{align*}
Analogously to Phase 2, we define for any $\ell \in [L]$ the sum over all family sizes at locus $\ell$ in host $m$ by $y_{1,m,\ell}^{(3), \Sigma} \defeq \sum_{n \in [N]} y_{1,m,n,\ell}$.
If $m$ has exactly one line at a single locus $\ell$, then $y_{1,m,\ell}^{(3), \Sigma}$ equals the value of this line.
Finally, the dual of host replacement maps is defined for any $\ell\in [L]$ as

\begin{align*}
    &\big(\death_{m_1, n_1, m_2}^{-1, 3} (y)\big)_{1, m',n',\ell} =
    \begin{cases}
        y_{1,m',n',\ell}           &\text{if $m_2$ has no lines}\\[0.5em]
        \dagger                     &\text{if $m_2$ has more than one line}\\ 
                                    &\text{or $m_1$ has at least one line}\\[0.5em] 
                                    &\text{otherwise}\\ 
        y_{1,m,\ell}^{(3), \Sigma}       &\hspace{0.3cm}\text{if }(m',n')=(m_1, n_1)\\
        y_{1,m',n',\ell}           &\hspace{0.3cm}\text{if }(m',n')\neq(m_1, n_1),\\
    \end{cases}
\end{align*}
or, in words:
\begin{itemize}
    \item If the replaced host has no ancestral lines, then no visible change occurs.
    \item Merging of lines due to host replacement is unwanted as well as gathering of two or more lines in one host due to host replacement.
    \item Otherwise, the ancestral particle is followed back.
\end{itemize}
Having defined duals of the third phase to all maps, one can now define $\bfY^{N,3}$ in the same manner as $\bfY^{N,1}$ and $\bfY^{N,2}$ by reversing time and replacing maps in the original PPP $\omega^N$ with their duals.
We denote the end of the third phase by
\begin{equation*}
    \tau_3^N \defeq \inf \Big\{ s \geq 0: \bfY^{N,3}_1 (s) \in \pig\{\dagger, 0^{[M] \times [N] \times [L]} \pig\} \Big\} .
\end{equation*}
This is the first time at which all ancestral particles have been hit by a mutation or at which the $\dagger$-state has been reached.
We denote the type frequencies observed at the single loci sampled at the beginning of Phase $3$ by $W^{(3), N} = (W^{(3), N}_\ell)_{\ell \in [L]}$, where
\begin{equation}\label{eq:W_ell}
    W^{(3), N}_\ell \defeq \frac{1}{\xi^N} \sum_{i=1}^{\xi^N} \delta_{X^N_{(\ell-1)\xi + i,1,\ell} (T-\tau_1-\tau_2)}
\end{equation}
is a random probability measure on the type space.
This corresponds exactly to the type frequencies at the single loci described in Remark \ref{rem:phase_2_final_positions}.

\begin{lemma}\label{lem:phase_3_frequencies}
For all $s \geq 0$ and $\ell \in [L]$ it holds that
\begin{equation*}
    \IP \Big( W_\ell^{(3),N} = \frac{1}{\xi^N} \sum_{\substack{(m,n) \in \\ [M] \times [N]}} \delta_{X^N_{m,n,\ell} (T- \tau_1 -\tau_2 - s)} Y_{1,m,n,\ell}^{N,3} (s)
    + \frac{1}{\xi^N} \sum_{\substack{(n,k) \in \\ Y^{N,3}_{1+\ell} (s)}} n \delta_k \; \big\vert \; \bfY^{N, 3} (s) \neq \dagger \Big) = 1.
\end{equation*}
In particular, this implies
\begin{equation}\label{eq:phase_3_frequencies}
    \IP \Big( W_\ell^{(3), N} = \frac{1}{\xi^N} \sum_{\substack{(n,k) \in \\ Y^{N,3}_{1+\ell} (\tau_3^N)}} n \delta_k \; \big\vert \; \bfY^{N, 3} (\tau_3^N) \neq \dagger \Big) = 1.
\end{equation}
\end{lemma}
Lemma \ref{lem:phase_3_frequencies} has the interpretation that if $\tau_3^N$ is finite and $\bfY^{N, 3} (\tau_3^N) \neq \dagger$, then every ancestral line has been hit by a mutation and hence type frequencies in the sample of the third phase can be written only in terms of $\bfY^{N, 3} (\tau_3^N)$. 

\begin{proof}[Proof of Lemma \ref{lem:phase_3_frequencies}]
   Just as in Phases 1 and 2, one can argue along the PPP.
\end{proof}
Equation \eqref{eq:phase_3_frequencies} of Lemma \ref{lem:phase_3_frequencies} will be the key tool of the third phase.
In order to make use of it, we need a lower bound for $\IP \big(\bfY^{N, 3} (\tau_3^N) \neq \dagger \big)$ as well as a characterization of the distribution of $\big(Y^{N, 3}_{1+\ell} (\tau_3^N)\big)_{\ell \in [L]}$.
The following Lemma provides the two.
Recall that $\bfY^{N,3}_{\ell+1}$ takes values in $\ctVs$ and tracks the typed family sizes of locus $\ell$.
The family sizes (without types) are accessed by $\bfY^{N,3,(1)}$ and take values in $\cVs$.
\begin{lemma}\label{lem:phase_3_whp}
    It holds that
    \begin{equation}\label{eq:phase_3_whp}
        \lim_{N \to \infty} \IP \pig( \bfY^{N,3} (\tau_3^N) \neq \dagger \pig) = 1 .
    \end{equation}
    Furthermore, there exist independent $\cVs$-valued random elements $\big(\mathcal{Z}^N_\ell \big)_{\ell \in [L]}$, where $\cZ_\ell^N$ has the distribution of the family sizes at absorption in a Hoppe's urn with $\xi^N$ initial particles and parameter $\frac{\mu_N M (1+\theta_N)}{2 \lambda_N}$ such that 
    \begin{equation*}
        \IP \Big( \cZ^N_\ell = \bfY^{N,3,(1)}_{\ell+1} (\tau_3^N), \ell \in[L] \; \big\vert \; \bfY^{N,3} (\tau_3^N) \neq \dagger \Big) = 1.
    \end{equation*}
\end{lemma}

\begin{proof}
    We prove Equation \eqref{eq:phase_3_whp} for $L=1$ first.
    In this case, there is no difference between (particle-wise) ancestral lines and locus-wise ancestral lines, so we use the two terms interchangeably.
    Denote by $\mathcal{N}$ the set of all multisets with elements from $\IN$.
    As before, we transform the process $\bfY^{N,3}$ to a simpler process.
    To this end, we define $b : S_\bfY^{N,3} \to \cN \times \cN \times \ctVs$ with
    \begin{equation*}
    b\big(y\big) = \pig(b_1 \big(y \big), b_2 \big(y \big), b_3 \big(y \big) \pig)   
    \end{equation*}
    where
    \begin{equation*}
        b_1 (y) = \bigmset{y_{1,m,n,1}: m \in [M], n \in [N], y^{(3), \#}_{1, m} = 1}
    \end{equation*}
    denotes the sizes of families that haven't been hit by a mutation and are the only ones in their host,
    \begin{equation*}
        b_2 (y) = \bigmset{y_{1,m,n,1}: m \in [M], n \in [N], y^{(3), \#}_{1, m} = 2}
    \end{equation*}
    denotes the sizes of families that haven't been hit by a mutation and are in host that contains two ancestral lines and
    \begin{equation*}
        b_3 (y) = y_2
    \end{equation*}
    tracks the families that have been hit by a mutation along with their type.\\
    Instead of studying $\big(\bfY^{N,3} (s)\big)_{s \geq 0}$, we can now study $B^N = b \big(\bfY^{N,3} (s)\big)_{s \geq 0}$.
    The initial state of $B^N$ is
    \begin{align*}
        B_1^N (0&) = b \big(\bfY^{N,3} (0)\big) = \smash{\mset{\underbrace{1,...,1}_{\xi^N}}},\\
        B_2^N (0&) = \varnothing,\\
        B_3^N (0&) = ().
    \end{align*}
    Note that
    \begin{equation*}
        \tau_3^N = \inf \Big\{ s \geq 0 : B^N (s) = \dagger \text{ or } B^N_1 (s) = B^N_2 (s) = \varnothing \Big\}.
    \end{equation*}
    and that by construction, $\big\lvert B_2^N \big\rvert \in \{0,2\}$.
    By counting the dual random maps, it can be verified that $B^N$ is a continuous time Markov chain making jumps from $(b_1, b_2, b_3)$ to

    \begin{align*}    
       &\Bigg(\svec{b_1 \setminus \{i,j\}}{b_2 \cup \{i,j\}}{b_3}\Bigg)^{\textrm{T}}
        &\text{a}   &\text{t rate}   & &2\tbinom{|b_1|}{2}\frac{\lambda_N(N-1)}{N^2 M}& &\text{Two ancestral lines gather in a}\\[-0.7em]
       &              &            &               & &      & &\text{common host via reinfection.}\\
        &\Bigg(\svec{b_1 \cup \{i+j\}}{b_2 \setminus \{i,j\}}{b_3}\Bigg)^{\textrm{T}} &\text{a}   &\text{t rate}  & &|b_2| \frac{\gamma_N}{N}   & &\text{Two ancestral lines in a common host}\\[-0.7em]
       &              &            &              & &                           & &\text{merge due to virus reproduction.}\\
       &\Bigg(\svec{b_1 \cup \{i,j\}}{b_2 \setminus \{i,j\}}{b_3}\Bigg)^{\textrm{T}} &\text{a}   &\text{t rate}  & &|b_2| \frac{\lambda_N \big(M-|b_1|-\tfrac{|b_2|}{2}\big)}{N M}   & &\text{Two ancestral lines in a common host}\\[-0.7em]
       &              &            &               & &      & &\text{are separated due to reinfection.}\\
       &\Bigg(\svec{b_1 \setminus \{i\}}{b_2}{b_3 \oplus (i,k)}\Bigg)^{\textrm{T}} &\text{a}   &\text{t rate}  & &(|b_1|+|b_2|) \frac{\mu_N}{N}\nu({k})                      & &\text{A line mutates to type $k \in [K]$.}\\[0.5em]
        &\hspace{0.8cm} \dagger &\text{a}   &\text{t rate}  & &                                                           & &\text{An unwanted event happens, i.e.}\\
        &                       &           &               & &2 \tbinom{|b_1| + |b_2|}{2}    \frac{\lambda_N}{N^2 M}    & &\text{- Coalescence due to reinfection,}\\[0.5em]
        &                       &           &               & +&\1_{b_2 \neq \varnothing}(|b_1|+|b_2|)                    & &\text{- other unwanted reinfection events},\\
        &                       &           &               &  &\cdot \big(|b_1|+\tfrac{|b_2|}{2}\big)\frac{\lambda_N}{N^2 M}     & & \\[0.5em]
        &                       &           &               & +&\frac{|b_2|}{2}                                          & &\text{- replacement of a host with two lines},\\[0.5em]
        &                       &           &               & +&|b_1| \big(|b_1|+\tfrac{|b_2|}{2} \big)\frac{1}{M}                 & &\text{- other visible host replacements.}
    \end{align*}
    By definition, $\dagger$ is an absorbing state. The typical behavior of the process $B^N$ as $N$ gets large is as follows:
    Starting in $B^N (0)$, the process stays constant until a line in $B_1^N$ is either hit by a mutation or by a particular host replacement or reinfection event:
    \begin{itemize}
        \item In case of a mutation, the line gets a type $k \in [K]$ drawn according to the mutation probabilities independently from everything else.
        The family size associated with the line is then removed from $B_1^N$ and placed in $B_3^N$, together with the type $k$ it received by the mutation.
        The second component stays constantly equal to $\varnothing$ during this time.
        \item Usually, if a line is hit by a reinfection or host replacement map, no change in $B^N$ happens. Some reinfection and host replacement maps, however, cause the affected ancestral line to be moved to a host in which there is already another ancestral line.
        The family sizes of the corresponding lines then move from the first component $B_1^N$ to the second component $B_2^N$, causing the number of elements $|B_2^N|$ to jump from $0$ to $2$.
    \end{itemize}
    After an event of the second type, the two lines in $B_2^N$ either coalesce due to viral reproduction or they are separated again by reinfection before any other change in $B^N$ happens.
    We define transitions of the kind
    \begin{align}
        \bigg(\svec{b_1}{\varnothing}{b_3}\bigg)^{\textrm{T}} \to \bigg(\svec{b_1 \setminus \{i,j\}}{\{i,j\}}{b_3}\bigg)^{\textrm{T}} \to \bigg(\svec{b_1 \cup \{i+j\}}{\varnothing}{b_3}\bigg)^{\textrm{T}}
    \end{align}
    as ``successful mergers'' and transitions of the kind
    \begin{align}
        \bigg(\svec{b_1}{\varnothing}{b_3}\bigg)^{\textrm{T}} \to \bigg(\svec{b_1 \setminus \{i,j\}}{\{i,j\}}{b_3}\bigg)^{\textrm{T}} \to \bigg(\svec{b_1 \cup \{i+j\}}{\varnothing}{b_3}\bigg)^{\textrm{T}}
    \end{align}
    as ``unsuccessful mergers''.
    A transition that is either a mutation or a successful merger is called a ``successful transition''.
    Similarly as in Phase 2, note that
    \begin{equation*}
        \tau_3^N = \min \big\{ S, S^\dagger \big\}
    \end{equation*}
    where
    \begin{equation*}
        S^\dagger \defeq \inf \pig\{s\geq 0: B^N(s) = \dagger \pig\}
    \end{equation*}
    and 
    \begin{equation*}
        S  \defeq\inf \pig\{s \geq 0: B_1^N(s) = B_2^N(s) = \varnothing \pig\} .
    \end{equation*}
    By Lemma \ref{Lemma:ControllSizeOfXi}, there is a constant $c_1>0$ such that w.h.p.
    \begin{equation}\label{eq:num_lines_phase_3}
        |B_1^N| + |B_2^N| \leq c_1 \log (\jN) .
    \end{equation}
    In the following we condition on the event that \eqref{eq:num_lines_phase_3} is fulfilled.
    Therefore, and due to the rate assumptions, $S^\dagger$ is asymptotically lower bounded by an exponentially distributed random variable with rate
    \begin{equation*}
        c_2\frac{\big(\log (\jN) \big)^2}{M} \max \Big\{ 1, \frac{\lambda_N}{N^2} \Big\}
    \end{equation*}
    for an appropriate $c_2 > 0$.
    On the other hand, $S$ is the waiting time for $\xi^N$ successful transitions.
    This time is upper bounded by the waiting time for $\xi^N$ mutations.
    Mutations  arrive on each ancestral line at exponentially distributed waiting times with rate greater than $c_3 \frac{\mu_N}{N}$ for an appropriate $c_3 > 0$.
    Therefore, $S$ is asymptotically upper bounded by an Erlang distribution with shape $c_1 \log (\jN)$ and rate $c_3\frac{\mu_N}{N}$.
    By Chebyshev's inequality, we get that
    \begin{align}\label{eq:sWHP}
        \IP \Big( S \geq 2 \frac{c_1 \log (\jN)N}{c_3\mu_N} \Big) \leq \big(\log (\jN)\big)^{-1} \to 0
    \end{align}
    as $N \to \infty$.\\
    On the other hand, by the CDF of the exponential distribution and the rate assumptions, for $N$ large enough,
    \begin{align}\label{eq:sdaggerWHP}
        \IP \Big( S^\dagger > 2&\tfrac{c_1 \log (\jN)N}{c_3\mu_N}\Big) = \exp{\Big(-\tfrac{(\log(\jN))^3 N}{\mu_N M}\Big)}\nonumber \\
        &\geq \exp{\Big(-\tfrac{(\log(\jN))^3 N}{\lambda_N} \Big)} \geq \exp{ \Big(-\tfrac{(\log(\jN))^3}{\jN}\Big)} \to 1
    \end{align}
    as $N \to \infty$.
    Equations \eqref{eq:sWHP} and \eqref{eq:sdaggerWHP} together imply that with high probability $S < S^\dagger$ which, in turn, implies equation \eqref{eq:phase_3_whp} for $L=1$.

    This concludes the proof of equation \eqref{eq:phase_3_whp} for $L=1$.
    We still have to show the second part of the Lemma, namely that $\bfY^{N,3}_2 (\tau_3^N)$ has essentially the distribution of the final state of a Hoppe's urn.
    In order to see this, note that all families in $i \in B_1^N$ are equally likely to be hit by a mutation and the type they get is independent of everything else.
    Also all pairs of families $(i,j)$ with $i,j \in B_1^N$ are equally likely to gather in a common host and coalesce due to reinfection.
    The ratio of mutation rate (per line) and coalescence (per unordered pair) is
    \begin{equation*}
     \frac{\frac{\mu_N}{N}}{2 \frac{\lambda_N}{N M}\frac{1}{1+\theta_N}}
    \end{equation*}

    The thereby described dynamics coincide with those of a reversed Hoppe urn with this parameter as long as the $\dagger$-state is not reached.

    Finally, we provide a sketch of proof for the case $L>1$.  We leave out the details to avoid excessive notation.
    For $L>1$ one can - similarly as in Phase 2 - study the restriction of $\bfY^{N,3}$ to single loci and apply the first part of the proof to see that the ancestry of each \textit{single} locus follows the dynamics of a Hoppe's urn.
    In order to realize that these $L$ Hoppe's urns are independent, note that dependence can only arise if a particle that is ancestral to sampled particles at more than one locus is hit by a mutation.
    At the beginning of Phase 3 there is no particle which is ancestral to more than one locus because all loci are unlinked due to recombination in the Phase 2.
    Due to reinfection and reproduction, it is possible (for finite $N$) that such a particle exists at some point during the third phase.
    Whenever such a particle arises, it is with high probability immediately hit by recombination event unlinking the linked loci again.
    This yields that an ancestral particle is ancestral to only one locus whenever it is hit by mutation with high probability.
    This shows the claimed independence.
\end{proof}

\begin{lemma}\label{lemma:dirichlet_phase_3}
    If $\big( \log(\jN) \big)^3 \frac{\lambda_N}{M} \in o(\mu_N)$, then $\big(W^{N, (3)}_\ell\big)_{\ell \in [L]}$ converges in distribution to the deterministic measures $(\nu_\ell)_{\ell \in [L]}$.
    If $\lim_{N \to \infty} \frac{\lambda_N}{M \mu_N} = C$, then for all $\ell \in [L]$,
    \begin{equation*}
        W_\ell^{N, (3)} \to W_\ell^{(3)}
    \end{equation*}
    in distribution as $N \to \infty$, where $W_\ell^{(3)}$ follows a Dirichlet distribution with base measure $\nu_\ell$ and concentration parameter $\frac{1+\theta}{C}$.
\end{lemma}

\begin{remark}
    The two cases considered in the lemma correspond to the two cases in Point \ref{assumptions:rates_mutation_cases} of Lemma \ref{assumptions:rates}.
\end{remark}

\begin{proof}
    By Lemma \ref{lem:phase_3_frequencies} and Lemma \ref{lem:phase_3_whp}
    \begin{equation*}
        W_\ell^{(3), N} \stackrel{d}{=} \frac{1}{\xi^N} \sum_{\substack{n\in \cZ_\ell^N}} n \delta_{k_n}
    \end{equation*}
    on an event that holds with high probability where $\cZ_\ell^N$
    has the distribution of the family sizes at absorption in a reversed Hoppe's urn with $\xi^N$ initial particles and parameter $\frac{\mu_N M (1+\theta_N)}{2 \lambda_N}$ and $k_n$ are i.id.\ draws according to $\nu_\ell$.

    If $\mu_N \gg \big( \log (\jN) \big)^3 \frac{\lambda_N}{M}$, then w.h.p.\ all lines left in the third phase are hit by a mutation before they have coalesced with any other line and therefore $\cZ_\ell^N$ only has ones as entries.
    To verify this, note that for a fixed locus $\ell$, the waiting time for the first coalescence to happen in the ancestry is the minimum of $\binom{\xi^N}{2}$ many independent Exp$\big(2 \frac{\lambda_N}{NM} \frac{1}{1+\theta_N} \big)$ distributed random variables and hence again exponentially distributed with parameter $\binom{\xi^N}{2} \cdot 2 \frac{\lambda_N}{NM} \frac{1}{1+\theta_N}$.    
    The time until all lines have been hit by a mutation is the maximum of $\xi^N$ many Exp$\big( \frac{\mu_N}{N}\big)$ random variables.
    Using $\mu_N \gg \big( \log (\jN) \big)^3 \frac{\lambda_N}{M}$ and the concentration of $\xi^N$ around $\theta \log (\jN)$ given in Lemma \ref{Lemma:ControllSizeOfXi} one can show by Chebyshev's inequality and the CDF of the exponential distribution that w.h.p.\ all lines have been hit by mutations before any coalescence happens.
    The types assigned to lines at locus $\ell$ at mutation events are i.i.d.\ with probabilities given by $\nu_\ell$ so, by the law or large numbers, the asymptotic type frequencies $\big(W^{(3)}_\ell \big)_{\ell \in [L]}$ are deterministic and given by $\nu_\ell$.

    If $\mu_N \sim \frac{\lambda_N}{M}$ then the ratio of the mutation rate (per line) and the coalescence rate (per unordered pair of lines)  approaches a constant for $N \to \infty$, namely
    \begin{equation*}
    \frac{\frac{\mu_N}{N}}{2 \frac{\lambda_N}{N M}\frac{1}{1+\theta_N}} \to \frac{1+\theta}{2C}.
    \end{equation*}
    In this regime it is well known that the type frequencies approach a Dirichlet distribution with concentration parameter $\frac{1+\theta}{C}$ and base measure $\nu_\ell$, see Lemma \ref{lem:random_colored_hoppe}. %
\end{proof}

\subsection{Proofs of the Main Theorems}
\subsubsection{Proof of Theorem \ref{thm:MainResult1}}
Lemma \ref{lemma:dirichlet_phase_3} characterized the asymptotic distribution of locus-wise type frequencies in a random sample taken from different hosts when the sample size is asymptotically much smaller than the size of the host population.
The next Lemma states that the locus-wise type frequencies in the \textit{whole} population converge to the same distribution.
This essentially proves Theorem \ref{thm:MainResult1}

\begin{lemma}\label{lem:type_frequencies_whol_pop}
    In the setting of Lemma \ref{lemma:dirichlet_phase_3}, the locus-wise type frequencies in the whole population $\Bar{\bfX}^N (T) = \big(\Bar{X}^N_\ell  (T)\big)_{\ell \in [L]}$ converge in distribution to the same limit as $\big(W^{N,(3)}_\ell\big)_{\ell \in [L]}$.
\end{lemma}

The proof relies on two different possible ways to interpret $W^{N, (3)}$.
One interpretation is that the components of $W^{N, (3)}$ arise from independent Hoppe's urns.
This interpretation was made use of in  Lemma \ref{lemma:dirichlet_phase_3}.
The other interpretation is that $W^{N, (3)}_\ell$ essentially arises as the type frequencies observed when making i.i.d.\ draws from $\Bar{X}^N_\ell$ at stationarity.
Combining these two interpretations we will see that if $W^{N,(3)}$ %
converges in distribution to some limit as $N \to \infty$, then $\Bar{\bfX}^{N}$ converges to the same limit.

\begin{proof}
    We first prove the result for one locus and two types (A and B).
    Then we discuss the extension to more types and more loci.
    Recall that in the two-type case $W^{N, (3)}$ is one-dimensional and denotes the frequency of type A observed in a sample of $\xi^N$ virus particles taken randomly from different hosts (that is, without replacement of hosts).
    Denote by $\widetilde{W}^{N}$ the type A frequencies observed in a random sample of size $\xi^N$ of virions from the whole population \emph{with} replacement.
    The sampling procedures for $W^{N, (3)}$ and $\widetilde{W}^N$ can be coupled in such a way that
    \begin{equation*}
        \IP \pig( \widetilde{W}^{N} = W^{N, (3)} \; \big\vert \; \xi^N \pig) \geq 1 \cdot \frac{M-1}{M} \cdot ... \cdot \frac{M-\xi^N + 1}{M}
    \end{equation*}
    where the r.h.s.\ is the probability that no two virions are sampled from the same host when drawing with replacement.
    Since $\xi_N \leq \jN$ a.s., by Assumption \ref{assumptions:rates}, Point \ref{ass:sample_size} the r.h.s.\ converges to $1$ as $N \to \infty$.
    Therefore, $\widetilde{W}^N$ converges in distribution to the same limit as $W^{N, (3)}$.
    This allows us to work with $\widetilde{W}^N$ instead of $W^{N, (3)}$ for the rest of the proof.

    Next, we want to express $\widetilde{W}^N$ in terms of $\Bar{X}^N (t)$.
    To this end, let $(U_i)_{i \in \IN}$ be i.i.d.\ Unif$([0,1])$ distributed random variables and see that
    \begin{equation*}
        \widetilde{W}^N \stackrel{d}{=} \frac{1}{\xi^N}\sum_{i=1}^{\xi^N} \1_{U_i \leq \Bar{X}^N (t)}.
    \end{equation*}
    Let $a \in [0,1]$ be a point of continuity of $\lim_{N \to \infty} \IP (\widetilde{W}^N \leq a)$.
    Then, by the law of total probability
    \begin{align*}
        \IP \Big( \widetilde{W}^N \leq a \Big) = &\int_0^1 \IP \big( \widetilde{W}^N \leq a \; \vert \; \Bar{X}^N (T) = x \big) d F_{\Bar{X}^N} (x).
    \end{align*}
    By definition, the integrand equals the CDF of a Binomial distribution with (random) parameters $\xi^N$ and $x$ evaluated at $a \xi^N$.
    Since $\xi^N$ converges w.h.p. to $\infty$, by Chebyshev's inequality there exists a sequence $\delta_N \searrow 0$ such that for any $N$
    \begin{equation*}
        (1-\delta_N) \1_{x \leq a-\delta_N}\leq \IP \big( \widetilde{W}^N \leq a \; \vert \; \Bar{X}^N (T) = x \big) \leq \1_{x \leq a + \delta_N} + \delta_N \1_{x > a + \delta_N}. 
    \end{equation*}
    Hence,
    \begin{equation*}
        (1-\delta_N) F_{\Bar{X}^N} (a-\delta_N) \leq \IP \Big( \widetilde{W}^N \leq a \Big) \leq F_{\Bar{X}^N} (a + \delta_N) + \delta_N
    \end{equation*}
    for any $N$ and thus also
    \begin{equation*}
        \limsup_{N \to \infty} (1-\delta_N) F_{\Bar{X}^N} (a-\delta_N) \leq \lim_{N \to \infty} \IP \Big( \widetilde{W}^N \leq a \Big) \leq \liminf_{N \to \infty} F_{\Bar{X}^N} (a + \delta_N) + \delta_N.
    \end{equation*}
    Since $\delta_N \to 0$ and since $F_{\bar{X}^N}$ is non-decreasing this implies
    \begin{equation*}
        \overline{F} (a-\varepsilon) \leq \lim_{N \to \infty} \IP \Big( \widetilde{W}^N \leq a \Big) \leq \underline{F} (a+\varepsilon)
    \end{equation*}
    for any $\varepsilon > 0$ small enough where $\overline{F}, \underline{F}$ denote the pointwise lim sup and lim inf of $F_{\bar{X}^N}$, respectively.
    If $\overline{F}$ is left continuous and $\underline{F}$ is right continuous in $a$, then taking $\varepsilon \to 0$ yields that
    \begin{equation*}
        \overline{F} (a) \leq \lim_{N \to \infty} \IP \Big( \widetilde{W}^N \leq a \Big) \leq \underline{F} (a)
    \end{equation*}
    If $\underline{F} (a + \varepsilon) \nrightarrow \underline{F} (a)$
    then either
    \begin{equation}\label{eq:caseone}
        \lim_{N \to \infty} \IP \Big( \widetilde{W}^N \leq a \Big) \leq \underline{F} (a) < \lim_{\varepsilon \searrow 0} \underline{F} (a + \varepsilon) \tag{i}
    \end{equation}
    or
    \begin{equation}\label{eq:casetwo}
        \underline{F} (a) < \lim \IP \Big( \widetilde{W}^N \leq a \Big) \leq  \lim_{\varepsilon \searrow 0} \underline{F} (a + \varepsilon). \tag{ii}
    \end{equation}
    The limit $\lim_{\varepsilon \searrow 0} \underline{F}(a +\varepsilon)$ exists due to monotonicity of $\underline{F}$.
    In Case \eqref{eq:casetwo}, since $\underline{F}$ is non-decreasing and $a$ is a point of continuity of $\lim_{N \to \infty} \IP (\widetilde{W}^N \leq a)$, this inequality must also hold for all $a''$ in a small interval $(a', a]$.
    But, being monotonic and bounded, $\underline{F}$ is continuous almost everywhere and thus it holds for almost all $a'' \in (a',a]$ that $\lim_{N \to \infty} \IP \big( \widetilde{W}^N \leq a''\big) \leq \underline{F} (a'')$, which is a contradiction. So Case (ii) does not occur.
    The same argument applies for $\overline{F}$.
    All together, we get that for all points of continuity $a \in [0,1]$ it holds that
    \begin{equation*}
        \lim_{N \to \infty} F_{\Bar{X}^N} (a) = \lim_{N \to \infty} \IP \big( \widetilde{W}^N \leq a \big) .
    \end{equation*}
    This proves the desired convergence in distribution for one locus and two types.

    The generalization to more than two types and one locus works analogously, with the only differences being that the CDF becomes a multivariate CDF and Chebyshev's inequality needs to be replaced by a multivariate analogue.
    When studying multiple loci, their independence can also be read off from the limiting multivariate CDF.
    
\end{proof}

\begin{proof}[Proof of Theorem \ref{thm:MainResult1}]
The proof now follows immediately from Lemma \ref{lemma:dirichlet_phase_3} and Lemma \ref{lem:type_frequencies_whol_pop}.
\end{proof}

\subsubsection{Proof of Theorem \ref{thm:MainResult2}}

We first prove a weaker variant of Theorem \ref{thm:MainResult2}.
\begin{lemma}\label{lem:weakerMainResult2}
    Theorem \ref{thm:MainResult2} holds when replacing $W^N (T) = W^{N,N} (T)$ with $W^{N, \jN} (T)$ where $(\jN)_{N \in \IN}$ is as given in Lemma \ref{assumptions:rates}.
\end{lemma}

\begin{proof}[Proof of Lemma \ref{lem:weakerMainResult2}]

Our aim is to find the limiting distribution of $W^{N,\jN} (T)$ as $N \to \infty$.
By Lemma \ref{lem:g2Y} and Lemma \ref{lem:phase_2_whp}, we know that the events
\begin{equation*}
    \pig\{\bfY^{N,1} \big(\tau_1^{N} \big) \neq \dagger \pig\}   \text{ and } \pig\{\bfY^{N,2} \big(\tau_2^{N} \big) \neq \dagger \pig\}
\end{equation*}
hold with high probability.
Since we only aim to prove convergence in distribution, we can argue on these two events for the rest of the proof.
Using Proposition \ref{prop:phase1_something} and Lemma \ref{lem:g2Y}, we can rewrite $W^{N, \jN}(T)$ as
\begin{align}\label{eq:proof_main_theorem_phase_1}
W^{N, \jN} (T) \stackrel{d}{=} W_g^N & \stackrel{d}{=} \frac{1}{\jN} \sum_{i=1}^{\xi^N} g^{(2)}_i \pig(\bfY^{N,1} \big(\tau_1^N \big)\pig) \delta_{X_{i,1}^N (T-\tau_1^N)}\notag \\
& \stackrel{d}{=} \frac{1}{\jN} \sum_{i=1}^{|Z^N|} Z^N_i \delta_{X_{i,1}^N (T-\tau_1^N)}
\end{align}
where $Z^N$ are the family sizes in a Hoppe's urn after $\jN$ draws which is independent of $\big(\bfX^N (t)\big)_{t \leq T -\tau_1^N}$. 
Thereby, we have expressed the distribution of $W^{N, \jN} (T)$ in terms of $Z^N$ and of
\begin{equation*}
\big(X_{i,1}^N (T-\tau_1^N)\big)_{i \in \{1, ... ,|Z^N| \}}
\end{equation*}
which are the types of the virus particles with label $1$ in the first $|Z^N|$ hosts.
The expression in \eqref{eq:proof_main_theorem_phase_1} can be understood as the type frequencies that arise when coloring the family sizes with the types of $X_{i,1}^N (T-\tau_1^N)$.
If these types are i.i.d.\ draws from some distribution, then \eqref{eq:proof_main_theorem_phase_1} converges to a Dirichlet distribution.
The rest of the proof is devoted to showing that the types are essentially i.i.d.\ and to describing their distribution. 

Using Proposition \ref{prop:dual_second_phase} and the type of exchangeability w.r.t.\ relabeling of $\bfX^N$ as defined and shown in Proposition \ref{prop:exchangeability}, the above expression \eqref{eq:proof_main_theorem_phase_1} is in distribution equal to
\begin{equation}\label{eq:eq:proof_main_theorem_phase_2}
    \frac{1}{\jN} \sum_{i=1}^{|Z^N|} Z^N_i \delta_{(X_{(i-1)L + 1,1,1}^N (T-\tau_1^N-\tau_2^N), ..., X_{iL,1,L}^N (T-\tau_1^N-\tau_2^N))}.
\end{equation}
This corresponds to the type frequencies that arise when the family sizes in a Hoppe's urn after $\jN$ draws are colored by
\begin{equation*}
    \delta_{(X_{(i-1)L + 1,1,1}^N (T-\tau_1^N-\tau_2^N), ..., X_{iL,1,L}^N (T-\tau_1^N-\tau_2^N))}.
\end{equation*}
It remains to be proven  that
\begin{equation*}
    \big(\delta_{(X_{(i-1)L + 1,1,1}^N (T-\tau_1^N-\tau_2^N), ..., X_{iL,1,L}^N (T-\tau_1^N-\tau_2^N))}\big)_{i \in [|Z^N|]}
\end{equation*}
are essentially i.i.d.\ draws from $\bar{X}^{N, \otimes} \big(T - \tau_1^N - \tau_2^N \big) \in \big(\cM_1 ([K]) \big)$, since then by Lemma \ref{lem:random_colored_hoppe} and Theorem \ref{thm:MainResult1} the Expression \eqref{eq:eq:proof_main_theorem_phase_2} converges in distribution to a 
$\Dir (\theta \nu^\otimes)$ distributed random variable in Case \eqref{eq:main_thm_cas_one} and to a $\Dir (\theta \psi^\otimes)$ distributed random variable in Case \eqref{eq:main_thm_case_two}.
To see this, let
\begin{equation*}
    (m_i, n_i)_{i \in [L \jN]}
\end{equation*}
be i.id.\ uniformly distributed on $[M] \times [N]$ and independent of everything else and define the event
\begin{equation*}
E^N \defeq \big\{ m_i \neq m_j \text{ for all } i \neq j \in [L \jN] \big\}.
\end{equation*}
Then, by Assumption \ref{assumptions:rates}, Point \ref{ass:sample_size} and Bernoulli's inequality, 
\begin{equation*}
    \IP (E^N) \geq \big( 1 - \tfrac{L \jN}{M}\big)^{\jN} \geq 1 - \tfrac{\jN^2}{M} \stackrel{N \to \infty}{\rightarrow} 1.
\end{equation*}
 It holds by the exchangeability stated in \eqref{eq:exchangeable} that
\begin{equation*}
    \pig(X_{(i-1)L + 1,1,1}^N \big(T-\tau_1^N-\tau_2^N\big), ..., X_{iL,1,L}^N \big(T-\tau_1^N-\tau_2^N\big)\pig)_{i \in [|Z^N|]}
\end{equation*}
is in distribution equal to
\begin{equation*}
    \pig(X_{m_{(i-1)L + 1}, n_{(i-1)L + 1},1}^N \big(T-\tau_1^N-\tau_2^N \big), ..., X_{m_{iL},n_{iL},L}^N \big(T-\tau_1^N-\tau_2^N \big) \pig)_{i \in [|Z^N|]}
\end{equation*} conditional on $E^N$.
By construction, these are i.i.d.\ draws from the distribution %
$\psi^{N, \otimes}$ which we define to be the unique stationary distribution of  $\Bar{X}^N$, and which converges to $\nu^\otimes$ in Case \eqref{eq:main_thm_cas_one} and to $\psi^\otimes$ in Case \eqref{eq:main_thm_case_two}.

\end{proof}
Now we can use Lemma \ref{lem:weakerMainResult2} to prove Theorem \ref{thm:MainResult2}.

\begin{proof}[Proof of Theorem \ref{thm:MainResult2}]
    Recall the definition
    \begin{equation*}
        W^{N, \jN} (T) = \frac{1}{\jN} \sum_{\tn \in \widetilde{\cI}} \delta_{X^N_{\tm, \tn} (T)} \in \cM_1 \big([K]^L \big)
    \end{equation*}
    where $\tilde{m}$ is a host index chosen uniformly at random and independent of everything from $[M]$ and
    \begin{equation*}
        \tilde{I} = \{ n_1, ..., n_{\jN} \}
    \end{equation*}
    are virus indices chosen without replacement from $[N]$.
    We can couple $W^{N, \jN} (T)$ with the \emph{exact} type frequencies when sampling \emph{all} virions from a random host by using the same random host index $\tilde{m}$ in the definition of
    \begin{equation*}
    W^N (T) = \frac{1}{N} \sum_{n=1}^N \delta_{X^N_{\tm, n} (T)} \in \cM_1 \big([K]^L\big).
    \end{equation*}
    For any $\bfk \in [K]^L$, the random variable $\jN \cdot \big(W^{N,\jN} (T)\big) ( \{\bfk\})$, which is the number of virions of sequence type $\bfk$ in the small sample of size $\jN$ from host $\tm$, follows a hypergeometric distribution with population size $N$, number of ``success'' items $N \cdot \big(W^N(T)\big)( \{\bfk\} )$ and number of draws $\jN$.
    Using Chebyshev's inequality we get that for any $\varepsilon > 0$,
    \begin{align*}
        \IP &\pig( \big\lVert W^{N, \jN} (T) - W^N(T) \big\rVert_\infty \geq \varepsilon \pig)\\[0.5em]
        \leq &\sum_{\bfk \in [K]^L} \IP \pig( \big\lvert \big(W^{N,\jN} (T)\big) ( \{\bfk\} ) - \big(W^N(T)\big) ( \{\bfk\} ) \big\rvert \geq \varepsilon \pig)
        \leq \; \frac{K^L}{4 \varepsilon^2 \jN} \stackrel{N \to \infty}{\to} 0.
    \end{align*}
    Therefore, for any closed $A \subseteq \cM_1 \big([K]^L\big)$ and $\varepsilon > 0$, we have
    \begin{align}\label{eq:proof_generalization}
    \limsup_{N \to \infty} \IP \pig(&W^N (T) \in A \pig)\\[0.5em]
    = \limsup_{N \to \infty} \IP &\Big( \big\{ W^N (T) \in A \big\} \cap \big\{ \big\lVert W^{N, \jN} (T) - W^N(T) \big\rVert_\infty < \varepsilon \big\} \Big) \notag \\[0.7em]
    \leq \limsup_{N \to \infty} \IP &\pig( W^{N, \jN} (T) \in A^\varepsilon\pig) \notag
    \end{align}
    where $A^\varepsilon$ denotes the open $\varepsilon$-surrounding of $A$.
    By Lemma \ref{lem:weakerMainResult2} and since $A^\varepsilon$ is a continuity set of the limiting distribution, the lim sup in the last line converges to the probability of $A^\varepsilon$ under the limiting distribution.
    
    Letting $\varepsilon \searrow 0$ we can upper bound the $\limsup$ of the first line by the probability of $A$ according to the limiting distribution (due to upper semicontinuity). The Portmanteau Theorem yields then the convergence of $W^N(T)$ to the limiting distribution of $W^{N,j_N}(T)$.
\end{proof}

\subsubsection{Proof of Theorem \ref{thm:MainResult3}}

Before we prove Theorem \ref{thm:MainResult3} we show one more lemma which says that the single locus type frequencies observed in the whole population do not vary too much on the time scale of the first two phases.
\begin{lemma}\label{lem:type_frequencies_constant}
Let $t' \defeq \frac{(\log (\jN))^2 N}{\rho_N}$.
Then,
\begin{equation*}
    \lim_{N \to \infty} \sup_{\bar{x} \in (\cM_1 ([K]))^L} \IP \Big( \sup_{t \in [0, t']} \big\lVert  \Bar{X}^N (t) - \bar{x} \big\rVert > c_N \; \pig\vert \;  \Bar{X}^N (0) = \Bar{x}\Big) = 0
\end{equation*}
where $c_N \geq \jN^{-0.5 + \varepsilon/2}$ and $\varepsilon \in \big(0,\frac{1}{4} \big)$ is arbitrary.\\
\end{lemma}

\begin{proof}
    Intuitively, the forces replication, reinfection, recombination and host replacement are ``symmetric'' in the sense that they do not favor one type over another.
    Only mutations introduce asymmetry.
    Formally, this implies that for any $\bar{x} \in \big(\cM_1 ([K])\big)^L$,
    \begin{equation}\label{eq:frequency_change}
        \big\lVert  \Bar{X}^N (t) - \bar{x} \big\rVert \leq \tfrac{1}{MN} \pig( 1 + \big\lvert U^N (t) \big\rvert + V^N (t) \pig) \text{ for all } t \geq 0
    \end{equation}
    almost surely, where $\big(U^N (t)\big)_{t \geq 0}$ is a martingale determined by all forces except mutation and $\big( V^N (t)\big)_{t \geq 0}$ is a non-decreasing process making jumps of size $+1$ whenever a mutation happens and furthermore $U^N (0) = V^N (0) = 0$.
    In order to obtain an upper bound for the r.h.s.\ of \eqref{eq:frequency_change}, we bound $|U^N (t)|$ and $V^N (t)$ separately.

    The process $\big(V^N (t)\big)_{t \geq 0}$ is a Poisson process with rate $M \mu_N$. 
    Using Chebyshev's inequality and the rate assumptions it is straightforward to check that its contribution is negligible.
    The contribution of $\big\lvert \big(U^N (t)\big)_{t \geq 0} \big\rvert$ can be bounded by realizing that for the quadratic variation it holds that
    \begin{equation*}
        \IE \pig[ \big\langle U^N \big\rangle_t \pig] \leq t \cdot \big( MN^2 + \gamma_N MN + \lambda_NM\big)
    \end{equation*}
    and by applying the Doob Inequality (9.2) of Chapter 1 in~\cite{Liptser1989} and the rate assumptions we arrive at
    \begin{align*}
        \mathbb{P} \bigg( \sup_{t \in [0,t']} \frac{\big\lvert U^N (t) \big\rvert}{NM} \geq c_N \bigg) &= \mathbb{P} \bigg( \sup_{t \in [0,t']} \big\lvert U^N (t) \big\rvert \geq c_N M N \bigg)\\[0.5em]
        & \leq \frac{1}{c_N^2 N^2 M^2} \IE \pig[ \big\langle U^N \big\rangle_{t'} \pig]\\[0.7em]
        & \leq \frac{t' \cdot \big( MN^2 + \gamma_N MN + \lambda_NM\big)}{c_N^2 N^2 M^2}.
    \end{align*}
    The term $t \gamma_N M N$ asymptotically dominates the rest of the numerator.
    Therefore, the whole expression is asymptotically dominated by
    \begin{align*}
    \frac{2 t' \gamma_N MN}{c_N^2 N^2 M^2} = \frac{2 \big(\log(\jN)\big)^2 N \gamma_N M N}{\rho_N c_N^2 N^2 M^2} = \frac{2 \big(\log(\jN)\big)^2 \gamma_N}{\rho_N c_N^2 M}.
    \end{align*}
    Using $j_N \log (j_N) \mu_N \in o (\rho_N)$, see Lemma \ref{assumptions:rates}, this is asymptotically dominated by
    \begin{align*}
    \frac{2 \big(\log(\jN)\big)^2 \gamma_N}{j_N \log (j_N) \mu_N c_N^2 M} = \frac{2 \log(\jN) \gamma_N}{j_N \mu_N c_N^2 M}.
    \end{align*}
    Since $\lambda_N$ and $\gamma_N$ are of the same asymptotic order and $\frac{\lambda_N}{M \mu_N} \lesssim C$, see Assumptions \ref{assumptions:main_rates}, this is asymptotically dominated by
    \begin{align*}
        2 C \frac{\log (\jN)}{\jN c_N^2} = 2C \frac{\log (\jN)}{\jN^\varepsilon} \stackrel{N \to \infty}{\longrightarrow} 0.
    \end{align*}
    This, proves the lemma for any $\bar{x}$, since none of the above bounds depends on $\bar{x}$.
\end{proof}

Again, we first prove a modified variant of Theorem \ref{thm:MainResult3} which we can then use to prove Theorem \ref{thm:MainResult3}.

\begin{lemma}\label{lem:weakerMainResult3}
    Theorem \ref{thm:MainResult3} holds when replacing $W^N (T)$ with $W^{N,\jN} (T)$ where $(\jN)_{N \in \IN}$ is as given in Lemma \ref{assumptions:rates}.
\end{lemma}

\begin{proof}[Proof of Theorem \ref{lem:weakerMainResult3}]
Our aim is to prove that in Case \eqref{eq:main_thm_cas_one},
\begin{equation}\label{eq:MainResult3ProofOne}
    \lim_{N \to \infty} \IP \pig( W^{N, \jN} (T) \in A, \bar{X}^{N,\mathfrak{s}}(T) \in B \pig) = \1_{\{\nu \in B\}}\int_A \fDir_{\theta, \nu^\otimes} (y) dy 
\end{equation}
and in Case \eqref{eq:main_thm_case_two},
\begin{equation}\label{eq:MainResult3ProofTwo}
    \lim_{N \to \infty} \IP \pig( W^{N, \jN} (T) \in A, \bar{X}^{N,\mathfrak{s}}(T) \in B \pig) = \int_A \int_B  \fDir_{\theta, x^\otimes} (y) \prod_{\ell=1}^L \fDir_{C, \nu_\ell} (x_\ell) dx dy
\end{equation}
for a large class of measurable $A \subseteq \cM_1 \big( [K]^L \big)$ and $B \subseteq \big( \cM_1 ([K])\big)^L$.
In Case \eqref{eq:main_thm_case_two}, it suffices to show this property for all sets
\begin{equation*}
    A \in \mathfrak{A} \defeq \pig\{ A^{(\zeta)} : \zeta \in \cM_1 \big( [K]^L \big) \pig\}
\end{equation*}
where
\begin{equation*}
    A^{(\zeta)} = \pig\{\psi \in \cM_1 \big( [K]^L \big) : \psi \big(\{s\}\big) \leq \zeta \big(\{s\}\big) \text{ for all } s \in [K]^L \setminus \{(K, ..., K)\} \pig\}
\end{equation*}
and $B \in \mathfrak{B} \defeq \pig\{B^{(\eta)} : \eta \in \big( \cM_1 ([K])\big)^L \pig\}$ where
\begin{equation*}
    B^{(\eta)} = \pig\{\varphi \in \big( \cM_1 ([K])\big)^L : \varphi_\ell \big(\{k\} \big) \leq \eta_\ell \big( \{k\} \big) \text{ for all } (\ell, k) \in [L] \times [K] \setminus \big\{ (K,L) \big\}\pig\}.
\end{equation*}
In Case \eqref{eq:main_thm_cas_one} it suffices to study $\mathfrak{B} \setminus \{B \in \mathfrak{B} : \nu^\otimes \in \partial B \}$.
This is due to the characterization of weak convergence via pointwise convergence of multivariate CDFs at all points of continuity as given in e.g.\ \cite{Durrett2019}, Theorem 3.10.2.

We first prove Case \eqref{eq:appendix_bound_two} by bounding the lim inf and lim sup in \eqref{eq:MainResult3ProofTwo}.
Let $A \in \mathfrak{A}$ and $B \in \mathfrak{B}$.
For the lim sup, note that
\begin{align}\label{eq:test}
    &\limsup_{N \to \infty} \IP \pig( \big\{ W^{N, \jN} (T) \in A \big\} \cap \big\{ \bar{X}^{N,\mathfrak{s}}(T) \in B \big\}\pig) \notag \\
    \leq &\limsup_{N \to \infty} \IP \Big( \big\{ W^{N, \jN} (T) \in A \big\} \cap \pig( \big\{ \bar{X}^{N,\mathfrak{s}}(T) \in B \big\} \cup \big\{\bar{X}^N (T) \in B_\varepsilon \big\} \pig)\Big)
\end{align}
for any $\varepsilon>0$ where $B_\varepsilon$ denotes the $\varepsilon$-neighborhood of $B$, that is
\begin{equation*}
    B_\varepsilon \defeq \pig\{ \varphi \in \big( \cM_1 ([K])\big)^L : d(\varphi, B) < \varepsilon\pig\}
\end{equation*}
with the metric $d : \big( \cM_1 ([K])\big)^L \times \big( \cM_1 ([K])\big)^L \to \IR_0^+$ defined by
\begin{equation*}
    d \big(\varphi^{(1)}, \varphi^{(2)} \big) = \max_{\ell \in [L], k \in [K]} \pig\lvert \varphi^{(1)}_\ell \big(\{k\} \big) - \varphi^{(2)}_\ell \big( \{k\} \big) \pig\rvert.
\end{equation*}
The expression in \eqref{eq:test} is lesser than or equal to
\begin{align*}
    &\limsup_{N \to \infty} \IP \pig( \big\{ W^{N, \jN} (T) \in A \big\} \cap \big\{\bar{X}^N (T) \in B_\varepsilon \big\}\pig)\\
    +& \limsup_{N \to \infty} \IP \pig( \big\{ W^{N, \jN} (T) \in A \big\} \cap \big\{\bar{X}^N (T) \notin B_\varepsilon \big\} \cap \big\{ \bar{X}^{N,\mathfrak{s}}(T) \in B \big\}\pig).
\end{align*}
The probability in the second term is dominated by $\IP \big(|\bar{X}^{N,\mathfrak{s}}(T) - X^N (T)| \geq \varepsilon \big)$ which tends to zero as $N \to \infty$ by assumption, and hence it remains to estimate the first term 
\begin{align*}
    &\limsup_{N \to \infty} \IP \pig( \big\{ W^{N, j_N} (T) \in A \big\} \cap \big\{\bar{X}^N (T) \in B_\varepsilon \big\}\pig)\\
    \leq&\limsup_{N \to \infty} \IP \pig( \big\{ W^{N, j_N} (T) \in A \big\} \cap \big\{\bar{X}^N (T) \in B \big\}\pig)\\
    &+\limsup_{N \to \infty} \IP \pig( \big\{\bar{X}^{N} (T) \in B_\varepsilon \setminus B \big\}\pig)
\end{align*}
Letting $\varepsilon \to 0$, we notice that in both Cases \eqref{eq:main_thm_cas_one} and \eqref{eq:main_thm_case_two} of Theorem \ref{thm:MainResult1}, the second summand vanishes (due to continuity of the limiting distribution).
Therefore, it suffices to study the first summand for which it holds that
\begin{align*}
    &\limsup_{N \to \infty} \IP \pig( \big\{ W^{N, \jN}(T) \in A \big\} \cap \big\{ \Bar{X}^N (T) \in B \big\} \pig)\\
    \leq &\limsup_{N \to \infty} \IP \pig( \big\{ W^{N, \jN}(T) \in A \big\} \cap \big\{ \Bar{X}^N (T) \in B \big\} \cap \big\{ \Bar{X}^N (T-\tau_1 - \tau_2) \in B \big\}\pig)\\
    + &\limsup_{N \to \infty} \IP \pig( \big\{ W^{N, \jN}(T) \in A \big\} \cap \big\{ \Bar{X}^N (T) \in B \big\} \cap \big\{ \Bar{X}^N (T-\tau_1 - \tau_2) \notin B_{\varepsilon'} \big\}\pig)\\
    +&\limsup_{N \to \infty} \IP \pig( \big\{ W^{N, \jN}(T) \in A \big\} \cap \big\{ \Bar{X}^N (T) \in B \big\} \cap \big\{ \Bar{X}^N (T-\tau_1 - \tau_2) \in B_{\varepsilon'} \setminus B \big\}\pig)\\
    \leq &\limsup_{N \to \infty} \IP \pig( \big\{ W^{N, \jN}(T) \in A \big\} \cap \big\{ \Bar{X}^N (T-\tau_1 - \tau_2) \in B \big\}\pig)\\
    + &\limsup_{N \to \infty} \IP \pig( \big\{ \Bar{X}^N (T) \in B \big\} \cap \big\{ \Bar{X}^N (T-\tau_1 - \tau_2) \notin B_{\varepsilon'} \big\}\pig)\\
    +&\limsup_{N \to \infty} \IP \pig( \big\{ \Bar{X}^N (T-\tau_1 - \tau_2) \in B_{\varepsilon'} \setminus B \big\}\pig)
\end{align*}
for any $\varepsilon' > 0$.
The second term equals zero due to Lemma \ref{lem:type_frequencies_constant}.
In the limit $\varepsilon' \to 0$, the last term vanishes as well since $\Bar{X}^N (T-\tau_1 - \tau_2)$ has the same distribution as $\Bar{X}^N (T)$. Hence, only the first term is relevant.
By the proof of Theorem \ref{thm:MainResult2}, it is clear that the first term yields the desired upper bound.

To obtain the correct asymptotic lower bound, first define the $\varepsilon$-interior of $B$, i.e.\ the set of points in $B$ that have at least distance $\varepsilon$ to the boundary by
\begin{equation*}
    B_{-\varepsilon} \defeq \pig\{ \varphi \in B : d \big(\varphi, B^c \big) > \varepsilon \pig\}
\end{equation*}
where $B^c \defeq \big( \cM_1 ([K])\big)^L \setminus B$.
Then, for any $\varepsilon > 0$ we have
\begin{align*}
    &\liminf_{N \to \infty} \IP \pig( \big\{W^{N, \jN} (T) \in A \big\} \cap \big\{\bar{X}^{N,\mathfrak{s}}(T) \in B \big\} \pig)\\
    \geq &\liminf_{N \to \infty} \IP \pig( \big\{W^{N, \jN} (T) \in A \big\} \cap \big\{\bar{X}^{N,\mathfrak{s}}(T) \in B \big\} \cap \big\{ \bar{X}^N (T) \in B_{-\varepsilon} \big\} \pig)\\
    \geq &\liminf_{N \to \infty} \IP \pig( \big\{W^{N, \jN} (T) \in A \big\}  \cap \big\{ \bar{X}^N (T) \in B_{-\varepsilon} \big\} \pig)\\
    - &\limsup_{N \to \infty} \IP \pig( \big\{\bar{X}^{N,\mathfrak{s}}(T) \notin B \} \cap \{ \bar{X}^N (T) \in B_{-\varepsilon} \big\} \pig).
\end{align*}
The last term equals zero since, by assumption, $|\bar{X}^{N,\mathfrak{s}}(T) -  \bar{X}^{N}(T)| \rightarrow 0 $ almost surely.
Therefore, the whole expression is greater than or equal to
\begin{align*}
    &\liminf_{N \to \infty} \IP \big(\{W^{N, \jN} (T) \in A\}  \cap \{ \bar{X}^N (T) \in B_{-\varepsilon}\} \big)\\
    \geq &\liminf_{N \to \infty} \IP \big(\{W^{N, \jN} (T) \in A\}  \cap \{ \bar{X}^N (T) \in B_{-\varepsilon}\} \cap \{ \bar{X}^N (T-\tau_1 -\tau_2) \in B_{-2\varepsilon}\} \big)\\
    \geq &\liminf_{N \to \infty} \IP \big(\{W^{N, \jN} (T) \in A\}  \cap \{ \bar{X}^N (T-\tau_1 - \tau_2) \in B_{-2\varepsilon}\}\big)\\
    - &\limsup_{N \to \infty} \IP \big( \bar{X}^N (T) \notin B_{-\varepsilon}\} \cap \{ \bar{X}^N (T-\tau_1 -\tau_2) \in B_{-2\varepsilon}\} \big).
\end{align*}
By Lemma \ref{lem:type_frequencies_constant}, the second term is zero.
In Case \eqref{eq:main_thm_case_two}, the first term equals
\begin{equation*}
    \int_A \int_{B_{-2\varepsilon}} \fDir_{\theta, x^\otimes} (y) \prod_{\ell=1}^L \fDir_{C, \nu_\ell} (x_\ell) dx dy .
\end{equation*}
Letting $\varepsilon \to 0$ yields the desired result.
One can argue analogously in Case \eqref{eq:main_thm_cas_one} where we can - as noted above - assume that $\nu^\otimes \notin \partial B$
\end{proof}

\begin{proof}[Proof of Theorem \ref{thm:MainResult3}]
    The proof of Theorem \ref{thm:MainResult3} now follows from Lemma \ref{lem:weakerMainResult3} by following along the steps in the proof of Theorem \ref{thm:MainResult2}.
    The only adaptation one has to make is to add a condition on $\bar{X}^{N,\mathfrak{s}}(T)$ in Equation \eqref{eq:proof_generalization}.
\end{proof}

\section{Appendix}

\subsection{Limiting Dirichlet Distribution with Random Parameter}\label{subsec:compound_dirichlet}
Let $C, \theta > 0$ and $\nu \in \cM_1 \big([K]^L\big)$ be as defined above. Recall that $\nu_\ell$ denotes the $\ell$-th marginal of $\nu$ for $\ell \in [L]$.
Denote by $f_{C, \nu_\ell}^\textnormal{Dir}: \cM_1 ([K]) \rightarrow \IR^+_0$ the density of a Dirichlet distribution with base measure $\nu_\ell$ and concentration parameter $C$.
Denote by $\cB \big( [0,1]^{K^L} \big)$ the Borel sigma algebra on the space of sequence type frequencies.
In Case (ii) of Theorem \ref{thm:MainResult2}, the distribution of the limiting type frequencies $W \defeq \lim_{N \to \infty} W^N$ can be represented by
\begin{equation*}
    \IP (W \in B) = \int_B \int_{(\cM_1 ([K]))^L} \fDir_{\theta, x^\otimes} (y) \prod_{\ell=1}^L \fDir_{C, \nu_\ell} (x_\ell) dx dy,
\end{equation*}
for $B \in \mathcal{B}([0,1]^{K^L})$.
Here, $x^\otimes$ denotes the product measure of $x_\ell, \ell \in [L]$.

\subsection{Type Frequencies in Randomly Colored Hoppe's Urns}
The next lemma describes the asymptotic type frequencies observed when coloring Hoppe's urns in a doubly stochastic manner.
To be precise, we consider the following setting:
Let $N \in \IN$ and let $E=\{e_1, .., e_k\}$ be a finite type space.
Let $p_N = (p_N (e))_{e \in E}$ be a random probability measure, independent of everything else.
First draw $p_N$ and then color the family sizes of a Hoppe's urn with parameter $\theta_N$ and  with $\jN$ draws i.i.d.\ according to $p_N$.
The resulting asymptotic type frequencies are given in the next lemma for $N \to \infty$ if $p_N \to p$ in distribution, where $p = (p_1,..., p_{e_k})$ is either deterministic or has a Lebesgue density.

\begin{lemma}\label{lem:random_colored_hoppe}
    For $N \in \IN$, let $W^N$ be the type frequencies observed when coloring the family sizes after $\jN$ draws from a $\theta_N$-Hoppe's urn i.i.d.\ with random colors $p_N$.
    If  $\jN \to \infty$, $\theta_N \to \theta>0$ and $p_N \to p$ in distribution where $p$ is either (a) deterministic or (b) has a Lebesgue density $f_p$, then in Case (a)
    \begin{equation*}
        W^N \stackrel{d}{\to} \textnormal{Dir}(\theta p_{e_1}, ..., \theta p_{e_k})
    \end{equation*}
    and in Case (b)
    \begin{equation*}
        W^N \stackrel{d}{\to} W
    \end{equation*}
    where $W$ has Lebesgue density $f_W (y_1, ... y_{e_k}) = \int_{\Delta} f^{\textnormal{Dir}}_{\theta, x} (y_1,...,y_{e_k}) f_p (x_1, ..., x_{e_k}) dx$.
\end{lemma}
The proof is based on the construction given in \cite{Tavare1987}, see also \cite{birkner2024}.
\begin{proof}
    We prove Case (a) first and then Case (b) via an approximation argument.
    Let $F^N_i (\jN)$ be the size of family $i$ after the $\jN$-th draw from Hoppe's urn with mutation rate $\theta_N/2$ per lineage. Consider the relative sizes of the families generated by Hoppe's urn
\begin{align}%
    \frac{F^N (\jN)}{\jN} = \left( \frac{1}{\jN} F_1^N (\jN), \frac{1}{\jN} F_2^N (\jN), ..., \frac{1}{\jN} F_{\jN}^N (\jN), 0, ...\right).
\end{align}
The random type frequencies $W^N$, which take values in $\cM_1 (E)$, can be expressed by summing over the relative family sizes of each type (or ``color''). That is,
\begin{equation*}
    W^N  =  \sum_{i=1}^\infty \frac{F_i^N (\jN)}{\jN} \cdot \delta_{V_i^N}
\end{equation*}
where the $V_i^N$ take values in $E$ and are independent (in $i$) draws from the same random distribution $p_N$ and $V_i^N \to V_i$ almost surely for each $i \in \mathbb{N}$ where $V_i$ has distribution $p$.
Such sequences can be defined using Lemma \ref{lem:bounding_V} below by taking i.i.d.\ Uniform$([0,1] \times [0,1])$ distributed random variables $U_1, U_2, ...$ and setting $V^N_i = v(U_i, p_N)$ with $v$ as in Lemma \ref{lem:bounding_V}.
In this construction, $p_N$ determines the (random) distribution of colors for the mutations and $U_i$ creates i.i.d.\ (in $i$) samples from this distribution.

A classical approach to study $W^N$ given in~\cite{Tavare1987} is to represent the family sizes $F_i^N$ via Yule processes with immigration.
In~\cite{Tavare1987}, only constant $\theta_N$ and constant, deterministic $p_N$ are  considered but the approach can be extended to our situation as follows:
Let $T_1, T_2, ...$ be the jump times of a time-homogeneous Poisson Process on $\mathbb{R}^+$ with rate one.
Define processes
\begin{equation*}
    \big(Z^N (t) \big)_{t \geq 0} = \big(Z_1^N (t), Z_2^N (t), ... \big)_{t \geq 0}
\end{equation*}
taking values in $\IN^\IN$ starting at $Z_0^N = (0, 0, ...)$ where $Z_i^N (t)$ denotes the size of family $i$ of the $N$-th process at time $t$.
At time $\frac{1}{\theta_N} T_i$ an immigrant appears and founds the $i$-th family which from then on grows independently of everything else as a rate $1$ Yule process.
Therefore, denote by $Y_1(t), Y_2(t), ...$ independent rate $1$ Yule processes with $Y_i(0) = 1$ for all $i$ and $Y_i(s) = 0$ for all $s < 0$.
We define
\begin{equation*}
    Z^N (t) = \left(Y_1 \big(t-\tfrac{T_1}{\theta_N}\big), Y_2 \big(t-\tfrac{T_2}{\theta_N}\big), ... \right) .
\end{equation*}
Let $S^N (t) \defeq \sum_{i=1}^\infty Z_i^N (t)$ be the number of individuals at time $t$ in the $N$-th process and $\tau^N \defeq \min \big\{ t: S^N (t) = \jN \big\}$ be the first time at which the $N$-th process reaches a total population size of $\jN$ individuals.
Then we have
\begin{equation*}
    \frac{F^N (\jN)}{\jN} \stackrel{d}{=} \left( \frac{1}{\jN} Z_1^N (\tau^N), \frac{1}{\jN} Z_2^N (\tau^N),  ...\right).
\end{equation*}
 Lemma \ref{lem:martingale_convergence} below yields
\begin{equation}%
    \left( e^{-\tau^N} Z_1^N (\tau^N), e^{-\tau^N} Z_2^N (\tau^N), ... \right) \stackrel{N \rightarrow \infty}{\to} \left(e^{-\frac{T_1}{\theta}}A_1, e^{-\frac{T_2}{\theta}}A_2, ... \right)
\end{equation}
almost surely coordinate-wise, where $A_1, A_2, ...$ are i.i.d.\ Exp(1) random variables.
It also holds that
\begin{equation}\label{eq:W_N_limit}
    W^N =  \frac{\sum_{i=1}^\infty e^{-\tau^N} Z_i^N (\tau^N) \delta_{V_i^N}}{\sum_{i=1}^\infty e^{-\tau^N} Z_i^N (\tau^N)} \stackrel{N \to \infty}{\longrightarrow} \frac{\sum_{i=1}^\infty e^{-\frac{T_i}{\theta}} A_i  \delta_{V_i}}{\sum_{i=1}^\infty e^{-\frac{T_i}{\theta}} A_i} \eqdef W
\end{equation}
almost surely since the numerator and denominator converge almost surely.
 We give the argument for the denominator, for the numerator the convergence follows analogously. 
We have
\begin{align*}
\sum_{i=1}^\infty e^{-\tau^N} Z_i^N (\tau^N) &
=\sum_{i=1}^\infty e^{-\frac{T_i}{\theta_N} } e^{-(\tau^N- \frac{T_i}{\theta_N})  } Y_i^N \Big(\tau^N - \frac{T_i}{\theta_N} \Big)  
\end{align*}
and estimate 
\begin{align*}
e^{-(\tau^N- \frac{T_i}{\theta_N})  } Y_i^N \Big(\tau^N - \frac{T_i}{\theta_N} \Big) 
\leq \sup_{t\geq 0} \sup_{N\geq 1} e^{-(t- \frac{T_i}{\theta_N})  } Y_i^N \Big(t - \frac{T_i}{\theta_N} \Big) : = M_i.
\end{align*}
The random variables $(M_i)_{i\geq 1}$ are identically distributed and $\mathbb{E}[M_1]<\infty.$ 
An application of the Borel-Cantelli Lemma yields 
$\limsup M_n/n =0$ a.s.
Furthermore,  
$$\frac{T_i}{i\theta_N} \xrightarrow{i\rightarrow \infty} \frac{1}{\theta_N} $$
almost surely. 
Hence, for $m\geq N_0$ and $N_0$ large enough we have 
\begin{align*}
\sup_{N} \sum_{i=m}^\infty e^{-\tau^N} Z_i^N (\tau^N) \leq  \sum_{i=m}^{\infty} e^{-\frac{i}{2\theta}} i. 
\end{align*}

In Case (a), the random variables $(V_i)_{i \in \IN}$ are i.i.d.\ and independent of $(A_i)_{i \in \IN}$ and $(T_i)_{i \in \IN}$.
Considering $\sum_i \delta_{(A_i, T_i/\theta)}$ as a Poisson point process on $\mathbb{R}^+ \times \mathbb{R}^+$ with intensity measure $e^{-x}dx \times \theta dt$ and regarding the multiplication by $\delta_{V_i}$ in the numerator as a Poisson coloring, one can show for the r.h.s.\ of \eqref{eq:W_N_limit} that
\begin{equation*}
    W(\{e\}) \stackrel{d}{=} \frac{G_e}{\sum_{r=1}^k G_r}
\end{equation*}
where $(G_r)_{r \in [k]}$ are independent and $G_r \sim$ Gamma($\theta p_r$).
It is well known that then
\begin{equation*}
    \big(W(\{e\})\big)_{e \in [k]} \sim \textnormal{Dirichlet} (\theta p_1, ..., \theta p_k).
\end{equation*}

In Case (b) the independence of $(V_i)_{i \in \IN}$ is lost, hence we argue as follows: first, note that $\sum_{e \in [k]} W (\{e\}) = 1$. Therefore, it suffices to study $W(\{e\})_{e \in [k-1]}$.
Using Lemma \ref{lem:bounding_V}, one can construct sequences $(\overline{V}_i^{(d)})$ and $(\underline{V}_i^{(d)})$ such that
\begin{equation*}
    \delta_{\underline{V}_i^{(d)}} (\{e\}) \leq \delta_{V_i} (\{e\}) \leq \delta_{\overline{V}_i^{(d)}} (\{e\}) 
\end{equation*}
for all $i, d \in \IN$ and $e \in [k-1]$ almost surely.
Letting
\begin{equation*}
    \underline{W}^{(d)} \defeq \frac{\sum_{i=1}^\infty e^{-\frac{T_i}{\theta}} A_i  \delta_{\underline{V}_i^{(d)}}}{\sum_{i=1}^\infty e^{-\frac{T_i}{\theta}} A_i} \quad \text{and} \quad \overline{W}^{(d)} \defeq \frac{\sum_{i=1}^\infty e^{-\frac{T_i}{\theta}} A_i  \delta_{\overline{V}_i^{(d)}}}{\sum_{i=1}^\infty e^{-\frac{T_i}{\theta}} A_i}
\end{equation*}
it holds that that
\begin{equation*}
    \underline{W}^{(d)} \big(\{e\} \big) \leq
    W \big(\{e\} \big) \leq \overline{W}^{(d)} \big(\{e\}\big)
\end{equation*}
for all $e \in [k-1]$ and $d \in \IN$.
We get that
\begin{align*}
    \IP &\big( W(\{1\}) \leq a_1, ..., W(\{k-1\}) \leq a_{k-1}\big)\\[0.7em]
    &\leq \IP \big( \underline{W}^{(d)}(\{1\}) \leq a_1, ..., \underline{W}^{(d)}(\{k\}) \leq a_k\big)\\
    &= \sum_{\beta \in \Delta^{(d)}} \underbrace{\IP \Big( \underline{W}^{(d)}(\{1\}) \leq a_1, ..., \underline{W}^{(d)}(\{k\}) \leq a_k \; \big\vert \; \lfloor B \rfloor^{(d)} = \beta \Big)}_{(*)} \IP \big( \lfloor B \rfloor^{(d)} =  \beta \big),
\end{align*}
where 
\begin{equation*}
\Delta^{(d)}: = \Big\{\beta= \big(\beta_1, ..., \beta_k \big): \beta_i \in \big\{0, \nicefrac{1}{2^d},..., 1 \big\}, \sum_{i=1}^k \beta_i =1 \Big\}.    
\end{equation*}
By Part (a) of the proof and Point 4 of Lemma \ref{lem:bounding_V}, we get that ($*$) is the CDF of a Dirichlet distribution with parameters $\theta$ and $\beta$.
Taking $d \to \infty$ and applying Lebesgue's differentiation theorem, the Vitali convergence theorem and continuity of ($*$) in $\beta$, it follows that
\begin{align*}
\IP \big( W(\{1\}) \leq a_1, ..., W&(\{k-1\}) \leq a_{k-1}\big)\\[0.7em]
\leq \int_0^{a_1}... \int_{0}^{a_{k-1}} \int_{\Delta} f^{\textnormal{Dir}}_{\theta, y} (x_{e_1},...,x_{e_k}) f_p&(y_{e_1}, ..., y_{e_k}) 
dy_{e_1} .. dy_{e_k} dx_{e_1} ... d_{x_{e_k}}.
\end{align*}
Applying the same reasoning for $\overline{W}$ on the event $\{B \in \cB^{d}\}$ from Lemma \ref{lem:bounding_V}, we get an analogous lower bound.
Together, this yields
\begin{align*}
\IP \big( W(\{1\}) \leq a_1, ..., W&(\{k-1\}) \leq a_{k-1}\big)\\
= \int_0^{a_1}... \int_{0}^{a_{k-1}} \int_{\Delta} f^{\textnormal{Dir}}_{\theta, y} (x_{e_1},...,x_{e_k}) f_p&(y_{e_1}, ..., y_{e_k}) 
dy_{e_1} .. dy_{e_k} dx_{e_1} ... d_{x_{e_k}}.
\end{align*}

\end{proof}

\begin{lemma}\label{lem:martingale_convergence}
    Let $\big(Y(t)\big)_{t \in \IR}$ be a rate $1$ Yule process with $Y(0)=1$ and $Y(s)=0$ for $s < 0$. Let $T$ be an independent random variable taking values in $[0, \infty)$. Let $\tau^N$ be a sequence of random times that converges to infinity almost surely. Let $c_N$ be a nonnegative sequence with $c_N \to c$ as $N \to \infty$. Then
    \begin{equation*}
        e^{-\tau^N} Y \left( \tau^N - T c_N\right) \to e^{-cT} A
    \end{equation*}
    almost surely where $A$ is Exp(1)-distributed and independent of $T$.
\end{lemma}
\begin{proof}
    Note that $e^{-t}Y(t)$ is an $L^2$ bounded martingale and converges almost surely to an Exp(1)-distributed random variable $A$ (see~\cite{birkner2024}).
    Note that
    \begin{align*}
        e^{-\tau_N} Y \left( \tau_N - T c_N\right) = e^{-T c_N} \cdot e^{-(\tau_N - T c_N)} Y \left( \tau_N - T c_N\right)
    \end{align*}
    Now see that $\tau_N - T c_N$ converges to $\infty$ a.s. so
    \begin{align*}
        \lim_{N \to \infty} e^{-(\tau_N - T c_N)} Y_1 \left( \tau_N - T c_N\right) = \lim_{t \to \infty} e^{-t}Y(t) = A
    \end{align*} almost surely.
    On the other hand $e^{-T c_N} \to e^{-T c}$ a.s., and since multiplication is compatible with almost sure convergence, we get the desired result.
\end{proof}

\begin{lemma}\label{lem:bounding_V}
    Fix $n \in \IN$ and let $(B_N)_{N \in \IN}$ be a sequence of random variables taking values in $\cM_1([n])$, converging almost surely to some limit $B$.
    Let $(U_i)_{i \in \IN} = (U_i^{(1)}, U_i^{(2)})_{i \in \IN}$ be a sequence of i.i.d.\ uniformly distributed random variables on $[0,1] \times [0,1]$ which are independent of $(B_N)_{N \in \IN}$.
    Define
    \begin{equation*}
        v: [0,1] \times [0,1] \times \cM_1 ([n]) \rightarrow [n]
    \end{equation*}
    by
    \begin{equation*}
        v(u, b) \defeq v \big(u^{(1)}, u^{(2)}, b\big)=
        \begin{cases}
            k, \; \; \text{ if } u^{(2)} \leq  b \big([n-1] \big
            ) \text{ and } u^{(1)} \in \pig[\tfrac{b([k-1])}{b([n-1])}, \tfrac{b([k])}{b([n-1])}\pig)\mybreak
            n, \quad \text{ otherwise.}
        \end{cases}
    \end{equation*}
    for $k \in [n-1]$.
    Assume that the distribution of $B_N$ has finite support for all $N \in \IN$ and that $\IP(B_N = b) > 0$ for some $b \in \cM_1([n])$.
    Then, conditional on $\{B_N = b\}$, the random variables $\big(v(U_i, B_N)\big)_{i \in \IN}$ are i.i.d.\ on $[n]$ with distribution $b$.
    Furthermore, define 
    \begin{equation*}
    \underline{v}: [0,1] \times [0,1] \times \cM_1 ([n]) \times \IN \rightarrow [n]    
    \end{equation*}
    via
    \begin{equation*}
        \underline{v}(u,b,d) \defeq
        \begin{cases}
            k, \quad \text{ if } u^{(2)} \leq  \frac{\lfloor b (\{k\}) \cdot 2^d \rfloor b([n-1])}{2^d b(\{k\})} \text{ and } u^{(1)} \in \big[\nicefrac{b([k-1])}{b([n-1])}, \nicefrac{b([k])}{b([n-1])}\big)\mybreak
            n, \quad \text{ otherwise.}
        \end{cases}
    \end{equation*}
    Note that $\underline{v}$ can be understood as a rounding down of $v$ and it holds that
    \begin{equation*}
        v(u,b) = k \implies \underline{v} (u,b,d) \in \{k, n\}.
    \end{equation*}
    for all $k \in [n]$.
    Likewise, we define a rounding up function $\overline{v}$. Here we need to make sure that when we are rounding up, we do not generate values larger than 1. %
    To this end, for $d \in \IN$ let
    \begin{equation*}
        \mathcal{B}^{(d)} = \Big\{b \in \cM_1 ([n]): \frac{\lceil b (\{k\}) \cdot 2^d \rceil b([n-1])}{2^d b(\{k\})} \leq 1 \text{ for all } k \in [n-1]\Big \}
    \end{equation*}
    and define $\overline{v} :[0,1] \times [0,1] \times \cB^{(d)} \IN \rightarrow [n]$ via
    \begin{equation*}
        \overline{v}(u,b,d) \defeq
        \begin{cases}
            k, \quad \text{ if } u^{(2)} \leq  \frac{\lceil b (\{k\}) \cdot 2^d \rceil b([n-1])}{2^d b(\{k\})} \text{ and } u^{(1)} \in \big[\nicefrac{b([k-1])}{b([n-1])}, \nicefrac{b([k])}{b([n-1])}\big)\mybreak
            n, \quad \text{ otherwise.}
        \end{cases}
    \end{equation*}
    The following hold:
    \begin{enumerate}
        \item $v(U_i, B_N) \to v (U_i, B)$ almost surely for all $i \in \IN$.
        \item $v(U_i, B)$ dominates $\underline{v}(U_i, B, d)$ for any $d \in \IN$ in the sense that
        \begin{equation*}
            v(U_i,B) = k \implies \underline{v} (U_i,B,d) \in \{k, n\}.
        \end{equation*}
        for all $k \in [n]$ almost surely.
        \item $v(U_i, B)$ is dominated by $\overline{v}(U_i, B, d)$ for any $d \in \IN$ in the sense that
        \begin{equation*}
            \overline{v} (U_i,B,d) = k \implies v(U_i,B) \in \{k, n\}
        \end{equation*}
        for all $k \in [n]$ on $\{B \in \cB^{(d)}\}$.
        \item Conditional on
        $$\Big\{B(\{1\}) \in \big[\tfrac{\beta_1}{2^d}, \tfrac{\beta_1 + 1}{2^d} \big), B(\{2\}) \in \big[\tfrac{\beta_2}{2^d}, \tfrac{\beta_2 + 1}{2^d}\big), ..., B(\{n-1\}) \in \big[\tfrac{\beta_{n-1}}{2^d}, \tfrac{\beta_{n-1} + 1}{2^d} \big) \Big\},$$
        where $\beta_1, ... \beta_{n-1} \in \{0,..., 2^d - 1\}$ with $\sum_{i=1}^{n-1} \beta_1 \leq 2^d$ it holds that
        \begin{equation*}
            \big(\underline{v}(U_i, B, d)\big)_{i \in \IN} \text{ are i.i.d.\ on $[n]$ with weights } \big(\tfrac{\beta_1}{2^d}, ..., \tfrac{\beta_{n-1}}{2^d}, 1 - {\textstyle\sum_{i=1}^{n-1}} \tfrac{\beta_i}{2^d} \big).
        \end{equation*}
        \item Conditional on $B \in \cB^{(d)}$ and
        $$\Big\{B(\{1\}) \in \big[\tfrac{\beta_1-1}{2^d}, \tfrac{\beta_1}{2^d} \big), B(\{2\}) \in \big[\tfrac{\beta_2-1}{2^d}, \tfrac{\beta_2}{2^d}\big), ..., B(\{n-1\}) \in \big[\tfrac{\beta_{n-1}-1}{2^d}, \tfrac{\beta_{n-1}}{2^d} \big) \Big\},$$
        where $\beta_1, ... \beta_{n-1} \in \{1,...,2^d\}$ with $\sum_{i=1}^{n-1} \beta_1 \leq 2^d$ it holds that
        \begin{equation*}
            \big(\overline{v}(U_i, B,d)\big)_{i \in \IN} \text{ are i.i.d.\ on $[n]$ with weights } \big(\tfrac{\beta_1}{2^d}, ..., \tfrac{\beta_{n-1}}{2^d}, 1 - {\textstyle\sum_{i=1}^{n-1}} \tfrac{\beta_i}{2^d} \big).
        \end{equation*}
        \item If $B$ has a Lebesgue density, then
    \begin{equation*}
        \lim_{d \to \infty} \IP \big(B \in \mathcal{B}^{(d)} \big) = 1.
    \end{equation*}
    \end{enumerate}
\end{lemma}

\begin{proof}
    \begin{enumerate}
    \item By definition we have that
        \begin{equation*}
        \lim_{N \to \infty} v(U_i, B_N) = v(U_i, B)
        \end{equation*}
        if 
        \begin{equation}\label{eq:condition_u2}
            U_i^{(2)} \neq B([n-1])
        \end{equation}
        and
        \begin{equation}\label{eq:condition_u1}
            U_i^{(1)} \neq \frac{B([k])}{B([n-1])} \text{ for all } k \in [n-2].
        \end{equation}
        Since $U_i$ is uniformly distributed on $[0,1] \times [0,1]$ and independent of $B$, conditions \eqref{eq:condition_u2} and \eqref{eq:condition_u1} hold together almost surely which proves the claim.
    \item Follows by definition.
    \item Follows by definition.
    \item\label{item:x} Note that for any fixed $b \in \cM_1 ([n])$, by independence of $U_i^{(1)}$ and $U_i^{(2)}$
    \begin{align*}
        &\IP \big(\underline{v}(U_i,b,d) = k\big)\\[0.7em]
        &= \IP \Big( U_i^{(2)} \leq \tfrac{\lfloor b (\{k\}) \cdot 2^d \rfloor b([n-1])}{2^d b(\{k\})} \text{ and } U_i^{(1)} \in \big[\nicefrac{b([k-1])}{b([n-1])}, \nicefrac{b([k])}{b([n-1])}\big)  \Big)\\[0.7em]
        &= \frac{\lfloor b (\{k\}) \cdot 2^d \rfloor b([n-1])}{2^d b(\{k\})} \cdot \frac{b(\{k\})}{b([n-1])} = \frac{\lfloor b(\{ k \}) \cdot 2^d \rfloor}{2^d}
    \end{align*}
    If $b(\{k\}) \in \big[\tfrac{\beta_k}{2^d}, \tfrac{\beta_k + 1}{2^d} \big)$ with $\beta_k \in \{0,...,2^d - 1\}$ and $k \in [n-1]$, then
    \begin{equation*}
        \IP \big(\underline{v}(U_i,b,d) = k\big) = \frac{\beta_k}{2^d}.
    \end{equation*}
    Independence in $i$ follows from that of the $(U_i)$.
    \item Works analogously to \ref{item:x}.  
    \item For any $\delta > 0$ define the event
    \begin{equation*}
        A_\delta \defeq \big\{ B (\{k\}) \geq \delta \text{ for all } k \in [n]\big\}.
    \end{equation*}
    Then it holds that
    \begin{align*}
        \IP \big(B \in \cB^{(d)} \big) &= 1 - \IP \big(B \notin \cB^{(d)} \big)\\
        &\geq 1 - \IP \big( \bar{A}_\delta \big) - \sum_{k=1}^{n-1} \IP \Big( \big\{ \tfrac{\lceil B (\{k\}) \cdot 2^d \rceil B([n-1])}{2^d B(\{k\})} > 1 \big\} \cap A_\delta\Big)\\
        &\geq 1 - \IP \big( \bar{A}_\delta \big) - \underbrace{\sum_{k=1}^{n-1} \IP \Big( (1 + \tfrac{1}{2^d \delta}) \cdot (1-\delta) > 1\Big)}_{\stackrel{d \to \infty}{\rightarrow} 0 }.
    \end{align*}
    Since $B$ has a Lebesgue density it holds that
    \begin{equation*}
        \IP \big( \bar{A}_\delta \big) \stackrel{\delta \to 0}{\rightarrow} 0.
    \end{equation*}
    Hence, $\lim_{d \to \infty} \IP \big(B \in \cB^{(d)}\big) = 1$.
    Note that the proof also works under weaker assumptions than the existence of a Lebesgue density.
    \end{enumerate}
\end{proof}

\begin{figure}[h!]
\centering
  \begin{tikzpicture}
    \draw[->] (-0.1,0) -- (5.2,0) node[right] {$u^{(1)}$};
    \draw[->] (0,-0.1) -- (0,5.2) node[above] {$u^{(2)}$};

    \draw[-] (3,3.5) -- (3,-0.15) node[below] {$\frac{b ([1])}{b([2])}$};
    \draw[-] (5, 3.5) -- (-0.15,3.5) node[left] {$\scriptstyle b([2])$};
    
    \fill[pattern=north west lines, draw=none] (0,0) rectangle (3,3.5);  
    \fill[pattern=north east lines, draw=none] (3,0) rectangle (5,3.5);
    \fill[pattern=vertical lines, draw=none] (0,3.5) rectangle (5,5);
      
    \draw[thick] (0,0) rectangle (5,5);
    
    \node[fill=white, inner sep=2pt] at (1.5,1.75) {$1$};
    \node[fill=white, inner sep=2pt] at (4,1.75) {$2$};
    \node[fill=white, inner sep=2pt] at (2.5,4.25) {$3$};

  \begin{scope}[shift={(7.3,0)}]
    \draw[->] (-0.1,0) -- (5.2,0) node[right] {$u^{(1)}$};
    \draw[->] (0,-0.1) -- (0,5.2) node[above] {$u^{(2)}$};

    \draw[-] (3,0.15) -- (3,-0.15) node[below] {$\frac{b ([1])}{b([2])}$};
    
    \draw[-] (0.15, 3) -- (-0.15,3) node[left, xshift=5pt, yshift=2pt] {$\frac{\lfloor b (\{1\}) 2^d \rfloor b([2])}{2^d b(\{1\})}$};
    \draw[-] (0.1, 2.4) -- (-0.1,2.4)node[left, xshift=5pt, yshift=-2pt]  {  $\tfrac{\lfloor b (\{2\}) \cdot 2^d \rfloor b([2])}{2^d b(\{2\})}$};

    \fill[pattern=north west lines,draw=black] (0,0) rectangle (3,3);  
    \fill[pattern=north east lines, draw=black] (3,0) rectangle (5,2.4);
    \fill[pattern=vertical lines] (0,3) rectangle (5,5);
    \fill[pattern=vertical lines] (3,2.4) rectangle (5,3);
      
    \draw[thick] (0,0) rectangle (5,5);
    
    \node[fill=white, inner sep=2pt] at (1.5,1.75) {$1$};
    \node[fill=white, inner sep=2pt] at (4,1.75) {$2$};
    \node[fill=white, inner sep=2pt] at (2.5,4.25) {$3$};
  \end{scope}

\begin{scope}[shift={(1,-5)}]
\draw[thick] (0,0) rectangle (3,3);
\fill[pattern=north west lines,draw=black] (0,0) rectangle (0.7,2);
\fill[pattern=north east lines,draw=black] (0.7,0) rectangle (1.7,2);
\fill[pattern=grid,draw=black] (1.7,0) rectangle (2,2);
\fill[pattern=horizontal lines,draw=black] (2,0) rectangle (3,2);
\fill[pattern=vertical lines,draw=black] (0,2) rectangle (3,3);
\end{scope}

\begin{scope}[shift={(8.3,-5)}]
\draw[thick] (0,0) rectangle (3,3);
\fill[pattern=north west lines,draw=black] (0,0) rectangle (0.7,1.6);
\fill[pattern=north east lines,draw=black] (0.7,0) rectangle (1.7,1.9);
\fill[pattern=grid,draw=black] (1.7,0) rectangle (2,1.5);
\fill[pattern=horizontal lines,draw=black] (2,0) rectangle (3,1.8);

\fill[pattern=vertical lines] (0,1.6) rectangle (0.7,3);
\fill[pattern=vertical lines] (0.7,1.9) rectangle (1.7,3);
\fill[pattern=vertical lines] (1.7,1.5) rectangle (2,3);
\fill[pattern=vertical lines] (2,1.8) rectangle (3,3);

\end{scope}

\end{tikzpicture}
    \caption{\textbf{Top:} illustration of the functions with $n=3$. Left: the function $v(u,b)$. Right: the corresponding $\underline{v}(u,b,d)$.
    \textbf{Bottom:} Left: sketch of an example of $v$ for $n=5$. Right: corresponding ``rounding down'' function $\underline{v}$.}
\end{figure}
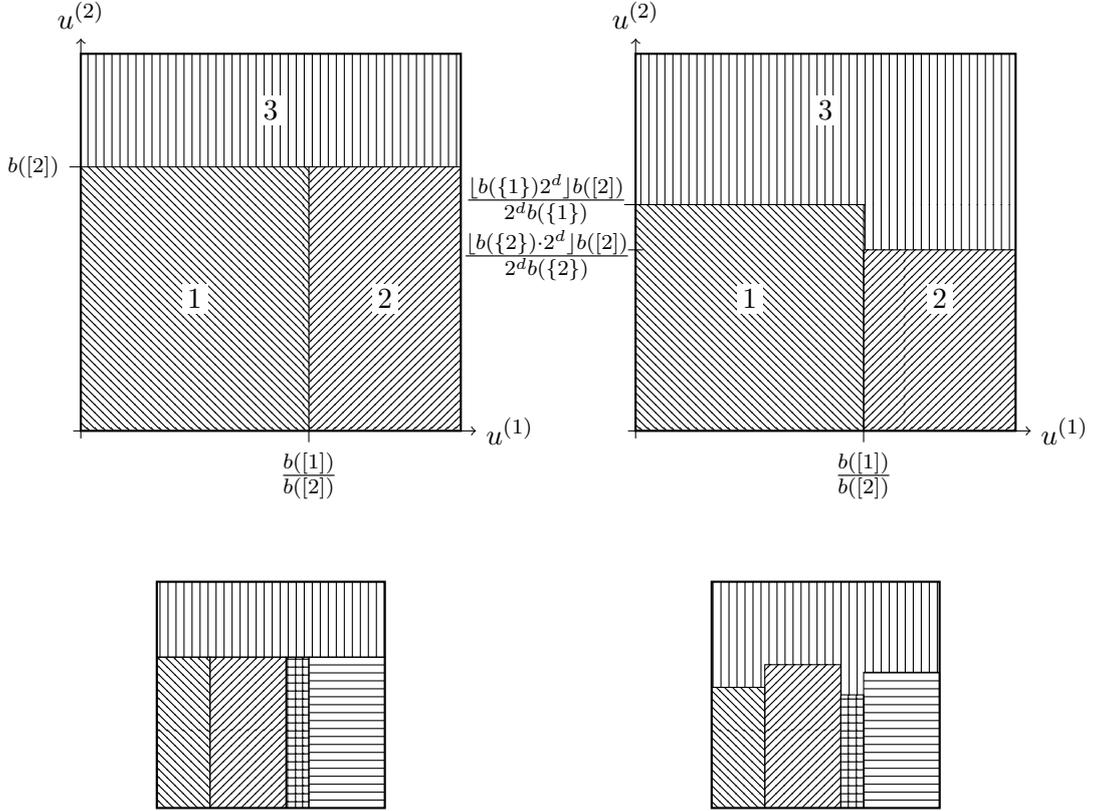

\subsection{Number of Families in the Hoppe's Urn}
\begin{lemma}\label{lem:hoppe_fam_sizes}
    Let $(\bfZ^N(t))_{t \geq 0}$ be a reversed Hoppe urn with parameters $(r, \theta_N)$ and $\jN$ initial particles where $\theta_N \to \theta$ as $N \to \infty$.
    Let $H^N = |\bfZ_\infty^{N,2}|$ be the size of the second component at absorption.
    Then
    \begin{align*}
        \IE [H^N&] = \sum_{i=1}^{\jN} \frac{\theta_N}{\theta_N+i-1}\\
        \textnormal{Var} [ H^N&] = \sum_{i=1}^{\jN} \frac{\theta_N}{\theta_N+i-1} \cdot \frac{i-1}{\theta_N + i -1}.
    \end{align*}
    Furthermore, for any $\alpha>0$
    \begin{equation}\label{eq:appendix_bound_one}
        (1-\alpha)\theta \log (\jN) \leq \liminf_{N \to \infty} \IE[H^N] \leq \limsup_{N \to \infty} \IE[H^N] \leq (1+\alpha)\theta \log (\jN) \tag{i}
    \end{equation}
    and
    \begin{equation*}\label{eq:appendix_bound_two}
        \limsup_{N \to \infty} \textnormal{Var} [H^N] \leq (1+\alpha)\theta \log (\jN). \tag{ii}
    \end{equation*}
\end{lemma}
\begin{proof}
    $H^N$ arises as the sum of $\jN$ independent Bernoulli random variables with success probabilities $(p_i^N)_{i \in [\jN]}$, $p_i^N = \frac{\theta_N}{\theta_N+\jN-i}$.
    To see this, let $(t_i)_{i=1,...j}$ be the jump times of $(\bfZ^N(t))_{t \geq 0}$.
    From Remark \ref{rem:hoppe} it follows that $H^N =|\bfZ_\infty^{N,2}|=|\bfZ^{N,2} (t_{\jN})|$.
    At the $i$-th jump, the value of $|\bfZ^{N,2}|$ increases by $1$ with probability $\frac{\theta_N}{\theta_N+\jN-i}$ or it stays constant, independently of everything else.
    By independence and the expectation and variance of Bernoulli random variables, one has
    \begin{equation*}
        \IE [H^N] = \sum_{i=1}^{\jN} p_i = \sum_{i=1}^{\jN} \frac{\theta_N}{\theta_N+\jN-i} = \sum_{i=1}^{\jN} \frac{\theta_N}{\theta_N+i-1}
    \end{equation*}
    and similarly
    \begin{equation*}
        \textnormal{Var} [H^N]
            = \sum_{i=1}^{\jN} p_i (1-p_i)
            = \sum_{i=1}^{\jN} \frac{\theta_N}{\theta_N+\jN-i}\frac{\jN-i}{\theta_N+\jN-i}
            = \sum_{i=1}^{\jN} \frac{\theta_N}{\theta_N+i-1} \cdot \frac{i-1}{\theta_N + i -1}.
    \end{equation*}
    For the asymptotic upper bounds, note that
    \begin{align*}
        \IE [H^N] = \sum_{i=0}^{\jN-1} &\frac{\theta_N}{\theta_N+i} = 1 + \sum_{i=0}^{\jN-2} \frac{\theta_N}{\theta_N+1+i}\\
        &\leq 1 + \theta_N\sum_{i=0}^{\jN-2} \frac{1}{1+i} \leq  1 + \theta_N \big(1+\log (\jN)\big)\\
        &= \Big(\frac{1 + \theta_N}{\log(\jN)} + \theta_N \Big)\log(\jN).
    \end{align*}
    Since $\jN \stackrel{N \to \infty}{\rightarrow} \infty$ and $\theta_N \stackrel{N \to \infty}{\rightarrow} \theta$, we have for $N$ large enough that $\frac{1 + \theta_N}{\log (\jN)} \leq \frac{\alpha}{2}$ and $\theta_N \leq (1+\frac{\alpha}{2})\theta$.
    Thus,
    \begin{equation*}
        \IE [H^N] \leq (1+\alpha)\theta \log(\jN)
    \end{equation*}
    for $N$ large enough, which proves the last inequality in \eqref{eq:appendix_bound_one}.
    The first inequality follows from a similar calculation.
    The Inequality \eqref{eq:appendix_bound_two} then follows immediately since $\textnormal{Var} [H^N] \leq \IE [H^N]$ for all $N \in \IN$.
\end{proof}

\subsection{Rate Calculations for Phase 2}

\subsubsection{Lower Bound on the Probability of an Unsuccessful Recombination}

Recall that we denoted by  $\beta_N(i, \xi)$ the probability for an unsuccessful recombination event when at the beginning of Phase 2 $\xi^N$ ancestral lines remained and already $i$ many of these lines where split up into two due to recombination. We have the following lower bound by $\beta_N(i, \xi)$: 

\begin{lemma}\label{lem:UnsucRec}
Given $\xi^N \in \Theta(\log(\jN))$ the probability $\beta_N(i, \xi)$ is lower bounded by 
$\frac{1}{1+\theta}- \varepsilon_N$, where $\varepsilon_N \searrow 0$. 
\end{lemma}

\begin{proof}
For shorter notation we write $\xi = \xi^N$.
Denote by 
\begin{align*}
    \beta_N &= \min_{i \in [\xi-1]} \beta_N(\xi, i)= \min_{i \in [\xi-1]} \IP \bigg(\Big(\svec{\xi-i}{0}{2i}\Big)^{\textit{T}} \to \Big(\svec{\xi-i-1}{1}{2i}\Big)^{\textit{T}}\bigg) \IP \bigg(\Big(\svec{\xi-i-1}{1}{2i}\Big)^{\textit{T}} \to \Big(\svec{\xi-i}{0}{2i}\Big)^{\textit{T}}\bigg)\\[1em]
    &\geq \underbrace{\min_{i \in [\xi-1]} \IP \bigg(\Big(\svec{\xi-i}{0}{2i}\Big)^{\textit{T}} \to \Big(\svec{\xi-i-1}{1}{2i}\Big)^{\textit{T}}\bigg)}_{\beta'_N} \underbrace{\min_{i \in [\xi-1]} \IP \bigg(\Big(\svec{\xi-i-1}{1}{2i}\Big)^{\textit{T}} \to \Big(\svec{\xi-i}{0}{2i}\Big)^{\textit{T}}\bigg)}_{\beta''_N},
\end{align*}
where, e.g.,
\begin{equation*}
\IP \bigg(\Big(\svec{\xi-i}{0}{2i}\bigg)^{\textit{T}} \to \Big(\svec{\xi-i-1}{1}{2i}\Big)^{\textit{T}}\Big)    
\end{equation*}
denotes the probability for a transition of the process $\big( H(s) \big)_{s\geq 0}$ from  $\Big(\svec{\xi-i}{0}{2i}\Big)^{\textit{T}}$ to $\Big(\svec{\xi-i-1}{1}{2i}\Big)^{\textit{T}}$.
We analyze the two factors separately.
For $\beta'_N$ we calculate 
\begin{align*}
        \IP &\bigg(\Big(\svec{\xi-i}{0}{2i}\Big)^{\textit{T}} \to \Big(\svec{\xi-i-1}{1}{2i}\Big)^{\textit{T}}\bigg) = \frac{(\xi - i) \frac{\rho_N}{N^2}(N-1)}{r\bigg( \Big(\svec{\xi-i}{0}{2i} \Big)^{\textit{T}}\bigg)}\\[1em]
    &= \frac{(\xi - i) \frac{\rho_N}{N^2}(N-1)}{(\xi - i) \frac{\rho_N}{N^2}(N-1)
    + (\xi+i)\frac{\mu_N}{N}
    + (\xi+i)^2 \frac{\lambda_N}{N^2 M}
    + \frac{(\xi + i)^2}{M}}\\[1em]
    &= \frac{1}{1
    + \frac{(\xi+i)\frac{\mu_N}{N}}{(\xi - i) \frac{\rho_N}{N^2}(N-1)}
    + \frac{(\xi+i)^2 \frac{\lambda_N}{N^2 M}}{(\xi - i) \frac{\rho_N}{N^2}(N-1)}
    + \frac{\frac{(\xi + i)^2}{M}}{(\xi - i) \frac{\rho_N}{N^2}(N-1)}}\\[1em]
    &\geq \frac{1}{1
    + \frac{2\xi\mu_N}{\frac{\rho_N}{N}(N-1)}
    + \frac{4\xi^2 \lambda_N}{\rho_N(N-1)M}
    + \frac{8\xi^2 N}{\rho_N M}}\\[1em]
    &\geq \frac{1}{1
    + \frac{4\xi\mu_N}{\rho_N}
    + \frac{8\xi^2 \lambda_N}{\rho_N M}
    + \frac{8\xi^2 N}{\rho_N M}}
    \geq 1
    - \varepsilon_N^{(1)} 
\end{align*}
with $\varepsilon^{(1)}_N \defeq \Big(\frac{4\xi\mu_N}{\rho_N}
    + \frac{8\xi^2 \lambda_N}{\rho_N M}
    + \frac{8\xi^2 N}{\rho_N M} \Big)\searrow 0$, where we used Lemma \ref{assumptions:rates},  and $\xi+i \leq 2\xi$ as well as $\xi-i \geq 1$.

Now we turn to $\beta''_N$. It holds that
\begin{align*}
    &\IP \bigg(\Big(\svec{\xi-i-1}{1}{2i}\Big)^{\textit{T}} \to \Big(\svec{\xi-i}{0}{2i}\Big)^{\textit{T}}\bigg) = \frac{2 \frac{\gamma_N}{N}}{r\bigg( \Big(\svec{\xi-i-1}{1}{2i} \Big)^{\textit{T}} \bigg)}\\
    &\geq \frac{2 \frac{\gamma_N}{N}}{\xi \frac{\rho_N}{N} + \frac{\lambda_N}{N} + 2 \frac{\gamma_N}{N}
        + 2\xi \frac{\mu_N}{N} + 4 \xi^2\frac{\lambda_N}{N^2 M} + 1}
    =\frac{1}
    {1
    + \frac{\xi \rho_N}{2 \gamma_N}
    + \frac{\lambda_N}{2 \gamma_N}
    + \frac{\xi \mu_N}{\gamma_N}
    + \frac{4 \xi^2 \lambda_N}{2 N M \gamma_N}
    + \frac{N}{2 \gamma_N}}\\[1em]
 &\geq \frac{1}{1+\theta} - \varepsilon^{(2)}_N
\end{align*}
for a sequence $\varepsilon^{(2)}_N\searrow 0 .$ 
Since $\varepsilon^{(1)}_N, \varepsilon^{(2)}_N \searrow 0$, we find a sequence $\varepsilon_N \searrow 0$, which fulfills the claim of the Lemma. 

\end{proof}

\textbf{Acknowledgement}
The authors were supported by the German Research Foundation (DFG), Project ID: 443227151, and by the Johanna Quandt Research Foundation.

\bibliographystyle{siam}
\bibliography{references} 
\end{document}